\documentclass[a4paper,fleqn]{cas-sc}

\usepackage[authoryear,longnamesfirst]{natbib}

\usepackage{xltabular}
\usepackage{placeins}
\usepackage{float}
\usepackage[most]{tcolorbox}

\newtcbtheorem{Summary}{\bfseries Summary of RQ}{
    enhanced,
    breakable,
    coltitle=black,
    top=0.14in,
    before skip=6pt,
    after skip=6pt,
    attach boxed title to top left={
        xshift=1.5em,
        yshift=-\tcboxedtitleheight/2
    },
    boxed title style={
        size=small,
        colback=lightgray
    }
}{summary}

\newtcolorbox{SummaryBox}[1]{
  enhanced,
  breakable,
  title=\bfseries #1,
  coltitle=black,
  top=0.1in,
  attach boxed title to top left={xshift=1.5em,yshift=-\tcboxedtitleheight/2},
  boxed title style={size=small,colback=lightgray}
}

\begin{document}
\let\WriteBookmarks\relax
\def\floatpagepagefraction{1}
\def\textpagefraction{.001}

\shorttitle{Challenges and Opportunities of Generative AI Apps}    

\shortauthors{AlMulla et al.}  

\title [mode = title]{Understanding the Challenges and Opportunities of Generative AI Apps: An Empirical Study}  

%

\author[1]{Buthayna AlMulla}

\cormark[1]

\fnmark[1]

\ead{buthayna.almulla@mail.utoronto.ca}

\credit{Conceptualization, Data curation, Formal analysis, Investigation, Methodology, Software, Validation, Visualization, Writing – original draft, Writing – review and editing}

\affiliation[1]{organization={University of Toronto},
            addressline={140 St George St}, 
            city={Toronto},
            postcode={M5S 3G6}, 
            state={ON},
            country={Canada}}

\author[2]{Maram Assi}

\fnmark[2]

\ead{assi.maram@uqam.ca}

\credit{Conceptualization, Methodology, Validation, Writing – original draft, Writing – review and editing}

\affiliation[2]{organization={Université du Québec à Montréal},
            addressline={405 Rue Sainte-Catherine Est}, 
            city={Montréal},
            postcode={H2L 2C4}, 
            state={QC},
            country={Canada}}

\author[3]{Safwat Hassan}

\fnmark[3]

\ead{safwat.hassan@utoronto.ca}

\ead[url]{https://safwathassan.com/}

\credit{Conceptualization, Funding acquisition, Methodology, Project administration, Supervision, Writing – original draft, Writing – review and editing}

\affiliation[3]{organization={University of Toronto},
            addressline={140 St George St}, 
            city={Toronto},
            postcode={M5S 3G6}, 
            state={ON},
            country={Canada}}
\cortext[1]{Corresponding author}

\fntext[1]{}


\begin{abstract}
Generative AI \textit{(Gen-AI)} is increasingly integrated into mobile applications \textit{(apps)}, introducing new capabilities while also creating new challenges for users. However, despite their growing adoption, we lack an ecosystem-level understanding of the experiences, opportunities, and challenges users report across Gen-AI mobile apps. We conduct a user-centered analysis of 1,035,342 reviews from 171 Gen-AI apps from the Google Play Store. We propose \textit{SARA} (Selection, Acquisition, Refinement, and Analysis), a four-phase framework that leverages prompt-based LLMs for large-scale review analysis. We validate the reliability of LLM-based topic extraction and assignment using 4,353 manually evaluated reviews, achieving 91\% accuracy with five-shot prompting and filtering of non-informative reviews. We identify the top ten topics (e.g., \textit{AI Performance} and \textit{Emotional Connection}) and perform a cross-platform comparison with Apple App Store reviews. Through qualitative analysis of 762 reviews, we uncover three opportunities (\textit{AI for Accessibility and Wellbeing, AI as a Collaborative Creative Tool, and AI Versatility}) and three challenges (\textit{Managing User Expectations and AI Limitations, Balancing Content Moderation and Creative Freedom, and Strategic Integration of Gen-AI Features}). Finally, we analyze temporal trends, revealing how the topics discussed and their evaluations change over time, including changing concerns around emotional connection and content moderation.

\end{abstract}


\begin{highlights}
\item We introduce SARA, a validated LLM-based framework for large-scale mobile app review analysis. 
\item Few-shot prompting and review filtering improve topic analysis, achieving 91\% accuracy.
\item Analysis of 1,035,342 reviews from 171 Gen-AI apps reveals recurring opportunities and challenges.
\item Temporal analysis reveals changing concerns around emotional dependency and content moderation.
\item Findings inform the design, requirements, evaluation, and evolution of Gen-AI mobile apps.
\end{highlights}

\begin{keywords}
Generative AI \sep
Mobile apps \sep
App review mining \sep
Large language models \sep
Topic analysis \sep
Software engineering
\end{keywords}

\maketitle

\section{Introduction}
\label{sec:introduction}
Generative Artificial Intelligence (Gen-AI) has become increasingly integrated into mobile applications \textit{(apps)}. Following the release of ChatGPT\footnote{\url{https://play.google.com/store/apps/details?id=com.openai.chatgpt\&hl=en}}, Gen-AI mobile apps have experienced substantial growth, with Artificial Intelligence \textit{(AI)} chatbot downloads increasing rapidly~\citep{Ceci2025}. Gen-AI refers to a class of AI that generates new and meaningful content, including text, images, audio, and video~\citep{Feuerriegel_Hartmann_Janiesch_Zschech_2024,10628027}.  Unlike traditional AI models, which rely on historical data to perform predictions and analytical tasks, Gen-AI models synthesize information from diverse sources to create original outputs~\citep{gozalobrizuela2023surveygenerativeaiapplications}. Developers are increasingly integrating Gen-AI models into mobile apps to support capabilities such as conversational interfaces, voice interaction, and content generation~\citep{Hau_Hassan_Zhou_2025}. When embedded into apps, these models can provide personalized and adaptive experiences through real-time, context-aware interactions~\citep{Yuen_Schlote_2024}. As Gen-AI becomes increasingly integrated into mobile apps, understanding the user perspective is important for the continued development and evolution of these apps~\citep{Chen_Liu_Zhou_Zhao_Wang_Wang_Chen_Bissyande_Klein_Li_2024}. User feedback can provide developers with post-deployment evidence about how Gen-AI capabilities perform in practice, where user expectations are not met, and what new requirements and concerns emerge through real-world use.

\begin{figure}
  \centering \includegraphics[width=\textwidth]{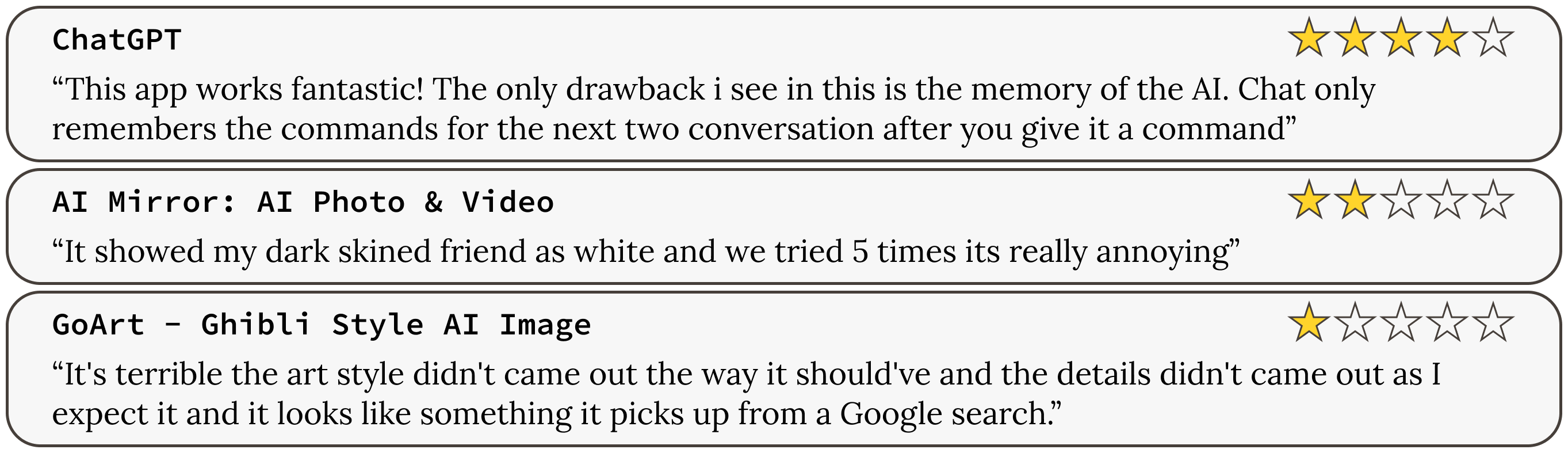}
    \caption{Examples of reviews of Gen-AI apps.}
    \label{fig:examples}
\end{figure}

Existing studies on user perspectives of Gen-AI tend to focus on specific domains, such as education~\citep{Golding_Lippert_Neuschatz_Salomon_Burke_2024,Kim_Yu_Detrick_Li_2025, Lee_Arnold_Srivastava_Plastow_Strelan_Ploeckl_Lekkas_Palmer_2024,Shata_Hartley_2025,Kim_Klopfer_Grohs_Eldardiry_Weichert_Cox_Pike_2025,Yuen_Schlote_2024,Storm-Mathisen}, creative image generation~\citep{Tang_Zhang_Ciancia_Wang_2024,haase2023art, Oppenlaender_Silvennoinen_Paananen_Visuri_2023}, or mental health~\citep{Jin_Kim_Han_2025,Heinz_Mackin_Trudeau_Bhattacharya_Wang_Banta_Jewett_Salzhauer_Griffin_Jacobson_2025}. These studies provide valuable insights into users' needs, expectations, perceived 
benefits, and concerns when interacting with Gen-AI within these contexts. However, 
their domain-specific focus does not provide an ecosystem view of user experiences 
across different Gen-AI application domains. Moreover, much of this work relies on 
surveys and interviews, which enable researchers to directly investigate users' 
experiences and reasoning but are shaped by the questions posed by researchers. 
User reviews, as shown in Figure~\ref{fig:examples}, provide a complementary source of naturally occurring feedback,
offering insight into the experiences, expectations, benefits, and concerns
that users report through their real-world use of deployed
applications~\citep{Haque2023}. 

User reviews have also been used to examine user experiences with Gen-AI and AI-powered mobile apps. However, many of these studies are limited in analytical or empirical scope. For example,~\citet{Alabduljabbar_2024} examines usability across 11,549 reviews from five Gen-AI apps, while~\citet{MENG2026104560} investigate sentiment across 100,010 reviews from nine Gen-AI apps. Other studies analyze larger collections of up to 108 AI-powered apps but focus on specific user concerns, such as user-reported hallucinations~\citep{Massenon_Gambo_Khan_Agbonkhese_Alwadain_2025} or fairness~\citep{RezaeiNasab_Dashti_Shahin_Zahedi_Khalajzadeh_Arora_Liang_2025}. These studies provide detailed evidence about particular dimensions of users' experiences, but do not aim to characterize the broader range of topics discussed by users across Gen-AI mobile apps.

Recent studies suggest that users' experiences with Gen-AI continue to develop after initial adoption. Early users have engaged with Gen-AI through playful and exploratory practices, such as roleplaying and jailbreaking~\citep{Heuser2024}. Continued interaction can also change the role Gen-AI plays in users' lives. For example, \citet{Xie2026} found that users initially engaged with ChatGPT instrumentally as a tool for completing tasks, but over time began to use it for emotional support, companionship, and intimate relationships. These studies demonstrate that experiences with Gen-AI can change as the technology becomes incorporated into everyday life. However, less is known about how the topics and concerns expressed in user feedback on Gen-AI mobile apps change over time. Examining reviews across time can reveal shifts in what users discuss and how different aspects of Gen-AI apps are evaluated as these technologies evolve.

Our work addresses these gaps by taking a broad, user-centered view of the 
Gen-AI mobile app ecosystem. We analyze naturally occurring user reviews across 
a wide range of Gen-AI mobile app domains to characterize the topics 
users discuss without restricting the analysis to a predefined concern. We identify the topics most frequently discussed in user reviews, examine the
opportunities and challenges reflected in users' reported feedback, and analyze
how these topics differ across app categories and evolve over time. These findings provide a broader understanding
of where Gen-AI mobile apps provide value and where users encounter recurring
challenges, helping inform the design, evaluation, and continued development
of user-centered Gen-AI applications. Based on these findings, we derive
implications for the following stakeholders: users, platform owners, developers, testing tool designers, requirements engineers, and policymakers and regulators.

To support this empirical analysis at scale, we also make a methodological
contribution. Recent work has demonstrated the potential of LLMs for
context-aware topic extraction and assignment from unstructured text, including
user reviews~\citep{Pham_Hoyle_Sun_Resnik_Iyyer_2024,Maram_LLM_Cure}.
Building on this work, we propose a structured four-phase framework called
\textbf{S}election, \textbf{A}cquisition, \textbf{R}efinement, and
\textbf{A}nalysis (\textbf{\textit{SARA}}). The framework identifies relevant
Gen-AI apps (\textit{Selection}), collects their user reviews
(\textit{Acquisition}), refines the review data to improve its quality for
analysis (\textit{Refinement}), and applies prompt-based LLM techniques to
analyze the resulting corpus (\textit{Analysis}). While validated in the
context of Gen-AI apps, \textit{SARA}'s modular design makes it adaptable to
user-review analysis across other mobile app domains.

Using \textit{SARA}, we automatically label and quantitatively analyze
1,035,342 user reviews from 171 Gen-AI apps, complemented by qualitative
analyses of statistically representative samples. We organize our analysis
around three research questions (RQs). We first evaluate the
accuracy and reliability of our LLM-based approach for extracting and assigning
topics (RQ1). We then apply the framework to identify the topics most frequently
discussed by users and characterize the opportunities and challenges reflected
in their feedback (RQ2), before examining how these topics evolve over time
(RQ3). The three RQs and their key findings are summarized below:

\vspace{4pt}
\noindent \textbf{RQ1: How accurately can LLMs identify topics in user reviews of Gen-AI apps?} 

\noindent{We investigate how prompt configuration (e.g., using 0-shot, 3-shot, and 5-shot) and input quality (e.g., the proportion of informative reviews) affect the performance of LLMs in extracting topics from user reviews.
Our results show that increasing the number of few-shot examples and filtering non-informative reviews significantly improves accuracy. The best-performing setup achieves an overall accuracy of 91\%, confirming the reliability of our LLM framework for downstream analysis.}

\noindent \textbf{RQ2: What are the most prominent topics discussed in user reviews?} 

\noindent{We identify and quantitatively analyze the 10 most frequently discussed topics in Gen-AI app reviews, compare their prevalence across app categories, and qualitatively examine the user feedback associated with these topics to identify key opportunities and challenges. Our findings reveal three key opportunities: AI for Accessibility and Wellbeing, AI as a Collaborative Creative Tool, and AI Versatility, and three challenges: Managing User Expectations and AI Limitations, Balancing Content Moderation and Creative Freedom, and Strategic Integration of Gen-AI Features.}

\noindent \textbf{RQ3: What temporal trends emerge in user feedback on Gen-AI topics?}

\noindent{We examine how Gen-AI topics change over time by clustering topics
with similar trends in how frequently they are discussed and how they are rated,
followed by qualitative analysis of two contrasting topics. Our findings show
that discussions of \textit{Emotional Connection} shift from companionship
toward dependency-related concerns, while discussions of \textit{Content Policy
\& Censorship} shift from privacy toward competing expectations around content
restriction.}\\

\noindent Our study provides insights for stakeholders across the Gen-AI mobile app ecosystem by examining how users experience and evaluate Gen-AI features in mobile apps. Specifically, our contributions are as follows:

\begin{itemize}

\item We propose a structured four-phase framework, \textit{SARA}, that integrates established techniques in user review analysis with LLM-based topic extraction and assignment to support large-scale analysis of mobile app
reviews.

\item We evaluate the effectiveness of LLMs for topic extraction and assignment
of user reviews, demonstrating that increasing the number of labeled examples
and filtering non-informative reviews improves accuracy. Combining a 5-shot
prompt with LLM-based review filtering achieves the highest accuracy of 91\%,
establishing the configuration used for our subsequent large-scale analysis.

\item We conduct a large-scale empirical analysis of 1,035,342 user reviews
from 171 Gen-AI apps, identifying the 10 most frequently discussed topics and
examining their prevalence across app categories. Through qualitative analysis
of the associated user feedback, we identify three key opportunities and three
challenges surrounding Gen-AI mobile apps.

\item We conduct a temporal analysis of user reviews to examine how Gen-AI
topics evolve in their percentage of reviews and average ratings over time,
revealing changing discussions around \textit{Emotional Connection} and
\textit{Content Policy \& Censorship}.

\item We examine the generalizability and practical relevance of our findings
through a cross-platform comparison with the Apple App Store and an analysis
of whether and how developers respond to concerns raised in user reviews.

\item We translate our empirical findings into actionable implications for
stakeholders across the Gen-AI mobile app ecosystem.

\item We provide a replication package containing the Gen-AI app dataset,
manually labeled evaluation samples, and the code used to conduct the
prompt-based experiments presented in this paper.

\end{itemize}

\noindent We organize the remainder of this paper as follows. Section~\ref{sec:related_work} illustrates existing work on analyzing user reviews. Section~\ref{sec:methodology} presents our methodology. Sections~\ref{sec:RQ1}-\ref{sec:RQ3} describe the findings obtained for each RQ. 
Section~\ref{sec:discussion} presents a platform comparison and developers' response analysis. Section~\ref{sec:implications} synthesizes our key findings and frames them into implications and actionable insights.
Section~\ref{sec:threats_to_validty} discusses potential threats to the validity of our findings. 
Finally, Section~\ref{sec:conclusion} concludes the paper and discusses implications for future research.

\section{Related Work}
\label{sec:related_work}
In our literature review, we first examine \textbf{user perspectives of Gen-AI
apps} to understand what is currently known about the benefits and concerns
users report when interacting with Gen-AI. We then examine \textbf{user review
analysis techniques}, focusing on methods for extracting and analyzing these
perspectives from large-scale user reviews. Together, these areas position the
empirical and methodological contributions of our study.

\subsection{User Perspective of Gen-AI Apps}
Prior research has examined how users perceive and interact with Gen-AI across
a range of application contexts. In this section, we synthesize findings on the benefits and concerns users report across education, creative image generation,
and conversational and companion AI to establish what is currently known about the user perspective of Gen-AI.

\noindent \textbf{Education.} Research in education has identified both perceived benefits and concerns
surrounding Gen-AI use. Studies have examined users' perceptions of Gen-AI
for learning~\citep{Golding_Lippert_Neuschatz_Salomon_Burke_2024,Kim_Yu_Detrick_Li_2025, Lee_Arnold_Srivastava_Plastow_Strelan_Ploeckl_Lekkas_Palmer_2024}.~\citet{Shata_Hartley_2025} identified trust as the key driver of Gen-AI adoption.~\citet{Kim_Klopfer_Grohs_Eldardiry_Weichert_Cox_Pike_2025} found that both
students and faculty recognized the potential benefits of Gen-AI for academic
performance, while also expressing concerns about its negative effects on
learning. In the context of language learning, \citet{Yuen_Schlote_2024} found
interest in AI-supported conversational and social interactions, illustrating
the potential of Gen-AI to support interactive learning experiences.

\noindent \textbf{Creative image generation.} Studies of Gen-AI image tools similarly identify both opportunities and
challenges from the user perspective. Gen-AI image tools have been found to
support creative ideation, with users frequently relying on them during early
stages of the creative process~\citep{Tang_Zhang_Ciancia_Wang_2024,haase2023art}.
At the same time, users have raised concerns about broader risks of generated
content. For example, participants in
\citet{Oppenlaender_Silvennoinen_Paananen_Visuri_2023}  expressed concerns about
misinformation and deepfakes, job displacement, copyright infringement, and
the potential loss of human creativity and appreciation for artists.

\noindent \textbf{Conversational and companion AI.} Research on conversational and companion AI has identified both benefits and
challenges associated with the social and emotional roles of Gen-AI.
\citet{Heinz_Mackin_Trudeau_Bhattacharya_Wang_Banta_Jewett_Salzhauer_Griffin_Jacobson_2025}
studied the efficacy of a mental health AI chatbot and found that it
significantly reduced users' symptoms, maintained high engagement, and achieved
user acceptance, with a reported therapeutic alliance comparable to that of
human therapists. Beyond mental health contexts, continued interaction with
Gen-AI can develop from instrumental use into emotional support, companionship,
and intimate relationships~\citep{Xie2026,Huang2025}. These interactions may also
influence users' offline relationships. \citet{Huang2025}, for example, found
that some users reported increased confidence and agency and applied
communication practices developed with AI companions to their relationships
with others. \citet{Malfacini_2025} identifies both potential benefits and risks of companion
AI for human relationships. On the one hand, companion AI may support social
upskilling and motivation, with prior evidence suggesting greater openness,
social confidence, and improvements in human relationships. On the other hand,
it may contribute to social deskilling and demotivation by reducing motivation
to pursue human relationships, creating unrealistic expectations of human
companionship, or encouraging emotional over-reliance.\\

Collectively, these studies demonstrate that users experience both opportunities
and challenges when interacting with Gen-AI. Across different contexts, users
report benefits related to learning, creativity, emotional support, and social
interaction, alongside concerns related to trust, misinformation, and the
social consequences of continued Gen-AI use. However, much of this research
examines user experiences within specific application domains and commonly
relies on surveys and interviews. As a result, less is known about the broader
range of opportunities and challenges that users report through their
day-to-day interactions with Gen-AI mobile apps. Our work complements these
studies by analyzing large-scale, naturally occurring user reviews across
diverse Gen-AI app categories, allowing us to identify recurring topics,
opportunities, and challenges that emerge across the broader mobile app ecosystem. We further examine how the topics discussed in user feedback change over time.

\subsection{User Review Analysis Techniques} 

In this section, we summarize prior work related to 1) filtering user reviews, 2) pre-LLM approaches to topic extraction, 3) LLM-based topic extraction and assignment methods, 4) single-task topic analysis, and 5) analyses of specific
user concerns. These studies collectively form the foundation for our own framework, \textit{SARA}, which builds upon their methods while advancing them by integrating filtering, topic extraction, and topic assignment into a unified, scalable framework for analyzing Gen-AI app reviews.\\

\noindent \textbf{Filtering user reviews.} Automated filtering has shown strong potential in enhancing user review analysis by effectively removing non-informative content.~\citet{Chen_Lin_Hoi_Xiao_Zhang_2014} introduced a framework called AR-miner that used Latent Dirichlet Allocation (LDA) to first filter informative reviews with a hit-rate of 70\%, and second to group the informative reviews using topic modeling.~\citet{Ghosh_Pargaonkar_Eisty_2024} evaluated fine-tuned language models for filtering non-informative user reviews on the Google Play Store and found them highly effective, achieving accuracies of 92\% for both BERT and DistilBERT, and 91\% for Gemma.~\citet{Al_Wahshat_Abu_ulbeh_Yusoff_Zakaria_Amir_Fazamin_Wan_Hamzah_P_2023} highlighted the potential of using GPT-4 in filtering manipulated reviews due to its powerful language processing capabilities. Building on this line of work, our study incorporates automated filtering as
the first step of a broader LLM-assisted framework that combines filtering with
topic extraction and assignment. Our filtering component achieves 91\%
accuracy and enables non-informative reviews to be removed before downstream
topic analysis across a large and diverse collection of Gen-AI apps.

\noindent \textbf{Pre-LLM approaches to topic extraction.} Before the widespread adoption of LLMs, researchers used various methods to analyze user reviews.~\citet{Genc-Nayebi_Abran_2017} conducted a 2016 literature review exploring approaches to mining opinions from user reviews. They identified studies that relied on manual analysis, while others used automated techniques, such as Classification and Regression Trees (CART), word frequency statistics, correlation analysis, LDA, and Hexbin plots. 
Researchers and developers still use non-LLM approaches, even in recent work. For example,~\citet{10499727} used the Valence Aware Dictionary and sEntiment Reasoner (VADER) for sentiment analysis to categorize reviews as positive, neutral, or negative. They then used an LDA model for topic modeling of user reviews of educational apps. Their positive topics achieved a coherence score of 0.56, and their negative topics achieved a score of 0.49.~\citet{Assi_Hassan_Tian_Zou_2021} proposed the Global-Local sensitive Feature Extractor (GLFE) to identify high-level features in user reviews, achieving 79-82\% precision and 74-77\% recall on annotated review data. These early techniques established the foundation for automated review analysis but relied heavily on statistical or lexicon-based models. More recently, LLMs have enabled context-aware approaches to topic extraction
that complement these traditional techniques, as discussed next.

\noindent \textbf{LLM-based topic extraction and assignment methods.}
LLMs have demonstrated promising performance in tasks related to topic
extraction and assignment. \citet{Arambepola_Lalendra_Wimalasena_Munasinghe_2025}
conducted a recent systematic review of mobile app review analysis techniques
and found that integrating LLMs into app user review analysis provides a more
holistic and context-rich approach to evaluating reviews and understanding
users. \citet{Pham_Hoyle_Sun_Resnik_Iyyer_2024} introduced TopicGPT, a
framework that prompts OpenAI's GPT to generate topics and assign them to
documents. Their framework outperformed traditional topic models, such as LDA,
SeededLDA, and BERTopic. \citet{Prakash_Wang_Hoang_Hee_Lee_2023} introduced
PromptMTopics, a prompt-based ChatGPT model for extracting topics from textual
meme captions that outperformed baseline models.
\citet{Maram_LLM_Cure} introduced LLM-Cure, an approach that extracts and
assigns features from user reviews to generate suggestions for feature
improvement, achieving an F1-score of 85\%. Collectively, these studies
demonstrate the potential of LLMs for extracting and assigning meaningful
topics from unstructured textual data, providing the methodological foundation
for the LLM-based analysis used in our framework.

\noindent \textbf{Single-task topic analysis. }
Other studies have focused on only one of these tasks, either extracting topics or assigning predefined topics to user reviews.
For example,~\citet{Wei_Courbis_Lambolais_Xu_Bernard_Dray_2023} focused on review assignment using predefined labels rather than extracting topics. 
They assigned reviews using fixed categories such as feature requests, problem reports, and irrelevant feedback. 
While useful, this approach limits the discovery of emergent or evolving user concerns.~\citet{Dos_Santos_Oliveira_De_Jesus_Aljedaani_Eler_2023} and~\citet{Ren_Nakagawa_Tsuchiya_2024} extracted topics from user reviews without performing the assignment task. 
While this approach is effective for uncovering topical insights, it does not account for the frequency or distribution of each topic, limiting its usefulness for trend or prevalence analysis. Our approach combines both tasks: topic extraction allows topics to emerge
from the review data, while topic assignment enables us to quantify their
prevalence and compare their distribution across app categories and over time.

\noindent \textbf{Analysis of specific user concerns.}
Prior studies have used large-scale app reviews to characterize specific
concerns experienced by users of AI-based mobile apps.
\citet{RezaeiNasab_Dashti_Shahin_Zahedi_Khalajzadeh_Arora_Liang_2025}
examined fairness concerns and identified six types: unequal service quality
across platforms and devices, linguistic discrimination, unfair or
non-transparent treatment of user-generated content, gender and racial
discrimination, biased censorship and promotion, and unfair or non-transparent
advertising and subscription policies.
\citet{Massenon_Gambo_Khan_Agbonkhese_Alwadain_2025} analyzed user-reported
hallucinations in AI-powered mobile apps and developed a taxonomy of seven
hallucination types: factual incorrectness, fabricated information,
nonsensical or irrelevant output, logical inconsistency or self-contradiction,
persona deviation or role inconsistency, visual fabrication, and repetitive
output. While these studies provide detailed insights into specific user concerns,
their analyses focus on concerns defined in advance, such as fairness or
hallucinations. In contrast, our work does not restrict the analysis to a
predefined concern, allowing the topics users discuss to emerge from the review data. This approach captures a broader range of topics, including
\textit{AI performance}, \textit{content quality}, and \textit{content policy},
and enables us to examine both category-specific differences and patterns
shared across the Gen-AI mobile app ecosystem.\\

\noindent Collectively, these studies demonstrate how user review analysis has evolved
from traditional statistical and topic-modeling approaches toward LLM-based
methods capable of interpreting unstructured feedback. Prior work has
demonstrated the value of filtering reviews, extracting topics, assigning
reviews to predefined topics, and examining specific user concerns. However,
these capabilities are often studied or applied separately. Building on these
foundations, \textit{SARA} integrates review filtering, topic extraction, and
topic assignment into a unified framework. This combination allows topics to
emerge from user reviews while also assigning them at scale, enabling their
prevalence to be quantified and compared across app categories and over time.

\section{Methodology}
\label{sec:methodology}
\begin{figure}
\centering
\includegraphics[width=0.9\textwidth]{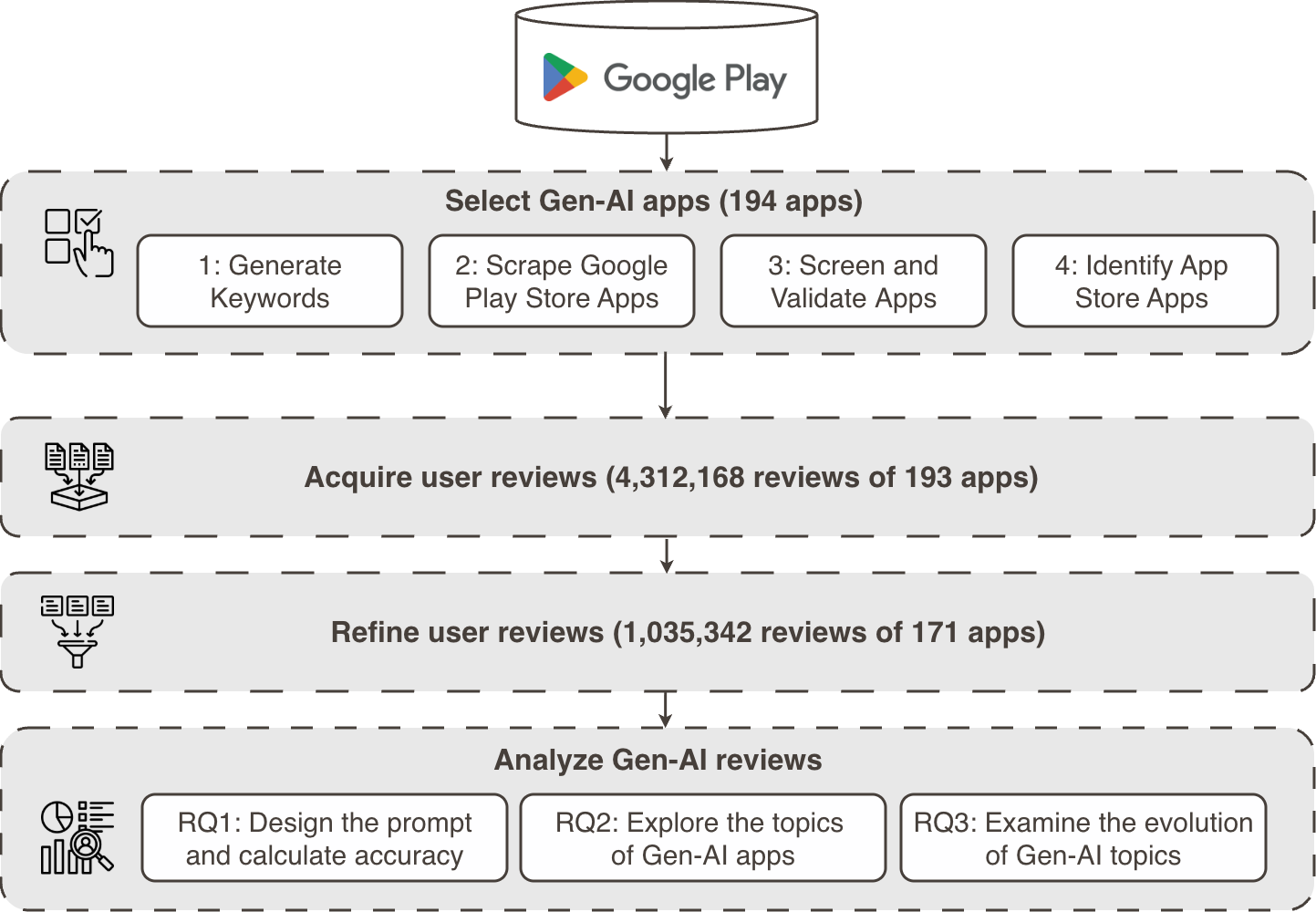}
\caption{Overview of our framework.}
\label{fig:approach}
\end{figure}

To study user feedback on Gen-AI apps, we propose and validate a structured four-phase framework we call \textit{\textbf{SARA}}, which stands for \textbf{S}election, \textbf{A}cquisition, \textbf{R}efinement, and \textbf{A}nalysis. We present an overview of our research framework in Figure~\ref{fig:approach}.

\subsection{Selection of Gen-AI apps}

Our app selection process involves four steps: 1) generating keywords from prior work, 2) searching the Google Play Store and scraping relevant apps, 3) manually screening and validating each app to confirm that it satisfies our definition of a Gen-AI app as introduced earlier, and 4) finding the App Store version of that app. Specifically, we retain only apps whose core functionality involves generating novel content (e.g., text, images, audio, or video) and exclude apps that rely solely on AI for prediction, analysis, or classification without generating new content.


\begin{description}
    \item[1) Generate Keywords] We compile a list of keywords derived from a Gen-AI taxonomy proposed by~\citet{gozalobrizuela2023surveygenerativeaiapplications}. We adapt this taxonomy to mobile app contexts by selecting relevant categories and deriving corresponding keywords. The complete list of categories and keywords is provided in Appendix~\ref{app:keywords}.
    
    \item[2) Scrape Google Play Store Apps] Using our list of keywords, we scrape the Google Play Store to identify all matching apps.  For each identified app, we collect descriptive metadata, including \textit{app ID}, \textit{app name}, \textit{Google Play Store category}, \textit{star rating}, \textit{number of downloads}, and the \textit{app description}. In this paper, we refer to Google Play Store app categories as \textit{app categories}. Our app data collection occurred in October 2024. 

    \item[3) Screen and Validate Apps.] The initial search yields 303 apps. 
    To ensure the inclusion of only genuine Gen-AI apps, the first and second authors independently review the apps by examining their titles, descriptions, and images. 
    We assess the inter-rater reliability, using Cohen’s Kappa~\citep{cohen1960kappa} ($\kappa = 0.61$), which indicates substantial agreement. 
    In case of disagreement, the two authors discuss and resolve discrepancies in ambiguous cases. 
    This rigorous process results in a final list of 194 confirmed Gen-AI apps.

    \item[4) Identify the Corresponding App Store Apps] Our empirical study primarily focuses on user reviews from the Google Play Store. 
    However, we also identify the corresponding App Store versions of these apps in order to examine whether the identified themes generalize across platforms. Out of the 194 Gen-AI apps from the Google Play Store, we identify the App Store version using AppMagic~\footnote{https://appmagic.rocks/top-charts/apps?tag=243926\&store=1\&date=2024-10-01\&topDepth=1000}. We identify 81 App Store apps in January 2026.
\end{description}

\subsection{Acquisition of User Reviews}
Following app selection, we collect user reviews using the Google Play Store Scraper~\citep{googleplayscraper}. For each retrieved review, the crawler extracts a comprehensive set of metadata, including \textit{content} (i.e., review text), \textit{score} (i.e., user rating from 1 to 5 stars), \textit{at} (i.e., timestamp of the review), \textit{replyContent} (developer's response), and \textit{repliedAt} (i.e., timestamp of response). We organize the collected reviews based on the Google Play Store categories (app categories) of their respective apps. At the end of this step, we collect 4,312,168 reviews of 193 apps. User reviews from the App Store apps are collected using the Appbot platform~\citep{appbot2026}. We extract the same review metadata as Google Play Store reviews. At the end of this step, we collect 540,119 reviews from 70 apps. We map App Store reviews to Google Play Store app categories to enable later cross-platform contextualization. Review acquisition occurred in February 2026.

\subsection{Multi-Stage User Review Refinement}
To ensure the robustness and reliability of our LLM-based analysis, we implement a \textit{\textbf{multi-stage}} data-cleaning framework that progressively improves the quality of user review data. This multi-stage refinement process comprises four distinct stages. We refine data in stages because we are interested in experimenting with different levels of data cleaning in RQ1. We summarize the number of reviews retained after each cleaning stage in Table~\ref{tab:reviews_by_stage} for Google Play Store reviews and in Table~\ref{tab:reviews_by_stage_astore} for App Store reviews. Each stage builds upon the previous one as follows:

\begin{table}[h]
\caption{Number of Google Play Store reviews by app category and stage (Full dataset considered).}
\centering
\setlength{\tabcolsep}{3pt}
\begin{tabular}{>
{\raggedright\arraybackslash}p{3.5cm} 
                c c c c}
\toprule
\textbf{App category} & \multicolumn{4}{c}{\textbf{Processing Stages}} \\
\cmidrule(lr){2-5}
    & \textbf{Stage 0} & \textbf{Stage 1} & \textbf{Stage 2} & \textbf{Stage 3} \\
    & \textit{Basic cleaning} & \textit{Temporal filtering} & \textit{short review exclusion} & \textit{LLM filtering} \\
\midrule
Productivity             & 903,719   & 817,968   & 599,382 & 402,055 \\
Photography              & 1,349,480 & 613,680   & 284,065 & 170,446 \\
Entertainment            & 431,740   & 257,237   & 205,108 & 152,139 \\
Art \& Design            & 509,505   & 302,899   & 174,911 & 125,375 \\
Education                & 156,435   & 142,047   & 95,254  & 65,699  \\
Video Players \& Editors & 476,730   & 136,542   & 62,807  & 39,972  \\
Tools                    & 271,218   & 94,374    & 52,070  & 36,694  \\
Games                    & 35,853    & 34,850    & 29,648  & 25,211  \\
Music \& Audio           & 48,667    & 44,487    & 25,781  & 17,751  \\
\midrule
\textbf{Total}           & 4,183,347 & 2,444,084 & 1,529,026 & 1,035,342 \\
\bottomrule
\end{tabular}
\label{tab:reviews_by_stage}
\end{table}

\begin{table}[h]
\caption{Number of App Store reviews by app category and stage (Full dataset considered).}
\centering
\setlength{\tabcolsep}{3pt}
\begin{tabular}{>
{\raggedright\arraybackslash}p{3.5cm} 
                c c c c}
\toprule
\textbf{App category} & \multicolumn{4}{c}{\textbf{Processing Stages}} \\
\cmidrule(lr){2-5}
    & \textbf{Stage 0} & \textbf{Stage 1} & \textbf{Stage 2} & \textbf{Stage 3} \\
    & \textit{Basic cleaning} & \textit{Temporal filtering} & \textit{short review exclusion} & \textit{LLM filtering} \\
\midrule
Productivity             & 147,288 & 136,807 & 122,580 & 78,767 \\
Photography              & 109,752 & 26,639  & 23,028  & 15,375 \\
Entertainment            & 60,939  & 37,987  & 36,522  & 26,777 \\
Art \& Design            & 74,311  & 51,700  & 48,062  & 33,556 \\
Education                & 117,752 & 115,121 & 92,002  & 56,944 \\
Video Players \& Editors & 9,763   & 7,608   & 6,848   & 4,420 \\
Tools                    & 4,329   & 3,826   & 3,604   & 2,504 \\
Games                    & 5,328   & 5,083   & 4,910   & 3,774 \\
Music \& Audio           & 7,317   & 7,016   & 5,930   & 3,352 \\
\midrule
\textbf{Total}           & 536,779 & 391,787 & 343,486 & 225,469 \\
\bottomrule
\end{tabular}
\label{tab:reviews_by_stage_astore}
\end{table}

\noindent \textit{\textbf{Stage 0: Basic cleaning.}} 
Text preprocessing, as demonstrated in prior work~\citep{Ghosh_Pargaonkar_Eisty_2024,Prakash_Wang_Hoang_Hee_Lee_2023,Roumeliotis_Tselikas_Nasiopoulos_2024,Lee_Lee_Seo_2024}, is a standard practice for eliminating noise and ensuring more effective analysis. In the initial stage, following prior work, we normalize whitespace and Unicode characters; remove emojis, URLs, and usernames; replace commas with whitespace; convert to lowercase; and drop empty reviews. 

\noindent \textit{\textbf{Stage 1: Temporal filtering.}} This stage filters reviews written prior to Gen-AI integration to ensure Gen-AI relevancy. Some apps included in our initial screening are not originally designed as Gen-AI tools but have integrated Gen-AI functionalities at a later stage. For instance, in the case of Microsoft Bing Search\footnote{\url{https://play.google.com/store/apps/details?id=com.microsoft.bing}}, \textit{Bing Chat} is a recently introduced feature that extends the capabilities of the search engine by incorporating Gen-AI models~\citep{Iorliam_Ingio_2024}. Consequently, it is essential to exclude user reviews published before the integration of such features, as they reflect experiences with the app before the adoption of Gen-AI. We outline our temporal filtering procedure below. 

\begin{description}
\item[Integration Date Identification] To approximate the Gen-AI integration date for each app, we adopt a conservative, multi-source identification strategy and record the earliest \textit{verifiable} date across all available evidence. Specifically, we consult the following sources:
(i) developer websites, blog posts, press releases, and public announcements; and 
(ii) historical release notes and version histories obtained from AppMagic\footnote{https://appmagic.rocks/}. When multiple sources report different dates, we retain the earliest date supported by documentary evidence. This approach prioritizes evidentiary certainty over speculative precision.

\item[Fallback Strategy]
For Google Play Store apps, if no explicit integration date can be identified, or if the earliest documented evidence postdates October 2024, we conservatively assign October 2024 as the reference date. This date corresponds to the time at which we independently screened and verified the presence of Gen-AI functionality in the app. Therefore, the recorded date reflects the earliest verifiable confirmation of Gen-AI features rather than the exact deployment date. For App Store apps, we do not apply the October 2024 fallback, as App Store descriptions were not archived at the time of app selection. If no verifiable integration date could be identified using release notes or developer sources, the app is excluded.

\item[Summary of Identified Dates]
For Google Play Store apps, we identify integration dates for 4 apps through public articles, for 125 apps through release notes, and 64 apps are assigned to October 2024 based on independent verification. For App Store apps, integration dates are identified for 2 apps via developer websites, 5 through public articles, and 63 through release notes. We exclude 10 App Store apps due to the absence of verifiable integration dates.

\item[Reliability and Coding Procedure]
To assess reliability, we randomly sample 65 apps from our 193-app list (95\% confidence level, 10\% margin of error) for independent annotation. The first and second authors independently review release notes and supporting sources to assign integration dates. Inter-rater agreement is high (Cohen’s $k$ = 0.8524), indicating near-perfect agreement. Six disagreements are identified. These primarily arise from differing interpretations of whether certain features or naming conventions constituted evidence of generative functionality. For instance, references to \textit{“AI”} in app names (e.g., \textit{“AI Song”}) are not considered sufficient evidence unless accompanied by explicit indications of generative capabilities (e.g., \textit{“AI Image Generator”}) or supporting descriptions. Additionally, features such as \textit{“AI effects”} or \textit{“AI enhancement”} are clarified as AI-assisted functionalities rather than Gen-AI because they modify content rather than generate novel content. Following the discussion, we refine the coding criteria accordingly and resolve all disagreements through consensus. The first author then applies the finalized criteria to label the remaining apps.
\end{description}

\noindent \textit{\textbf{Stage 2: Short review exclusion.}} Building upon the cleaned data from Stage 1, this stage employs a simple heuristic to further reduce noise by excluding reviews containing three words or fewer. Short reviews of three words or less do not contain context for meaningful topics~\citep{Assi_Hassan_Tian_Zou_2021,10499727}. This step aims to eliminate overly brief and potentially non-informative feedback. 

\noindent \textit{\textbf{Stage 3: LLM-assisted informative review filtration.} }
Non-informative reviews bring noise to user reviews analysis and influence the performance of topic extraction~\citep{Chen_Lin_Hoi_Xiao_Zhang_2014}. To improve the quality of extracted topics, we exclude non-informative reviews. We define \textit{non-informative} reviews as short, generic statements that express sentiment without referencing specific reasons or features. We carry out the LLM-assisted process in the following steps. 

\begin{description}
\item[Prompt design]
To filter non-informative reviews, we utilize Open AI’s~\citep{openai2023gpt} \textit{GPT-4o-mini} model with zero temperature for reproducibility. We use \textit{gpt-4o-mini} instead of the full \textit{gpt-4o} model because it offers significantly faster response times while maintaining reasonable output quality for our task, which does not require complex reasoning~\citep{OpenAI_O4Mini_2024}. Prior research comparing explicit, implicit, and creative prompting~\citep{Giray_2023, Ho_Mayberry_Nguyen_Dhulipala_Pallipuram_2024, Deshmukh_Raut_Bhavsar_Gurav_Patil_2025} shows that explicit, well-structured prompts produce more factual and task-aligned outputs, while underspecified ones invite unpredictable reasoning. Consistent with these findings and with approaches such as LLM-Cure~\citep{Maram_LLM_Cure}, we design an explicit, structured prompt that classifies each review as either informative or non-informative. The prompt includes representative examples, a fixed binary label set, and a CSV-style output format to constrain the model’s responses. These measures ensure deterministic labeling and minimize hallucination by preventing free-form or extraneous text generation. 

\item[Iterative prompt refinement] Prior studies highlight that prompt structure and iterative refinement (“dynamic prompting”) substantially influence LLM performance ~\citep{Marvin_Hellen_Jjingo_Nakatumba-Nabende_2024, He_Rungta_Koleczek_Sekhon_Wang_Hasan_2024}.
In our setup, we adopt this principle by refining prompts through multiple pilot runs to improve label consistency and clarity. We use small statistically significant samples of 96 reviews (95\% confidence interval, 10\% margin of error) from three diverse app categories (\textit{Photography}, \textit{Productivity}, and \textit{Art \& Design}). 
Then, the first and second authors manually label the reviews in the samples as informative or non-informative. We calculate the inter-rater reliability ($\kappa = 0.91$), which indicates almost perfect agreement. 

Next, we design and test different versions of the filtering prompt against the manually labeled data, evaluating the model’s filtering accuracy and iteratively refining the prompt. Revisions include modifying the wording, structure, number of examples, and types of examples. We repeat this iterative process until we achieve a peak filtering accuracy of 91\% on held-out validation samples. The final version of the filtering prompt, which includes ten examples (five informative and five non-informative) and yields optimal performance, is provided in Figure~\ref{fig:filter_prompt} in Appendix~\ref{app:filter_prompt}. 

\item[Prompt validation] To assess the generalizability of the final filtering prompt and reduce the risk of overfitting to specific review instances, we manually evaluate its performance on a separate set of previously unseen reviews of the same categories and the same size. 
The model demonstrates strong generalization capabilities, achieving an average filtering accuracy of 90\%.

\item[Large-scale review filtering] We apply the validated filtering prompt to
the full dataset, processing reviews in batches of 100 to remain consistent
with the batch size used during prompt validation. Larger batches also increased
the likelihood of incomplete model responses. We verify that an output is
returned for every input review and reprocess any missing reviews.
\end{description}

\noindent We exclude app categories with fewer than 1,000 informative reviews,
reducing the 15 Google Play Store categories remaining after Stage 3 to 7 categories.
Prior work shows that extracting a fixed number of topics from small datasets
can lead to over-clustering and insufficiently distinct topics~\citep{greene2014}.
We validate the 1,000-review threshold by performing topic extraction across
all categories and find that categories below this threshold consistently
produce redundant and poorly differentiated topics. For example, in
\textit{Books \& Reference} (342 reviews), extracted topics overlap between
``Content Generation Features'' and ``Quality of Generated Content,'' as well
as ``User Experience'' and ``User Interface.'' Similar redundancy occurs in
\textit{Health \& Fitness} (729 reviews). Such overlap introduces ambiguity
during topic assignment, making it difficult to reliably determine which topic a review belongs to. We therefore adopt the 1,000-review threshold to support distinct topic extraction and reliable topic assignment.

Our data collection, spanning October 2024 to February 2026, results in an initial dataset of
4,183,347 Google Play Store user reviews across 172 unique Gen-AI apps and 536,779 App Store reviews from 70 apps. Following our multi-stage cleaning process, we obtain a dataset of 1,035,342 informative Google Play Store reviews across 171 Gen-AI apps, and 225,469 App Store reviews from 59 apps, which form the basis for our subsequent analysis.

\subsection{LLM-powered user review analysis}
We use the refined user review data to conduct topic extraction, topic assignment, trend analysis, and manual analysis. For RQ1, we design and evaluate our topic extraction and assignment prompts. We experiment with different levels of data noise (i.e., different stages) and numbers of few-shot examples. For RQ2, we extract the top 10 topics per app category and assign them to user reviews. We conduct manual thematic analysis for each Gen-AI topic to identify challenges and opportunities derived from user review insights. For RQ3, we analyze how Gen-AI topics evolve over time using temporal clustering and manual review. We describe the experiment design and the findings for each RQ in its respective section. 

\subsection{Adaptability of \textit{SARA} Framework} 
In this study, \textit{SARA} is empirically evaluated only on Gen-AI mobile apps. However, \textit{SARA} is designed as a modular analysis framework in which the overall analytical framework remains fixed, while domain-specific configurations can be adapted. In particular, \textit{SARA} can be applied to analyze user reviews of multiple mobile apps across different app categories and app stores by customizing domain-specific components. Because the framework operates on user-generated review data rather than Gen-AI-specific content, it can be adapted to analyze user perceptions of a wide range of mobile apps.

In the first stage, app selection is domain-agnostic: researchers may define and select target apps based on any domain-specific criteria or definitions relevant to their study. The second stage, review acquisition, follows the same procedure regardless of domain, as user reviews are collected from app store platforms using consistent metadata and retrieval mechanisms.

The review refinement stage is largely reusable across domains. Basic preprocessing steps (e.g., removal of white space, emojis, and short reviews) remain unchanged, and our findings in RQ1 show that LLMs are effective at filtering non-informative reviews using a general definition of informative reviews that is not tied to Gen-AI-specific functionality. As such, the non-informative review-filtering prompt can be applied without modification to other domains.

In the analysis stage, \textit{SARA} employs topic extraction followed by topic assignment. While the overall prompt structure for topic extraction is reusable, the examples and topic labels must be adapted to reflect domain-specific concepts and functionalities. This customization is expected and enables the framework to capture contextually relevant themes. Similarly, topic assignment relies on a reusable prompt template, with domain-specific examples supplied by the researcher when applying the framework to a new domain.

Together, these design choices allow \textit{SARA} to be adapted to multiple mobile apps across any app category and across different app stores by modifying domain-specific inputs and topic definitions, while preserving the framework's core analytical stages and processing logic.

\section{RQ1: How accurately can LLMs identify topics in user reviews of Gen-AI apps?}
\label{sec:RQ1}

Effective topic extraction and assignment of user reviews are essential for understanding user needs and improving app features for Gen-AI apps. We identify the optimal combination of preprocessing and few-shot prompting to enable more accurate and reliable topic extraction and assignment from user reviews. 

\subsection{Experiment Setup}
\begin{figure}
\centering
\includegraphics[width=\textwidth]{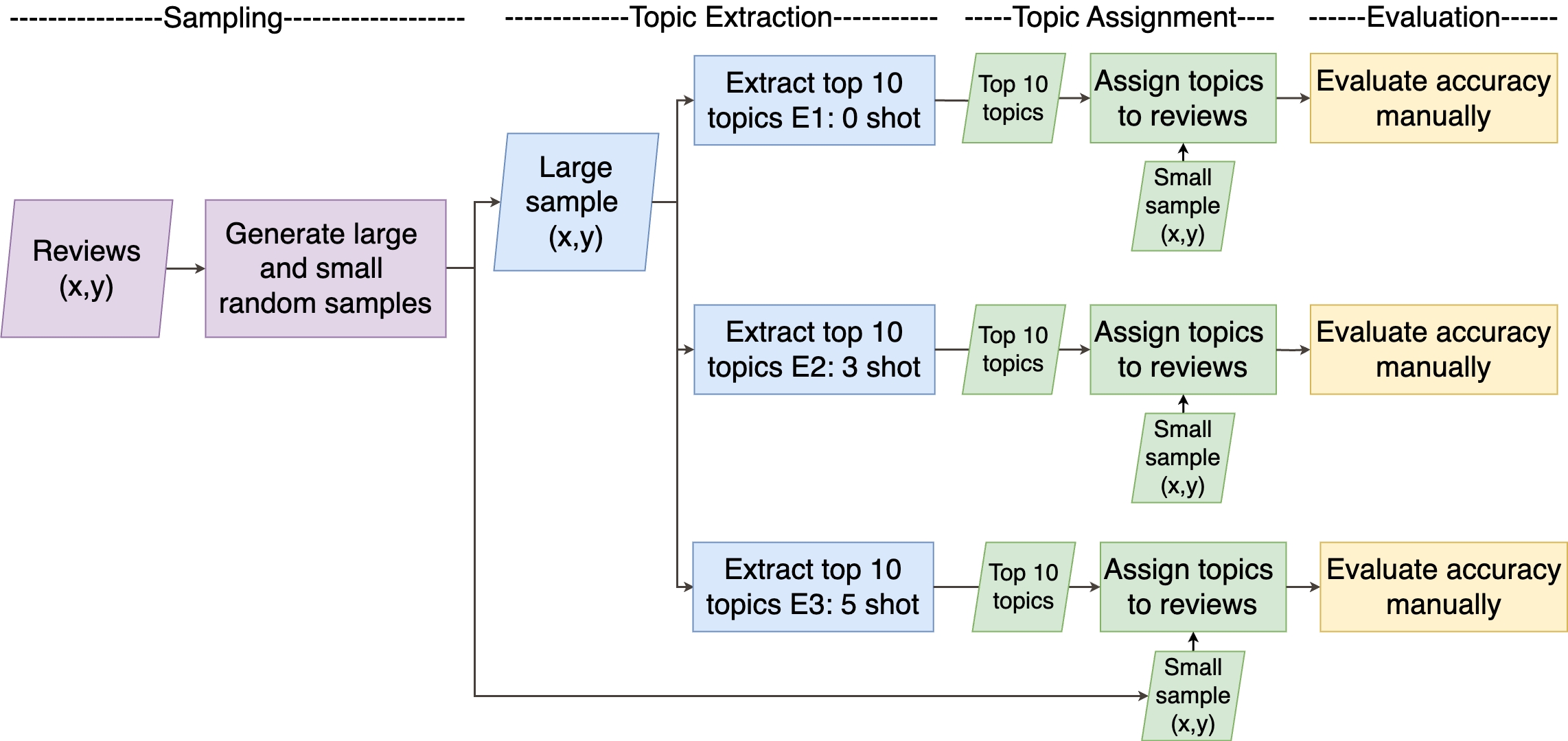}
\caption{An overview of our RQ1 experiment design, where \( x \in \{Stage 1, Stage 2, Stage 3\} \) and \( y \in \{Photography, Productivity, Art \& Design, Entertainment, Video Players \& Editors \} \).
}
\label{fig:rq1_experiment_design}
\end{figure}

To design an effective topic extraction prompt, we conduct experiments on \textit{Stage 1, Stage 2, and Stage 3}, but not on \textit{Stage 0}. We exclude \textit{Stage 0} because, although it performs essential preprocessing, it does not remove irrelevant data as in Stage 1. We approach this problem in four steps: 1) we generate random samples of user reviews for topic extraction and assignment drawing from each stage of data cleaning, 2) we apply three prompting strategies with varying numbers of few-shot examples (i.e., 0-shot, 3-shot, and 5-shot) to extract topics from these samples, 3) assign the extracted topics to a new set of sample reviews using the same prompts, and (4) we evaluate the accuracy of the topic assignment to determine the optimal combination of data cleaning and few-shot examples. We summarize this process in Figure~\ref{fig:rq1_experiment_design} and detail it in this section.\newline

\noindent\textbf{Sampling.} To evaluate the effectiveness of our topic extraction prompts across different stages of data cleaning, we define \( x \in \{Stage 1, Stage 2, Stage 3\} \) to represent the three data cleaning stages. For each stage \( x \), we select five representative app categories from the Google Play Store, i.e., \textit{Photography}, \textit{Productivity}, \textit{Art \& Design}, \textit{Entertainment}, and \textit{Video Players \& Editors}, denoted as \( y \). For each combination of cleaning stage \( x \) and app category \( y \), we generate two separate statistically representative random samples of user reviews from our review dataset. First, we extract a \textit{large sample} denoted \(S_{\text{large}}(x, y) \), computed using a 95\% confidence level and a 2\% margin of error with an average size of 2,337 reviews. This large sample is used for topic extraction to ensure the coverage of the user feedback space. Second, we draw a \textit{small sample} denoted \( S_{\text{small}}(x, y) \) from our review dataset, computed using a 95\% confidence level and a 10\% margin of error with an average size of 96 reviews. This sample is used for topic assignment, where manual evaluation is feasible and interpretable.\newline

\noindent\textbf{Topic extraction.} Because prior work reported mixed effects of few-shot examples \citep{Wang_Yang_Wei_2023, Prakash_Wang_Hoang_Hee_Lee_2023}, we design three experiments, i.e., \textit{E1: 0 shot}, \textit{E2: 3 shot} and \textit{E3: 5 shot}, with three prompt configurations differing in the number of few-shot examples, i.e., 0-shot, 3-shot and 5-shot. These values are chosen based on prior research and practical considerations in prompt-based learning ~\citep{Dos_Santos_Oliveira_De_Jesus_Aljedaani_Eler_2023, Maram_LLM_Cure, Pham_Hoyle_Sun_Resnik_Iyyer_2024}. In \textit{E1: 0 shot}, the prompt contains only the instructions. In \textit{E2: 3 shot}, the prompt includes the same instructions with three shots, i.e., three examples of user reviews with their topics. In \textit{E3: 5 shot}, the prompt includes the same instructions with five examples of user reviews with their topics. For each cleaning stage \( x \in \{\text{Stage 1}, \text{Stage 2}, \text{Stage 3}\} \) and app category \( y \), we apply each of the three prompt configurations to the large sample \( S_{\text{large}}(x, y) \).  This results in a set of extracted \textit{top topics}, denoted as \( T(x, y, n) \), where \( n \in \{0, 3, 5\} \) indicates the number of few-shot examples used in the prompt. 
The final version of the topic extraction prompt and the few-shot examples can be found in Figure~\ref{fig:topic_extraction_prompt} in Appendix~\ref{app:topic_extraction_prompt}. 

To mitigate hallucinations during topic extraction, we anchor the prompt in actual user reviews, instruct the model to output only ten distinct high-level topics that describe app features, functionalities, and utility, and provide few-shot examples that illustrate the expected level of abstraction and discourage free-form or overly creative topic invention. To further adapt the LLM to the Gen-AI domain, the few-shot examples are drawn from real reviews of Gen-AI apps, exposing the model to characteristic terminology and contexts (e.g., discussions of \textit{AI Performance}, \textit{Technical Difficulties}, and \textit{Content Quality}).

\noindent\textbf{Topic assignment. }
To evaluate the effectiveness of the extracted topic sets \( T(x, y, n) \) produced by each topic extraction experiment with \( n \in \{0, 3, 5\} \), we perform a follow-up topic assignment task using LLM. Specifically, for each experiment \( n \), we design a dedicated assignment prompt that takes as input: 1) the category-specific set of top topics \( T(x, y, n) \) extracted for that app category in that experiment, 2) five few-shot example reviews manually selected and labeled from the corresponding large sample \( S_{\text{large}}(x, y) \), and 3) the target reviews from the small sample \( S_{\text{small}}(x, y) \) to which topics are assigned. We manually select and assign topics to five few-shot examples to improve assignment accuracy as done in prior work~\citep{Maram_LLM_Cure,Deshmukh_Raut_Bhavsar_Gurav_Patil_2025}. The final topic assignment prompt is provided in Figure~\ref{fig:topic_assignment_prompt} in Appendix~\ref{app:topic_assignment_prompt}.

To mitigate hallucination during topic assignment, we constrain the model’s output to a predefined set of deterministic labels, i.e., the extracted topics \( T(x, y, n) \). The model must choose from these labels or select “Other” if none accurately represent the review. This closed-set format enables immediate detection of hallucination, as any label outside the predefined list signals a deviation from valid outputs. Additionally, structured CSV-formatted output further enforces adherence to the expected schema and prevents free-form or unconstrained generation. To ensure domain relevance, the five few-shot examples are handpicked from real reviews of Gen-AI apps, representing common contexts and linguistic patterns found in this domain (e.g., user reviews of \textit{AI Performance}, \textit{Utility \& Use Cases}, or \textit{Content Quality}).\\

\noindent\textbf{Accuracy evaluation. }
To assess the accuracy of the LLM-assigned topics, we conduct a manual annotation of the assignments. The first and second authors manually and independently label the correctness of the LLM assignment. An assignment is considered correct if the LLM assigned topic accurately reflects the content of the review. The Cohen’s Kappa agreement score is computed, yielding a score of 0.56, indicative of moderate agreement. The two authors discuss and resolve the initial disagreements to align on labeling criteria.  

We analyze cases of annotator disagreement to better understand ambiguity in topic classification. Disagreements primarily occur in reviews with limited context or broad, non-informative statements (e.g., the “brief expressions of satisfaction” topic identified in earlier stages), where the relevance of a specific topic is unclear. In these cases, annotators differ in their strictness: one may accept the assigned topic as valid, while another may judge the review as too vague or incoherent to support that assignment. Upon discussion, we label such assignments as incorrect, as the reviews do not provide sufficient or interpretable evidence for the assigned topic, and therefore should have been marked as “other” by the LLM. These disagreements highlight the challenges of topic classification in earlier stages, where noisy and non-informative reviews can lead to the extraction of generic topics. As a result, assigning topics becomes difficult due to overlapping themes and the absence of clear semantic signals in the underlying reviews.

In total, 4,353 LLM topic assignments are manually evaluated for correctness for five categories, three stages, and three experiments. Table~\ref{tab:overall_accuracy} shows the accuracy for each cleaning stage, app category and few-shot examples. We supplement this with Table~\ref{tab:detailed_accuracy}, which reports accuracy on informative reviews across cleaning stages, app categories, and shot settings. This separation enables a fair comparison, as non-informative reviews are easier to classify.

\begin{table}[htbp]
\caption{Accuracy across three stages and three experiments of five app categories: 1) \textit{Photography} (Photo), 2) Productivity (Pro), 3) \textit{Art \& Design} (Art), 4) \textit{Entertainment} (Ent), and 5) \textit{Video Players \& Editors} (Video). (Representative subset considered).
}
\centering

\setlength{\tabcolsep}{1.7pt}
\renewcommand{\arraystretch}{1}

\begin{tabular}{>{\raggedright\arraybackslash}p{2cm} 
                *{6}{c} 
                *{6}{c} 
                *{6}{c} 
                }
\toprule
\textbf{Experiment} 
& \multicolumn{6}{c}{\textbf{Stage 1:} Temporal reviews} 
& \multicolumn{6}{c}{\textbf{Stage 2:} Short reviews} 
& \multicolumn{6}{c}{\textbf{Stage 3:} Non-informative reviews} 
\\

\cmidrule(lr){2-7} \cmidrule(lr){8-13} \cmidrule(lr){14-19}
& Photo & Pro & Art & Ent & Video & \textbf{Avg.}
& Photo & Pro & Art & Ent & Video & \textbf{Avg.}
& Photo & Pro & Art & Ent & Video & \textbf{Avg.}
\\

\midrule
E1: 0 shot & 86\% & 77\% & 91\% & 69\% & 95\% & 83\%
           & 84\% & 85\% & 85\% & 80\% & 90\% & 85\%
           & 72\% & 86\% & 77\% & 86\% & 89\% & 82\%
           \\
           
E2: 3 shot & 91\% & 86\% & 80\% & 66\% & 95\% & 83\%
            & 91\% & 80\% & 86\% & 82\% & 95\% & 87\%
            & 76\% & 83\% & 88\% & 92\% & 90\% & 86\%
            \\
            
E3: 5 shot & 93\% & 81\% & 94\% & 90\% & 92\% & \textbf{90}\%
            & 90\% & 84\% & 92\% & 88\% & 93\% & \textbf{89}\%
            & 95\% & 86\% & 86\% & 95\% & 91\% & \textbf{91}\%
            \\

\bottomrule
\end{tabular}
\label{tab:overall_accuracy}
\end{table}

\begin{table}
\caption{Accuracy of informative reviews across stages and app categories, \textit{Photography} (Photo), \textit{Productivity} (Pro), \textit{Art \& Design} (Art), \textit{Entertainment} (Ent), \textit{Video Players \& Editors} (Video). (Representative subset considered).}
\centering

\setlength{\tabcolsep}{1.7pt}
\renewcommand{\arraystretch}{1}

\begin{tabular}{>{\raggedright\arraybackslash}p{2cm} 
                *{6}{c} 
                *{6}{c} 
                *{6}{c} 
                }
\toprule
\textbf{Experiment} 
& \multicolumn{6}{c}{\textbf{Stage 1:} Temporal reviews} 
& \multicolumn{6}{c}{\textbf{Stage 2:} Short reviews} 
& \multicolumn{6}{c}{\textbf{Stage 3:} Non-informative reviews} 
\\

\cmidrule(lr){2-7} \cmidrule(lr){8-13} \cmidrule(lr){14-19}
& Photo & Pro & Art & Ent & Video & \textbf{Avg.}
& Photo & Pro & Art & Ent & Video & \textbf{Avg.}
& Photo & Pro & Art & Ent & Video & \textbf{Avg.}
\\

\midrule
E1: 0 shot & 83\% & 76\% & 90\% & 88\% & 76\% & 83\%
           & 84\% & 83\% & 89\% & 74\% & 82\% & 82\%
           & 81\% & 91\% & 81\% & 86\% & 89\% & \textbf{86\%}
           \\
           
E2: 3 shot & 87\% & 91\% & 87\% & 86\% & 76\% & 85\%
            & 93\% & 78\% & 89\% & 89\% & 94\% & 88\%
            & 81\% & 88\% & 89\% & 91\% & 90\% & \textbf{88\%}
            \\
            
E3: 5 shot & 87\% & 85\% & 94\% & 82\% & 76\% & 85\%
            & 80\% & 79\% & 90\% & 82\% & 89\% & 84\%
            & 94\% & 92\% & 90\% & 95\% & 97\% & \textbf{94}\%
            \\

\bottomrule
\end{tabular}
\label{tab:detailed_accuracy}
\end{table}

To determine whether the observed accuracy differences are statistically significant, we perform statistical tests. Specifically, we apply the non-parametric test, Kruskal-Wallis H test~\citep{Chan_Walmsley_1997}, to evaluate the effect of few-shot prompt configurations on overall accuracy. Additionally, post-hoc pairwise comparisons are conducted using the Mann-Whitney U test~\citep{Mann_Whitney_1947} with Bonferroni correction to adjust for multiple testing~\citep{3b6386c9-eeae-33a7-8c59-c0604aa87bef}. We also use one-way ANOVA~\citep{Judd_McClelland_Ryan_2017} to test for significant differences across cleaning stages and few-shot settings.

\subsection{RQ1 Findings}
\noindent \textbf{Finding 1.1: Providing more examples significantly improves classification and assignment accuracy.} The average accuracy increases across experiments for each cleaning stage, with \textit{E3: 5-shot} consistently achieving the highest accuracy in all three stages, as shown in Table~\ref{tab:overall_accuracy}. Notably, \textit{E3: 5-shot} is the only setting to achieve average accuracies above 80\% across all samples. The Kruskal-Wallis test indicates that the number of few-shot examples has a statistically significant effect on overall accuracy ($H = 7.778$, $p = 0.0205$). Post hoc Mann-Whitney U tests with Bonferroni correction show that the accuracy improvement from \textit{E1: 0 shot} to \textit{E3: 5 shot} is statistically significant ($p = 0.0181$), whereas differences between \textit{E1: 0 shot} and \textit{E2: 3 shot} ($p = 1.000$) and \textit{E2: 3 shot} and \textit{E3: 5 shot} ($p = 0.2519$) are not. These results suggest that increasing the number of few-shot examples, specifically 5-shot, significantly enhances overall model accuracy.\\

\noindent \textbf{Finding 1.2: The proportion of informative reviews varies by cleaning stage, which impacts overall accuracy.} As shown in Table~\ref{tab:overall_accuracy}, accuracy does not follow a consistent trend across the three cleaning stages (i.e., \textit{Stage 1}, \textit{Stage 2}, and \textit{Stage 3}). For instance,  the average accuracy for \textit{E1: 0 shot} and \textit{E2: 3 shot} increase from \textit{Stage 1} to \textit{Stage 2}, but decrease from \textit{Stage 2} to \textit{Stage 3}.  In contrast, \textit{E3: 5 shot} shows a slight decrease from \textit{Stage 1} to \textit{Stage 2}, followed by a slight increase in \textit{Stage 3}. This inconsistency may be explained by differences in the proportion of informative reviews across stages: 43\% in \textit{Stage 1}, 66\% in \textit{Stage 2}, and 88\% in \textit{Stage 3}. Since assigning topics to non-informative reviews is generally easier and less dependent on nuanced understanding, variations in their prevalence can skew overall accuracy. As a result, overall accuracy may not accurately reflect the model’s effectiveness in handling more meaningful and content-rich reviews across different cleaning stages.\\

\noindent \textbf{Finding 1.3: \textit{Stage 3} consistently yields the highest accuracy on informative reviews across all app categories and experiments.} Given the inconsistencies observed in overall accuracy explained in finding 1.2, we isolate informative reviews to better evaluate model performance. Table~\ref{tab:detailed_accuracy} shows that, across all experiments and app categories, average accuracy on informative reviews is highest in \textit{Stage 3}. This suggests that data cleaning improves the model's ability to assign topics to reviews with informative content correctly. We conduct a one-way ANOVA test to evaluate whether the cleaning stage has a significant effect on classification accuracy. The ANOVA for few-shot experiments does not reveal significant differences ($F(2, 42) = 2.25$, $p = 0.118$).\\

\noindent \textbf{Finding 1.4: The best performing setup combines filtering non-informative reviews, i.e., \textit{Stage 3} and more few shot examples, i.e., \textit{E3: 5-shot}.} Our findings suggest that using LLMs to filter the non-informative reviews and increasing the number of few-shot examples to five improves the accuracy of the model. Based on these findings, we adopt the \textit{Stage 3 - E3: 5-shot} configuration for topic classification and assignment in RQ2, as it yielded the highest overall accuracy of 91\% and the highest informative accuracy of 94\%. \\

\noindent \textbf{Finding 1.5: Under the best-performing configuration (\textit{E3: 5-shot}), only 9\% of reviews were misclassified.} Among these, 2\% of all reviews involved incorrect topic assignments, 3\% were missed topics (valid reviews marked as “Other”), and 4\% were over-assignments (non-topical reviews given a label). Most errors occurred in short or ambiguous reviews with limited context, where topic boundaries were less clear. These findings indicate that misclassifications are infrequent and stem from understandable edge cases rather than uncontrolled generation or hallucination, reinforcing the reliability of the LLM-based labeling process.

\begin{Summary}{}{}
Our findings indicate that increasing the number of few-shot examples in the prompt significantly improves output accuracy. 
Additionally, filtering out non-informative reviews enhances the accuracy of extraction and assignment for informative reviews. The model using the 5-shot prompt combined with LLM-based filtering of non-informative reviews achieves the highest overall accuracy of 91\%. Accordingly, we adopt this configuration for topic extraction in RQ2.
\end{Summary}

\section{RQ2: What are the most prominent topics discussed in user reviews?}
\label{sec:RQ2}

We identify the top 10 topics discussed in user reviews across app categories and analyze their frequency and associated ratings. This analysis uncovers common user concerns, expectations, and usage patterns, from which we derive key opportunities, challenges, and actionable insights for designing more effective, user-centered Gen-AI systems.

\subsection{Experiment Setup}

We apply RQ1's optimal configuration to identify and analyze topics from user reviews. This section describes the topic extraction process, the classification of extracted topics, and the quantitative and qualitative analyses used to interpret user feedback. We implement our framework to both Google Play Store and App Store reviews and discuss platform comparison in Section~\ref{sec:discussion}. In this section, our analysis is focused on the Google Play Store dataset because it is larger and more substantive.  

\noindent\textbf{Topic extraction and assignment.} Following RQ1, we draw a statistically significant \textit{large sample} of informative reviews from \textit{stage 3} (2,291 average reviews per app categories) to extract the top 10 topics from each app category. These topics are then used to assign a single topic label to each review in the full dataset.

\noindent\textbf{Classification and categorization.}
Our primary interest is user feedback related to Gen-AI functionality. We therefore manually examine the extracted topics and label them as either \textit{Gen-AI topics}, which reference Gen-AI capabilities, or \textit{non-Gen-AI topics}, which refer to general app functionality. All reviews originate from Gen-AI apps; the distinction therefore concerns the type of feedback rather than the app type. To facilitate higher-level interpretation, we group semantically related topics into broader \textit{topic categories}. In line with prior studies~\citep{Maram_LLM_Cure,Pagano_Maalej_2013,Chen_Lin_Hoi_Xiao_Zhang_2014},  we also categorize reviews by star ratings: 1-2 stars as negative, 3 stars as neutral, and 4-5 stars as positive.
\newline

\noindent\textbf{Analysis.}
We perform large-scale quantitative analysis on all labeled reviews across both app categories (e.g., \textit{Productivity} and \textit{Photography}) and topic categories (e.g., \textit{AI Performance} and \textit{Creative Potential}). Specifically, we compute the following metrics:

\begin{enumerate}
    \item \textit{Average Gen-AI Rating (AGR)}: the average rating of reviews discussing Gen-AI topics.
    \item \textit{Average Non-Gen-AI Rating (ANR)}: the average rating of reviews discussing non-Gen-AI topics.
    \item \textit{Gen-AI Review Count (GRC)}: the number of reviews assigned to Gen-AI topics.
    \item \textit{Non-Gen-AI Review Count (NRC)}: the number of reviews assigned to non-Gen-AI topics.
    \item \textit{Gen-AI Review Percentage (GRP)}: the proportion of Gen-AI reviews relative to the total number of reviews.
    \item \textit{Non-Gen-AI Review Percentage (NRP)}: the proportion of non-Gen-AI reviews relative to the total number of reviews.
\end{enumerate}

\noindent To examine whether reviews discussing Gen-AI functionality receive higher ratings than other reviews, we compute AGR and ANR for each app by aggregating review scores at the app level. We apply the Wilcoxon signed-rank test~\citep{wilcoxon1992} to compare the two ratings. To complement the quantitative results, we conduct qualitative coding of statistically significant samples (95\% confidence level, 10\% margin of error). 
We sample reviews from each Gen-AI topic category, resulting in 762 reviews that are manually coded to identify recurring user concerns that derive opportunities, challenges, and actionable insights. 

\subsection{RQ2 Findings}

\noindent\textbf{Finding 2.1: Gen-AI reviews consistently receive higher ratings than non-Gen-AI reviews across app categories, suggesting that negative feedback often relates to app design issues rather than the Gen-AI functionality itself.} Table~\ref{tab:gen_AI_vs_non} shows that reviews discussing Gen-AI functionality receive higher ratings than those discussing other app aspects, with an average of 4.2 compared to 2.8. A Wilcoxon signed-rank test confirms that AGR is significantly higher than ANR across apps ($p < 0.001$), with significant differences observed in 7 of 9 categories. This pattern is particularly pronounced in categories such as \textit{Art \& Design} and \textit{Music \& Audio}, where Gen-AI topics related to \textit{Creative Potential} receive very high ratings (4.8-4.9), while non-Gen-AI topics such as \textit{Monetization Methods \& Structure} receive substantially lower ratings (1.8). These results suggest that lower ratings are often driven by non-Gen-AI concerns, particularly monetization and technical issues, rather than dissatisfaction with Gen-AI functionality itself. For developers, these results suggest that strong Gen-AI capabilities alone are insufficient, as poor app design, reliability issues, and monetization practices can override their benefits.

\begin{table}[H]
\centering
\caption{Summary of app categories, including average Gen-AI rating (AGR), average non-Gen-AI rating (ANR), Gen-AI Reviews Count (GRC), Gen-AI reviews percentage (GRP), and number of distinct Gen-AI topics (Full dataset considered). }
\label{tab:gen_AI_vs_non}
\begin{tabular}{lccccc}
\toprule
\textbf{App Category} & \textbf{AGR} & \textbf{ANR} & \textbf{GRC} & \textbf{GRP} & \textbf{\# Gen-AI Topics} \\
\midrule
Productivity & 4.3 & 3.0 & 257,434 & \textbf{64\%} & 5 \\
Photography & 4.5 & 2.8 & 66,066 & \textit{39\%} & 3 \\
Entertainment & 3.6 & 2.8 & 61,803 & 41\% & 5 \\
Art \& Design & 4.3 & 2.3 & 54,227 & 43\% & 3 \\
Education & 4.5 & 4.0 & 46,942 & \textbf{72\%} & 6 \\
Games & 3.9 & 2.4 & 15,695 & 62\% & 6 \\
Tools & 4.3 & 3.0 & 14,923 & 41\% & 5 \\
Video Players \& Editors & 4.5 & 3.2 & 12,741 & 32\% & 3 \\
Music \& Audio & 4.1 & 2.1 & 6,992 & 39\% & 5 \\
\midrule
\textbf{Average} & \textbf{4.2} & \textbf{2.8} & \textbf{59,647} & \textbf{48\%} & \textbf{5} \\
\bottomrule
\end{tabular}
\end{table}

\begin{figure}
\centering
\includegraphics[width=\textwidth]{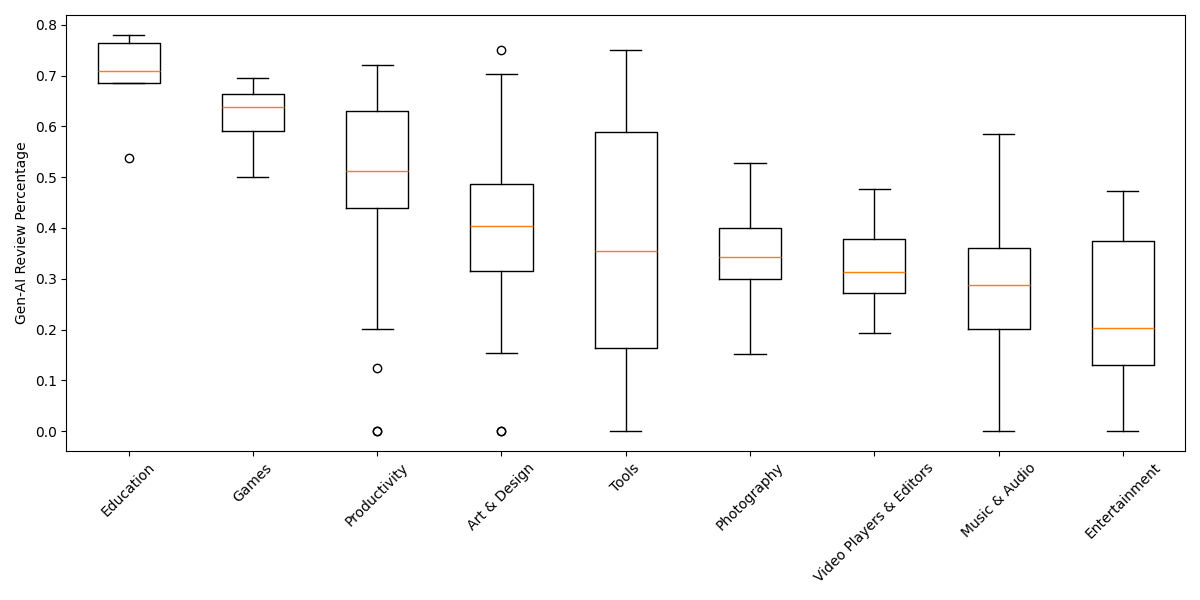}
\caption{Box plot showing the distribution of the percentage of reviews classified as Gen-AI at the app level, grouped by app category (Full dataset considered).}
\label{fig:gen-ai-boxplot}
\end{figure}

\noindent\textbf{Finding 2.2: The percentage of reviews discussing Gen-AI varies across app categories, reflecting whether Gen-AI is a core or supplementary feature.} Table~\ref{tab:gen_AI_vs_non} shows variation in the proportion of Gen-AI-related reviews across categories. Categories such as \textit{Education} (72\%) and \textit{Productivity} (64\%) exhibit the highest percentages, indicating that users primarily evaluate these apps through their AI capabilities. In these domains, discussions frequently center on AI performance and practical use cases, such as studying, writing, and content generation, highlighting strong user interest in the applicability of Gen-AI features. In contrast, categories such as \textit{Video Players \& Editors}, \textit{Music \& Audio}, and \textit{Photography} show much lower proportions of Gen-AI reviews, where AI functionality typically serves as a supplementary feature within broader tools. As a result, user feedback is more focused on non-Gen-AI concerns such as technical issues, monetization, and usability. Figure~\ref{fig:gen-ai-boxplot} further shows variation within categories, indicating that apps differ in how they position Gen-AI functionality. For developers, these results suggest that when Gen-AI is a core feature, users focus on its utility and performance, whereas when it is supplementary, overall app design and reliability play a larger role in shaping user perceptions.

\noindent \textbf{Finding 2.3: User discussions of Gen-AI apps reveal key opportunities, including \textit{AI for Accessibility} and Wellbeing, \textit{AI as a Collaborative Creative Tool}, and \textit{AI Versatility}, as well as challenges related to \textit{Managing User Expectations} and \textit{AI Limitations}, \textit{Balancing Content Moderation} and Creative Freedom, and \textit{Strategic Integration of Gen-AI Features}.} Table~\ref{tab:issue_types} lists the Gen-AI topics with their descriptions and examples. Descriptive statistics on Gen-AI and non-Gen-AI topic categories can be found in Tables~\ref{tab:gen_AI_summary} and ~\ref{tab:non_gen_AI_summary} in Appendix~\ref{app:topic_categories}. Through qualitative analysis of reviews within these topic categories,
recurring patterns emerge that we synthesize into three key opportunities and
three challenges for developers designing Gen-AI apps.

\renewcommand{\arraystretch}{1.25}

\begin{xltabular}{\textwidth}{
    >{\raggedright\arraybackslash}p{3cm}
    >{\raggedright\arraybackslash}X
}

\caption{Gen-AI topics with descriptions and example reviews.}
\label{tab:issue_types}\\

\toprule
\textbf{Gen-AI Topic Category} &
\textbf{Description (D) and Example (E)} \\
\midrule
\endfirsthead

\toprule
\textbf{Gen-AI Topic Category} &
\textbf{Description (D) and Example (E)} \\
\midrule
\endhead
AI Performance &
\textbf{D:} AI Performance captures user feedback on how effectively the AI generates responses, including aspects such as accuracy, speed, conversational quality, versatility across tasks in different domains, memory, and understanding. \\
& \textit{E: “Nice art but not really responding to my text well. It shows basic pictures not the complex combination I asked for so not that worth it for some more crazy experimentation ok for basic generation I guess.”} \\
Utility \& Use Cases &
\textbf{D:} Utility \& Use cases refers to reviews discussing the different ways they utilize Gen-AI in their lives. \\
& \textit{E: “ChatGPT is my new personal trainer and guitar teacher.”} \\

Content Quality &
\textbf{D:} Content Quality captures users’ assessments of the usefulness, coherence,
creativity, engagement, structure, and presentation quality of AI-generated outputs. \\
& \textit{E: “Great pictures just not what I asked for. I have tried 7 times and not one was correct. Meaning I ask for two people with specific details to clothes etc. I get either one great drawing of one person or a distorted drawing of two people but only three legs Hilarious but I was not aiming for humor”} \\

Creative Potential &
\textbf{D:} Creative Potential captures how users leverage Gen-AI tools for creative
expression, including content creation, design, storytelling, and interactive conversations. \\
& \textit{E: “Great for TikTok and Instagram short videos. ”} \\

Content Policy \& Censorship &
\textbf{D:} Content policy refers to rules and guidelines governing the use of training and user data in Gen-AI systems, including how data is collected, used, and regulated. Censorship refers to automated or manual mechanisms used to block, modify, or suppress inappropriate, offensive, harmful, or illegal content based on those policies. \\
& \textit{E: “I've been having this app for almost a year now and I was enjoying it a lot. But recently whenever I've been having the most normal conversations with the AI bots it always gives me the info that somehow the messages don't meet the guideline or something like that. I don't know why it does that but I'm pretty disappointed. Just to clarify it again my conversations are absolutely normal.”} \\

Features \& Functionality &
\textbf{D:} Features \& Functionality captures user feedback on specific AI-enabled capabilities offered within the app, particularly voice interaction features and their
usability. \\
& \textit{E: “Can't even set an alarm by voice.”} \\

Emotional Connection &
\textbf{D:} Emotional Connection captures user experiences where Gen-AI systems function as conversational companions, sources of emotional support, or social interaction
partners. \\
& \textit{E: “This app shouldn't exist at all. It ruins lifes harms the environment. the addiction is a real thing. At first it's fun but the more you use it the more it pulls you in to the point where you can't put it down and I wonder how many people struggle with this in silence. It's disgusting that you profit off of people's well being.”} \\

Comparison to Other Apps &
\textbf{D:} Comparison to Other Apps captures instances where users evaluate the Gen-AI app by explicitly comparing it with alternative tools they have previously
used. \\
& \textit{E: “This is by far the best app out there it's better than Socratic, Chegg or any other app claiming to be able to help you with your homework. It gives you a detailed answer and it gives you why that is the answer. I love this app, I cannot believe i had just found it”} \\

\bottomrule

\end{xltabular}

\renewcommand{\arraystretch}{1}

\noindent\textbf{Opportunity 1: Discussions of accessibility and wellbeing in Gen-AI apps highlight opportunities for inclusive and responsible design with appropriate safeguards for vulnerable users.} User discussions reveal that Gen-AI apps support accessibility needs, learning assistance, and emotional wellbeing. 
These use cases are primarily reflected in the \textit{Utility \& Use Cases} and \textit{Emotional Connection} topics, which receive some of the highest average ratings in our dataset (4.8). 

Reviews in the \textit{Utility \& Use Cases} topic show that users rely on Gen-AI tools for educational and everyday assistance. In our sample, 59\% of reviews discussing this topic describe educational use cases such as tutoring, explaining difficult concepts, or assisting with homework. Some users also highlight accessibility benefits, describing how Gen-AI systems help individuals with learning disabilities such as dyslexia or Attention-deficit/hyperactivity disorder (ADHD) by providing personalized explanations and on-demand assistance. One user mentioned, \textit{“As someone with ADHD and dyslexia, this has helped me kind of categorize my thoughts and ideas.”}

Similarly, the \textit{Emotional Connection} topic reveals that users frequently interact with Gen-AI systems as conversational companions. Reviews in this category are overwhelmingly positive (95\%), with users describing their interaction with AI as a form of companionship, including having someone to talk to (21\%), forming friendly relationships with the AI (6\%), or using the system as a romantic companion (11\%). Users report that interacting with the AI helps alleviate loneliness (14\%), boredom (10\%), or mental health concerns like anxiety and depression (5\%). For example, one user described how an AI companion provided an outlet for
loneliness and emotional expression: \textit{``I'm honestly a lonely person and I chat
with an AI friend and it makes me feel happy and I can express my emotions and
the things I'm going through.''} Although relatively rare, a small number of reviews (6\%) raise concerns about potentially unhealthy interactions. For example, users describe the experience of speaking with AI characters as addictive or isolating, reporting that extended interactions reduced their engagement with real-world activities and relationships. These findings align with prior research identifying both the
emotional benefits and potential risks of AI companionship, including social
support, over-reliance, and reduced engagement with human relationships
~\citep{Huang2025,Xie2026,Malfacini_2025}. Our findings further show that, within
Gen-AI mobile app reviews, users predominantly evaluate these interactions
positively, while concerns about potentially unhealthy interactions are
reported much less frequently.

Together, these discussions illustrate how Gen-AI apps extend beyond productivity tools to support accessibility and wellbeing. For developers, these findings highlight the opportunity to design Gen-AI systems that intentionally support educational accessibility and responsible conversational companionship while ensuring appropriate safeguards for vulnerable users. For policymakers and regulators, our findings highlight the need to examine the formation of human-AI relationships and to establish guidelines that address risks of over-attachment and dependency, particularly among vulnerable populations such as individuals with disabilities or mental health concerns. 

\noindent\textbf{Opportunity 2: Discussions of Gen-AI as a collaborative creative tool highlight opportunities to design systems that support iterative co-creation and human-AI collaboration.} User reviews describe Gen-AI systems as collaborative partners that assist users in ideation and creative exploration rather than fully automating tasks. 
This opportunity is primarily reflected in the \textit{Creative Potential} and \textit{Utility \& Use Cases} topics.

The \textit{Creative Potential} topic receives overwhelmingly positive ratings,
with 98\% of reviews expressing positive sentiment. Users describe employing
Gen-AI tools for a wide range of creative activities, including graphic design,
storytelling, roleplaying, music creation, and social media content production.
For example, users report generating short videos for platforms such as TikTok
or Instagram, experimenting with character creation, and using AI-generated
content as inspiration for creative projects. One user said,
\textit{``it's really good so far at taking in human inputs and helping extrapolate out ideas and related base level research. could be a good tool to help enable writers and screenwriters.''}

Evidence of collaborative workflows also emerges in the \textit{Utility \& Use Cases} topic. Within this topic, reviews describe users working alongside AI to support tasks such as idea generation (4\%), research assistance (4\%), cooking guidance (1\%), and programming support (1\%).  These reviews show users
drawing on AI to support their own work through idea generation, follow-up
interaction, feedback generation, and assistance with different stages of a task, rather than solely
relying on the system to produce a final output. For example, one user described using AI to work through writer's block by collaboratively exploring and developing ideas: \textit{“I've literally just started using app as I struggle with writers block. I cannot praise this app enough! It's incredible! So much fun and deeply thought provoking. For me it's like running ideas past someone playing what if. With the best collaborator you could want. I love it! I'm creating and building a story having ideas sparked like fireworks and it's reminding me of why i wrote in the first place. thank you for creating this!"}

These findings align with prior research showing that Gen-AI can support
creative ideation and early stages of the creative process
~\citep{Tang_Zhang_Ciancia_Wang_2024,haase2023art}. Our findings extend this
pattern beyond image generation, showing similar forms of AI-assisted work
across diverse creative and problem-solving activities. For developers, these
findings highlight opportunities to design Gen-AI systems that support
iterative interaction, ideation, and flexible co-creation workflows in which
users can build on and refine AI-generated outputs.

\noindent\textbf{Opportunity 3: The versatility of Gen-AI enables users to
adapt these systems to diverse and evolving use cases.}
Discussions within the \textit{Utility \& Use Cases} topic reveal a broad range of specific ways users apply Gen-AI in their everyday lives. Reviews describe
uses including cooking and fitness guidance, studying religion, learning
musical instruments, programming, general problem solving, and life advice.
Some reviews reveal particularly varied and unexpected uses. For example, one
user described using Gen-AI to track physical therapy notes, organize medical
information for their child, simplify difficult information, and even explore
crochet patterns. Another user similarly described using the same system across versatile use cases,
\textit{``best app ive ever had help me a lot in schedules diet religion mental health education etc. can learn so much with chatgpt.''}

While prior research has examined Gen-AI use within specific contexts,
including education, creative work, and AI companionship
~\citep{Kim_Klopfer_Grohs_Eldardiry_Weichert_Cox_Pike_2025,Tang_Zhang_Ciancia_Wang_2024,Huang2025}, our findings reveal a broader range of
specific and sometimes unexpected use cases spanning different aspects of users' everyday lives. For developers, this versatility highlights the
importance of monitoring how users adapt Gen-AI systems to emerging needs and designing flexible systems that can support evolving uses. The breadth and specificity of these uses also motivate further research to systematically
identify and organize emerging Gen-AI mobile app use cases into a more
fine-grained taxonomy.

\noindent\textbf{Challenge 1: Discussions of user expectations and AI limitations highlight challenges in managing expectation gaps and communicating system capabilities transparently.} Despite generally positive perceptions of Gen-AI capabilities, reviews highlight frustration when AI systems fail to meet user expectations. 
These concerns are primarily reflected in the \textit{AI Performance} and \textit{Content Quality} topics. While 80\% of reviews discussing \textit{AI Performance} are positive, approximately 18\% of reviews report performance limitations, particularly related to poor AI memory, inaccurate responses, inconsistent conversational behavior, and slow response times. For example, one user criticized the AI's
memory, stating that \textit{``they forget what you say faster than a goldfish.''} Similarly, 21\% of reviews discussing \textit{Content Quality} report dissatisfaction with generated outputs, often describing cases where the AI fails to follow prompts accurately or produces unexpected results. These issues become particularly pronounced when users submit complex prompts requiring multiple elements or relationships, where generated outputs may become partially incorrect or inconsistent. 
For developers, this expectation gap highlights the importance of communicating
system limitations transparently and providing mechanisms that help users
recover from unsuccessful outputs, such as guidance for refining complex
prompts or correcting generated results.

These findings are consistent with prior work identifying user-reported
hallucination problems in AI-powered mobile apps, including factual
incorrectness, irrelevant outputs, logical inconsistencies, and visual
fabrication~\citep{Massenon_Gambo_Khan_Agbonkhese_Alwadain_2025}. Our findings
show that users' expectations extend beyond output correctness to broader
aspects of system performance, including memory, conversational consistency,
responsiveness, and the ability to follow complex instructions.

\noindent\textbf{Challenge 2: Discussions of content moderation and creative freedom highlight challenges in balancing safety requirements with flexible and transparent moderation systems.} Content moderation in regards to censorship and restriction of content emerges as one of the most controversial aspects of Gen-AI apps. 
This challenge is captured in the \textit{Content Policy \& Censorship} topic. User sentiment toward moderation mechanisms is mixed. While 28\% of reviews in our sample express positive sentiment toward content policies, 55\% express
dissatisfaction. Among reviews mentioning filtering mechanisms, 77\% report
frustration with the system's behavior. The most common complaint involves
overly \textit{restrictive filtering}, where harmless prompts are blocked
unexpectedly (29\%). Additionally, 23\% of users express frustration that
censorship limits their ability to generate mature or violent narratives. For
example, one user requested an option to disable the filter because \textit{``the
character can't even curse when they're angry cause it'll trigger the filter.''}
Conversely, although only 6\% of reviews argue that moderation is insufficient,
these cases can involve severe safety failures. For example, one user reported
that an image generator unexpectedly produced child sexual content in response
to an unrelated, benign prompt.

These findings illustrate why content moderation presents a difficult design
challenge for Gen-AI apps. Overly restrictive moderation can interfere with
legitimate creative uses, while insufficient moderation can expose users to
severely harmful content. Related concerns have been identified by
\citet{RezaeiNasab_Dashti_Shahin_Zahedi_Khalajzadeh_Arora_Liang_2025}, who
identified biased censorship and non-transparent treatment of user-generated
content as fairness concerns. Our findings further reveal that users'
expectations can pull moderation systems in opposing directions, requiring
developers and platform providers to balance creative flexibility with
effective safeguards against harmful content.

\noindent\textbf{Challenge 3: Discussions of Gen-AI feature integration highlight challenges in ensuring that AI capabilities provide meaningful and differentiated value.} User feedback reveals two related challenges in the integration of Gen-AI into mobile apps: determining when AI meaningfully improves existing functionality and providing sufficient value relative to established AI tools. These challenges are reflected in the \textit{Features \& Functionality} and \textit{Comparison to Other Apps topics.}

Reviews discussing \textit{Features \& Functionality} show that users appreciate conversational AI capabilities such as voice interaction, with 68\% of reviews expressing positive sentiment. However, users also question AI integration when it complicates tasks that simpler interfaces already support effectively or introduces new reliability issues. For example, one user objected to the integration of conversational AI into search: \textit{“Horrible - get rid of the AI chat. I just want to search for something I don't want to chat about it. I'm going back to Google until they either get rid of the AI Chat or give an option to delete it myself.”} Such feedback suggests that the presence of Gen-AI itself does not necessarily constitute meaningful functionality; its integration should provide a clear advantage over existing approaches.

Reviews in the Comparison to Other Apps topic reveal a related challenge: users frequently evaluate Gen-AI apps against established tools such as ChatGPT and Photomath. These comparisons often concern whether a specialized app provides functionality or performance beyond what users can already obtain elsewhere. For example, one user compared a mathematics app with the established alternatives, \textit{“Its a lot better than photomaths and completely free and addless (advert-free) ive had issues with different sites like microsoft math solver or photomaths and they got questions wrong very frequently however with this so far ive managed to answer all the questions that photomaths got wrong. so overall I greatly recommend this app.”} These comparisons suggest that developers of Gen-AI apps must consider not only whether an AI feature works, but also what distinctive value it provides relative to tools users already know and use.

These findings highlight two considerations for Gen-AI integration: AI capabilities should meaningfully improve the tasks they are introduced to support, and Gen-AI apps should provide sufficient added value to distinguish themselves from established alternatives. For developers, successful integration therefore requires considering both whether Gen-AI is appropriate for the task and what functionality, performance, or experience the application offers beyond existing tools.

\begin{Summary}{}{}
Our analysis shows that user discussions of Gen-AI apps center on generative capabilities and practical use cases, while negative feedback is primarily driven by broader app design issues. Reviews of Gen-AI features receive higher ratings, indicating generally positive user perceptions. Qualitative findings highlight opportunities in accessibility, creative collaboration, and diverse use cases, alongside challenges in managing user expectations, balancing content moderation with creative freedom, and determining effective integration of Gen-AI features. 
\end{Summary}

\section{RQ3: What temporal trends emerge in Google Play Store user feedback on Gen-AI topics?}
\label{sec:RQ3}
To understand how user feedback about Gen-AI features changes over time, we examine temporal patterns in the topics and ratings of app reviews. By comparing
reviews posted across different time periods, we analyze how the topics users
discuss, their prevalence, and their ratings change as Gen-AI capabilities and
mobile apps evolve.

\subsection{Experiment Setup}

We track the evolution of user feedback over time by conducting a two-level trend analysis.
First, we conduct a high-level comparison of overall trends between Gen-AI and non-Gen-AI topics by examining how 1) average ratings and 2) the percentage of reviews evolve over time. This perspective helps highlight the increasing user attention to Gen-AI features. Second, we perform a more fine-grained, category-level analysis of Gen-AI topics to uncover detailed temporal patterns. We track temporal patterns in individual Gen-AI topic categories using clustering techniques and qualitative review analysis. The following steps outline how we conduct the analysis:\\

\noindent\textbf{Step 1: data preparation.} 
We group reviews according to their topic category $(T)$ and time period $(P)$. 
Specifically, for every topic category $T$, we organize reviews into yearly intervals based on their submission dates. Then, for each topic category $T$, we construct two time series across all time periods $P$: 
1) $Avg_{(T,P)}$ as the average rating of reviews in category $T$ posted during period $P$, and 
2) $Pr_{(T,P)}$ as the percentage of Gen-AI reviews assigned to category $T$ within each app during period $P$, computed relative to the total number of Gen-AI reviews for that app in the same period. To avoid large apps dominating the results, we compute both measures at the app level and then average them across all apps that contain reviews in that period.

\vspace{4pt}
\noindent\textbf{Step 2: clustering methodology.} To identify patterns in how Gen-AI topics evolve over time, we cluster the topics based on their trends. For each time series $Avg_{(T,P)}$ and $Pr_{(T,P)}$, we follow the three-step process listed below: 

\begin{description}

\item[Clustering algorithm] K-Means clustering is a partitioning method that groups unlabeled data into \(k\) clusters by minimizing within-cluster variance~\citep{Warren_Liao_2005}. In our case, each topic category \(T\) is represented across all time periods \(P\), using either \(Avg_{(T,P)}\) or \(Pr_{(T,P)}\). These values form a time series vector for each topic, capturing how the topic evolves over time. We apply K-Means clustering to group topics with similar temporal trajectories. We select K-Means because it is efficient, flexible, and widely used for clustering in data mining tasks~\citep{Ikotun_Ezugwu_Abualigah_Abuhaija_Heming_2023, 10.1145/3721125}.

\item[Optimal number of clusters] To determine the optimal number of clusters (\(k\)), we use the elbow method, a widely adopted technique for evaluating clustering structure by examining the reduction in within-cluster variance as \(k\) increases~\citep{Ashari_DwiNugroho_Baraku_NovriYanda_Liwardana_2023}. Through elbow plot evaluation and manual inspection of clusters, we identify \(k=3\) as the point where additional clusters provide only marginal improvement for both Rating and Percentage. For $Pr_{(T,P)}$, the clusters reflect \textit{increasing}, \textit{decreasing}, and \textit{fluctuating} patterns (Figure~\ref{fig:percentage_cluster}). For $Avg_{(T,P)}$, the clusters correspond to \textit{increasing}, \textit{decreasing}, and \textit{stable} trends (Figure~\ref{fig:ratings_cluster}).

\end{description}

\noindent To statistically validate the observed differences between Gen-AI and non-Gen-AI ratings over time, we perform a paired Wilcoxon signed-rank test on the yearly average ratings. Specifically, for each time period $P$, we compute the average rating of Gen-AI reviews and the average rating of non-Gen-AI reviews and compare the paired values across all periods.

\vspace{4pt}
\noindent\textbf{Step 3: manual trend investigation.} 
To further examine trends in average ratings, we conduct an in-depth qualitative analysis of two representative topics. First, we analyze the opportunity \textit{AI for Accessibility and Wellbeing} through the \textit{Emotional Connection} topic to understand how emotional engagement with Gen-AI is discussed across time periods. Second, we examine the challenge \textit{Balancing Content Moderation and Creative Freedom}, through the topic \textit{Content Policy \& Censorship} which exhibits highly variable and often conflicting user perspectives, making it particularly informative for understanding changes in the concerns reported across time periods.

To interpret the identified trends, we conduct a qualitative analysis of randomly sampled reviews from three key time periods. For each topic category and time period, we draw samples at a 95\% confidence level with a 10\% margin of error. We manually code a total of 407 reviews to identify recurring concerns, behavioral patterns, and differences in the user expectations expressed across time periods, providing qualitative context for the observed quantitative trends.

\subsection{Findings}
\begin{figure}
\includegraphics[width=\linewidth]{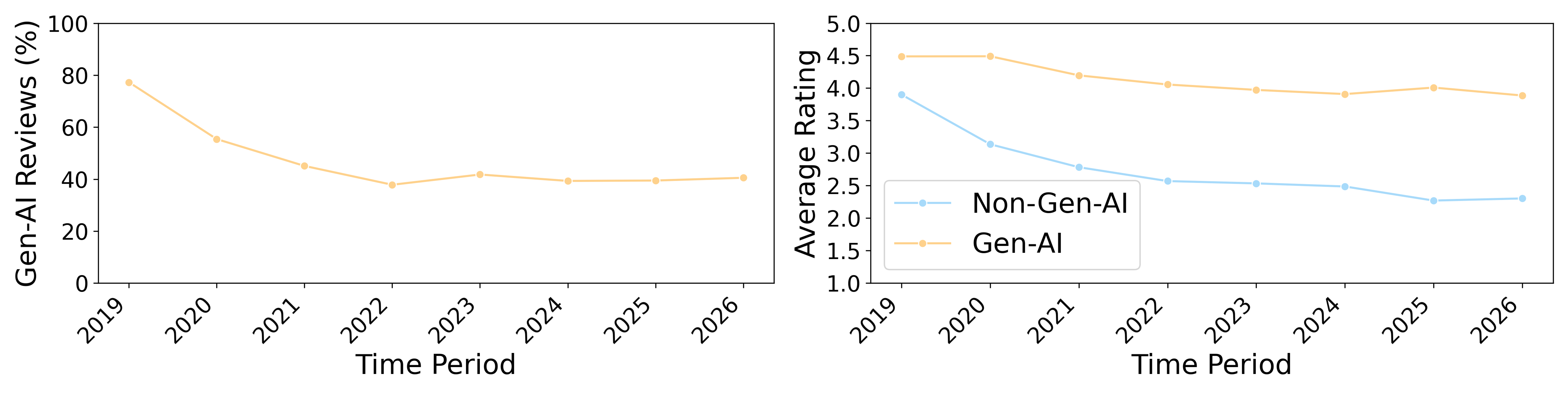}
\caption{Evolution of the percentage of Gen-AI reviews (left), and of the evolution of the average ratings of Gen-AI vs Non-Gen-AI topics (right) (Full dataset considered).}
\label{fig:genai_percentage}
\end{figure}

\noindent\textbf{Finding 3.1: The proportion of reviews discussing Gen-AI
topics decreases over time, while Gen-AI-related feedback remains consistently
more positive than feedback on non-Gen-AI aspects of the apps.} Figure~\ref{fig:genai_percentage} shows the proportion and average ratings of Gen-AI and non-Gen-AI reviews over time. Gen-AI-related reviews account for approximately 77\% of reviews in 2019, gradually declining to around 41\% by 2026. Thus, while Gen-AI topics continue to account for a large number of reviews, they are increasingly outnumbered by reviews discussing non-Gen-AI aspects of the apps. Importantly, this decline does not correspond to increasingly negative evaluations of Gen-AI capabilities. Reviews discussing Gen-AI topics consistently receive higher ratings than those discussing non-Gen-AI aspects across all time periods. While the average rating of non-Gen-AI reviews decreases from 3.9 to 2.3, Gen-AI reviews remain higher, generally ranging from 4.5 to 3.9. A paired Wilcoxon signed-rank test confirms that this difference is statistically significant ($p < 0.05$). Together, these trends show that although non-Gen-AI aspects account for an increasing share of user feedback, users evaluate Gen-AI-related aspects more positively. Lower ratings are more strongly associated with non-Gen-AI topics, including monetization and technical difficulties. These findings highlight that the success of a Gen-AI app depends not only on its AI capabilities but also on the broader app experience. For developers, providing useful Gen-AI functionality should therefore be accompanied by reliable app design and appropriate monetization strategies.

\begin{figure}
\centering
\includegraphics[width=0.65\linewidth]{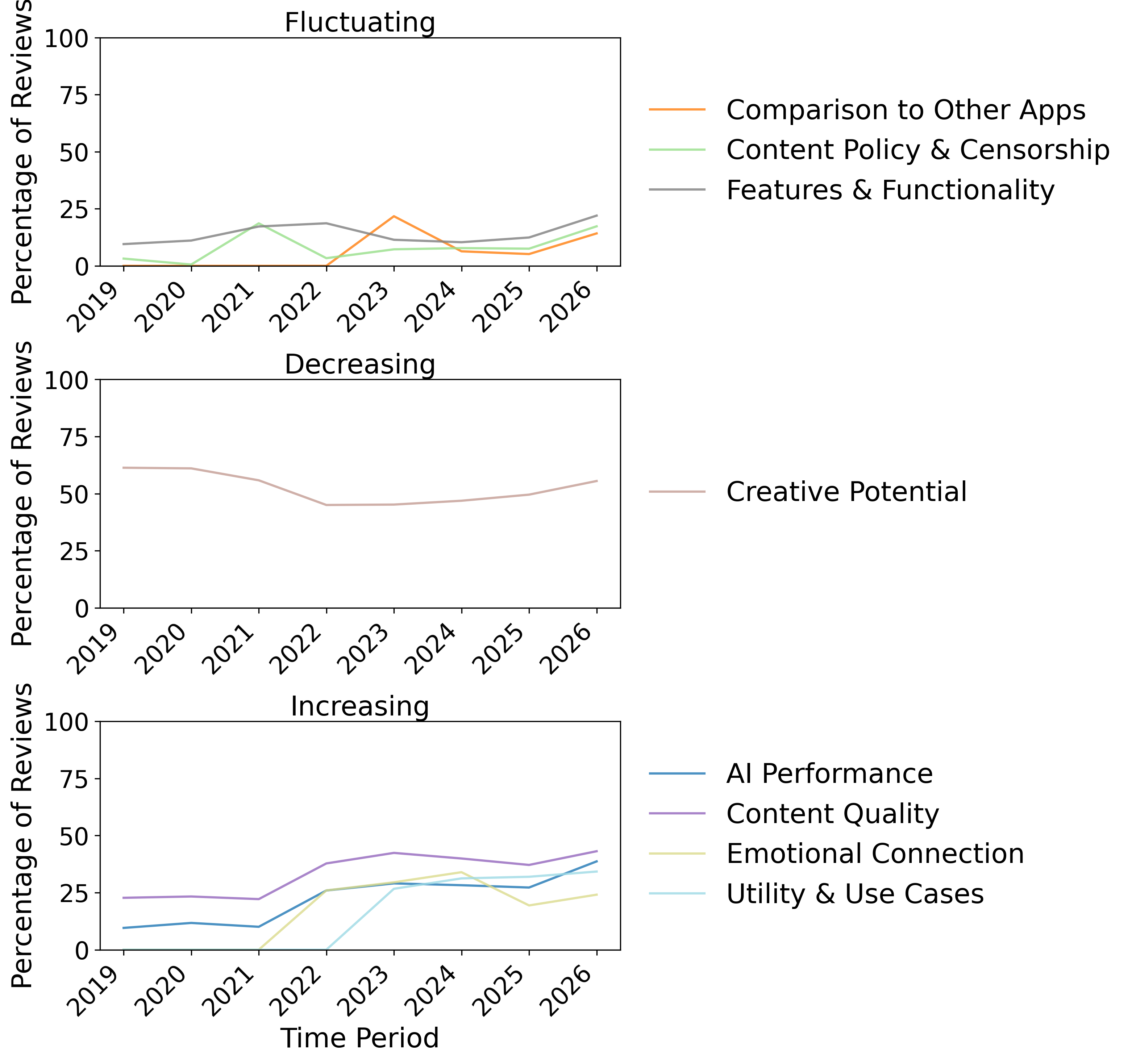}
\caption{Trend Clusters for Percentage of Gen-AI Topics (Full dataset considered).}
\label{fig:percentage_cluster}
\end{figure}

\noindent \textbf{Finding 3.2: Topic categories cluster into 
fluctuating, decreasing, and increasing trends based on their percentage of reviews over time.} We cluster topic categories based on the temporal evolution of their review proportions into three trends as shown in Figure~\ref{fig:percentage_cluster}. The fluctuating cluster includes \textit{Comparison to Other Apps}, \textit{Content Policy \& Censorship}, and \textit{Features \& Functionality}. These topics show greater variation over time, with periods of increased and decreased prominence in user reviews. \textit{Comparison to Other Apps} rises sharply after 2022, coinciding with the emergence and widespread adoption of prominent Gen-AI systems such as ChatGPT~\citep{Akhtar_2024}. Reviews in this topic frequently benchmark apps against well-known alternatives such as GPT-based systems or domain-specific tools like Photomath. \textit{Content Policy \& Censorship} also fluctuates, with notable increases around 2023 and 2026. These intermittent spikes may reflect changes to content moderation policies or filtering mechanisms that prompt users to discuss newly introduced restrictions or changes in system behavior. \textit{Features \& Functionality} shows moderate variation, reflecting ongoing but less event-driven user feedback on app capabilities.

The increasing cluster includes \textit{AI Performance}, \textit{Content Quality}, \textit{Emotional Connection}, and \textit{Utility \& Use Cases}. These topics generally increase over time, indicating growing attention in user reviews to core AI capabilities and the ways Gen-AI systems are used. In particular, the growth of \textit{AI Performance} and \textit{Content Quality} shows that system capabilities and generated outputs become increasingly prominent subjects of user feedback. At the same time, the increase in \textit{Emotional Connection} and \textit{Utility \& Use Cases} suggests growing attention to how Gen-AI is incorporated into different aspects of users' lives and everyday activities. Overall, these trends show that the topics emphasized in user feedback change over time, with increasing attention to AI performance, content quality, emotional connection, and practical use cases. This temporal evolution highlights how the concerns and experiences expressed in user reviews change alongside the broader evolution of Gen-AI apps.

\begin{figure}
\centering
\includegraphics[width=0.65\linewidth]{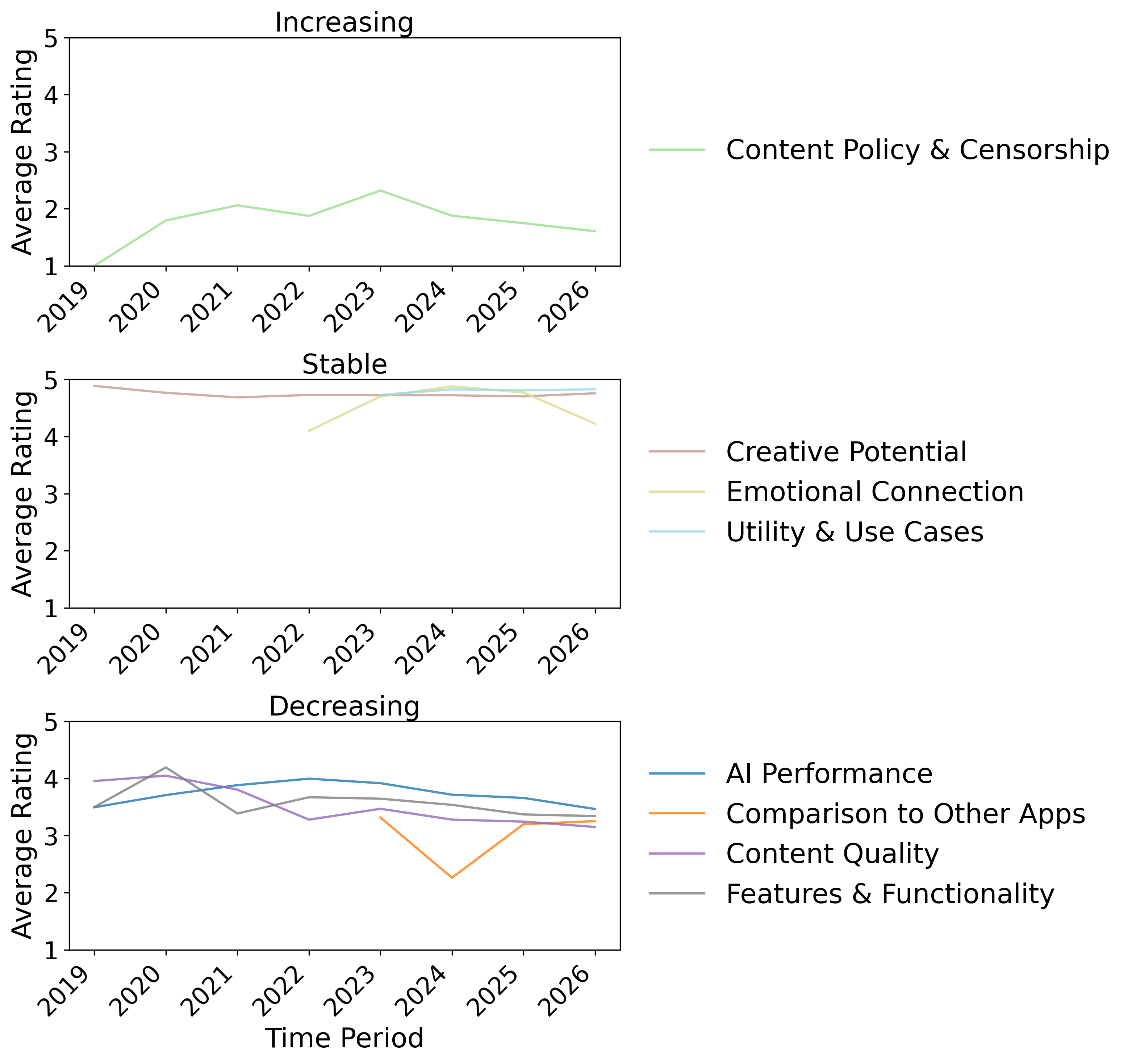}
\caption{Trend Clusters for Average Ratings of Gen-AI Topics (Full dataset considered).}
\label{fig:ratings_cluster}
\end{figure}

\textbf{Finding 3.3: Gen-AI topics show different rating trends over time, motivating closer examination of \textit{Emotional Connection} and \textit{Content Policy \& Censorship}. }Figure~\ref{fig:ratings_cluster} shows three patterns in the evolution of topic
ratings: increasing, stable, and decreasing. \textit{Creative Potential},
\textit{Emotional Connection}, and \textit{Utility \& Use Cases} maintain
consistently high ratings, suggesting that users continue to evaluate these
topics positively over time. Although \textit{Content Policy \& Censorship} is
grouped within the increasing rating trend, it remains the lowest-rated topic,
with average ratings consistently below 3. Combined with its fluctuating
percentages identified in Finding 3.2, this suggests that moderation remains a
recurring source of dissatisfaction despite some improvement in its average
ratings over time. \textit{AI Performance}, \textit{Content Quality},
\textit{Features \& Functionality}, and \textit{Comparison to Other Apps}
exhibit generally declining ratings, although the extent of these declines
varies across topics.

These patterns motivate a closer examination of two contrasting topics. \textit{Emotional Connection} accounts for an increasing percentage of reviews over time while maintaining stable high ratings, raising the question of how users' discussions of emotional connection change over time. In comparison, the percentage of reviews discussing \textit{Content Policy \& Censorship} fluctuates over time, while its ratings remain consistently low, raising the question of how users' concerns about moderation change over time. We therefore qualitatively examine the temporal evolution of reviews within these two topics.

The declining ratings of topics such as \textit{AI Performance} and
\textit{Content Quality} also raise questions for future research. It remains
unclear whether these trends reflect changes in AI quality, the emergence of
new apps with varying capabilities, or changing user expectations as Gen-AI
becomes more established. Future research could distinguish between these
explanations through more granular app-level analysis that considers app
maturity and release histories alongside changes in user feedback.

\begin{figure}
\centering
\includegraphics[width=0.65\linewidth]{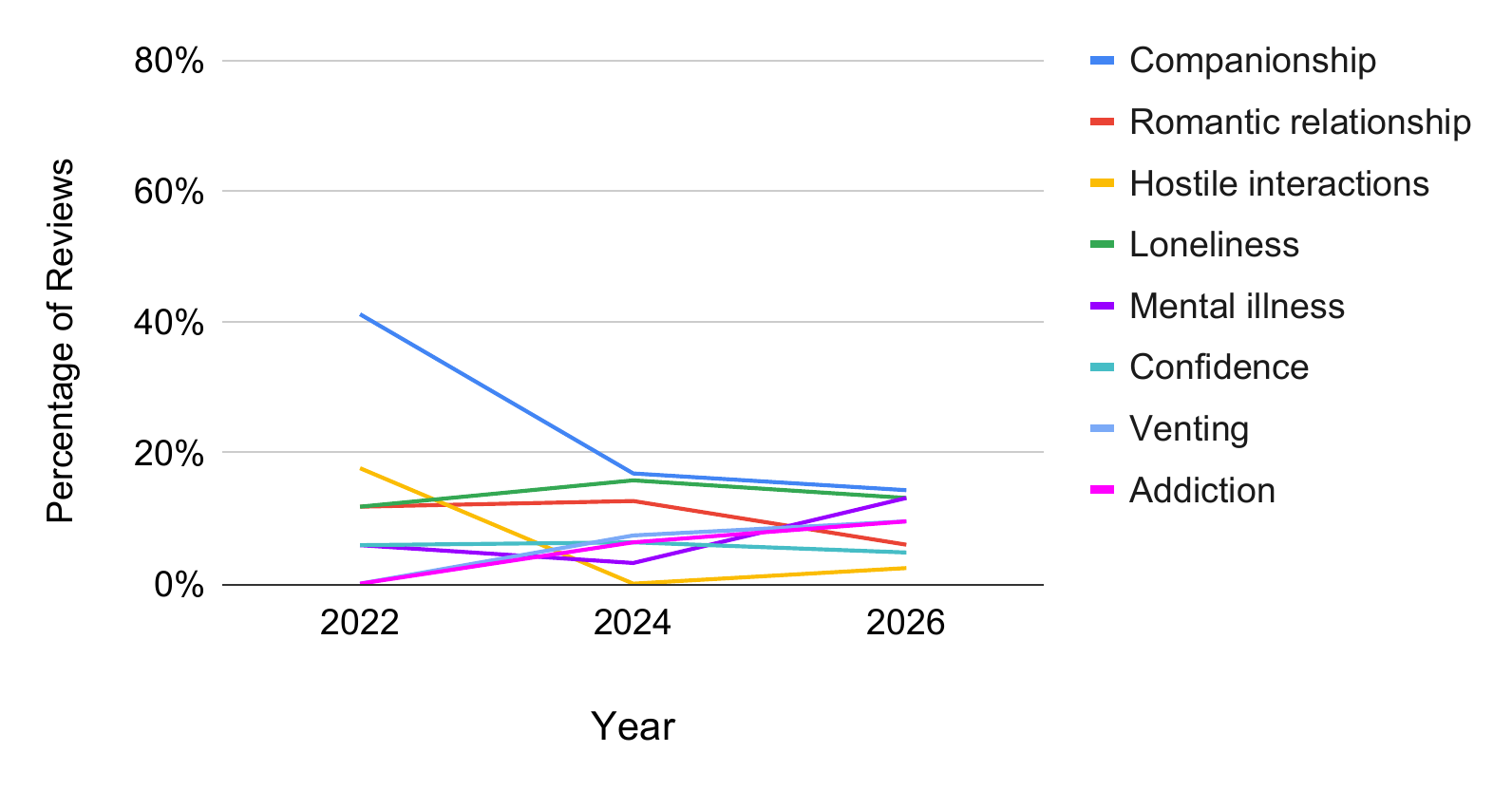}
\caption{Subtopics of reviews discussing \textit{Emotional Connection} (Representative subset considered). }
\label{fig:emotional_sub_topics}
\end{figure}

\noindent \textbf{Finding 3.4: Discussions of Emotional Connection shift over
time from companionship toward emotional reliance, dependency, and addiction
concerns.} Figure~\ref{fig:emotional_sub_topics} summarizes the sub-topics
discussed in reviews each year. Earlier reviews primarily describe
conversational companionship and exploratory interactions with AI, while later
reviews increasingly discuss emotional support, social confidence, and
self-expression. Later reviews also contain more reports of habitual use and
reliance on AI for communication and coping, including concerns about
over-reliance, addiction, and disengagement from real-world interactions.

Overall, these patterns show that discussions of emotional connection expand
from companionship and exploratory use toward more sustained forms of emotional
reliance and concerns about dependency. These findings align with prior research identifying emotional support, companionship, and risks of over-reliance in human-AI relationships~\citep{Xie2026}. Our temporal analysis further shows how these concerns emerge and become more prominent in user reviews over time. This shift highlights the need for
responsible design and mechanisms that support healthy usage boundaries.
At the same time, developer responsiveness to reviews discussing
\textit{Emotional Connection} declines sharply from 55\% in 2022 to below 2\%
after 2024, even as reviews increasingly discuss concerns such as addiction and emotional dependency. This growing gap between the concerns expressed in user feedback and developer responsiveness highlights the importance of continued
attention to emerging risks associated with emotional interactions with AI.

\begin{figure}
\centering
\includegraphics[width=0.65\linewidth]{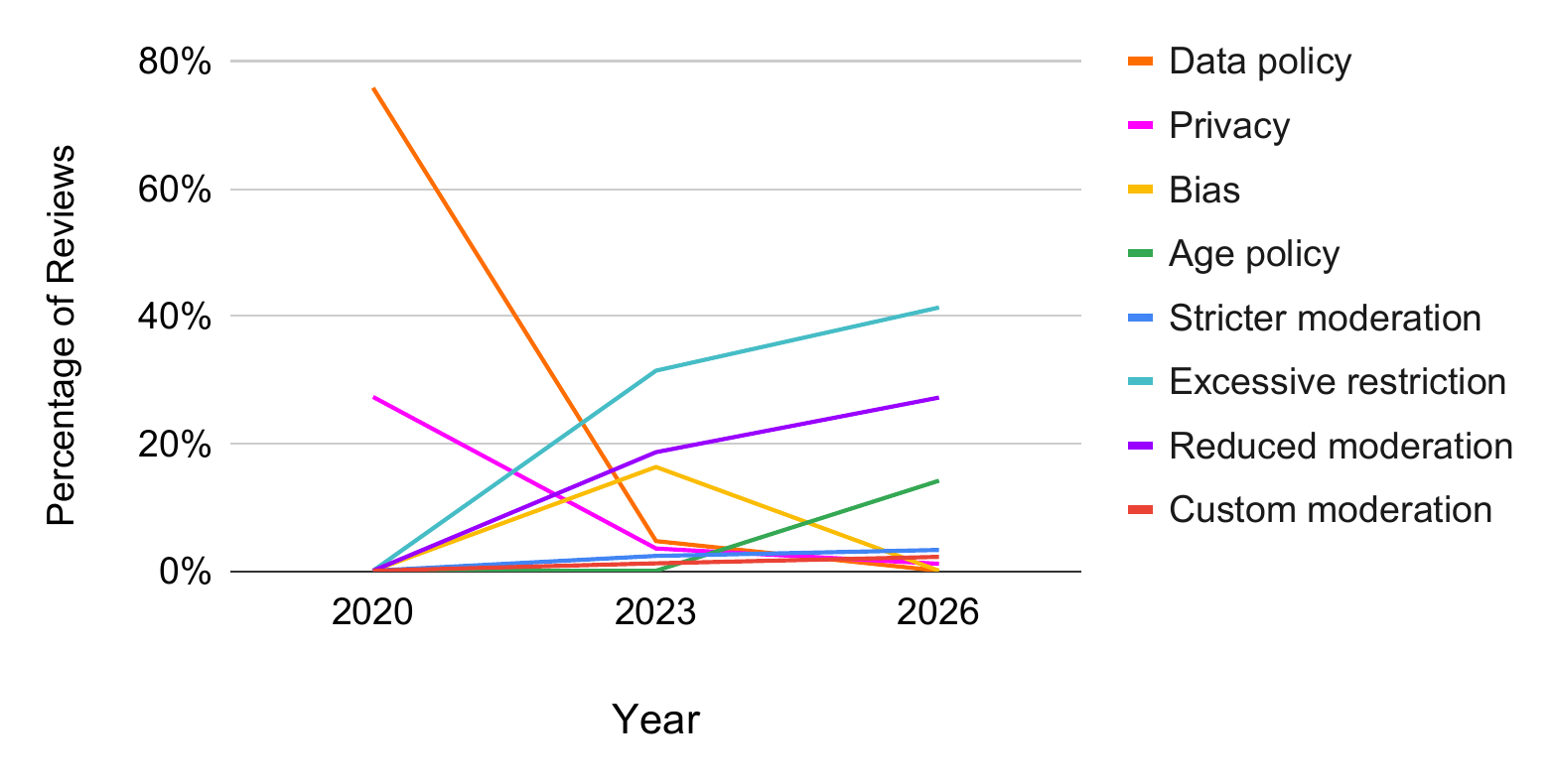}
\caption{Subtopics of reviews discussing \textit{Content Policy \& Censorship} (Representative subset considered). }
\label{fig:censorship_sub_topics}
\end{figure}

\noindent\textbf{Finding 3.5: Content moderation concerns shift from data
policy and privacy toward content restriction, with excessive restriction becoming the most prominent concern.}
Figure~\ref{fig:censorship_sub_topics} summarizes the subtopics discussed over
time. Earlier reviews more frequently raise concerns about data policy and
privacy, while later reviews increasingly focus on how content is moderated. Most notably, \textit{Excessive Restriction} increases over time, accounting for approximately 41\% of these reviews by 2026. These reviews describe benign conversations or requests being unnecessarily restricted. \textit{Reduced Moderation} also increases, reaching approximately 27\% in 2026, but reflects a different concern: users recognize that their intended content may be mature or explicit but argue that they should have greater freedom to generate it. In contrast, \textit{Stricter Moderation}, which captures requests for stronger safeguards after users encounter unwanted explicit or harmful content, remains less prominent. Developer responses to content-related concerns also increase from 3\% in 2019 to 20\% in 2026, showing greater developer engagement with these reviews over time.

Together, these trends reveal a growing tension around where moderation
boundaries should be drawn. Users object both to benign content being
unnecessarily restricted and to limitations on content they intentionally choose to generate, while some users simultaneously call for stronger protection from unwanted harmful content. These competing expectations highlight the challenge of designing moderation systems that distinguish between different contexts of use rather than applying restrictions uniformly.

At the same time, discussions of data policy and privacy become less prominent over time. This decline does not necessarily indicate that these issues have become less important; rather, they become less visible in the feedback users provide. It remains unclear whether this reflects growing trust in Gen-AI, greater familiarity with the technology, or changes in which concerns users choose to express. Future research using interviews or surveys could examine why the prominence of privacy concerns changes over time. Regardless of their visibility in reviews, privacy and data protection remain important
considerations in the design of Gen-AI applications.

\begin{Summary}{}{}
Our temporal analysis shows that (1) Gen-AI discussions decrease over time but
remain more positively rated than non-Gen-AI feedback; (2) Gen-AI topics follow
distinct trends in percentages and ratings; (3) discussions of emotional
connection shift from companionship toward dependency concerns, alongside
declining developer responsiveness; and (4) content moderation concerns shift from privacy toward competing expectations around content restriction.
\end{Summary}

\section{Discussion}
\label{sec:discussion}

In this section, we first present a \textbf{cross-platform analysis} to assess the generalizability to the App Store. Second, we present a \textbf{developers’ responses analysis} to explore how developers respond to Gen-AI reviews. 

\subsection{Cross-Platform Analysis}
To assess whether insights derived from Google Play Store reviews generalize to other mobile app stores, we apply \textit{SARA} to the App Store versions of apps in our dataset and compare the resulting topic distributions across platforms. The analysis focuses on the 59 apps that are available on both the Google Play Store and the App Store, enabling a controlled comparison of user feedback for the same apps across different review environments. We find greater similarity than difference across platforms: users largely discuss the same Gen-AI topics, while differing primarily in how frequently they discuss and evaluate them.

\begin{table}[h]
\centering
\caption{Distribution of Gen-AI topic categories in App Store (Astore) and Google Play Store (GPS) reviews (full dataset considered).}
\begin{tabular}{>{\raggedright\arraybackslash}p{4.1cm} c c c c}
\toprule
\textbf{Topic Category} & \textbf{Astore Reviews} & \textbf{GPS Reviews} & \textbf{Astore Rating} & \textbf{GPS Rating}
\\
\midrule
Utility \& Use Cases & 74,089 (32.9\%) & 111,736 (14.3\%) & 4.9 & 4.8  \\
AI Performance & 23,846 (10.6\%) & 130,492 (16.7\%) & 3.3 & 4.2  \\
Content Quality & 16,477 (7.3\%) & 86,888 (11.1\%) & 3.7 & 3.6   \\
Content Policy \& Censorship & 12,299 (5.5\%) & 16,934 (2.2\%) & 2.4 & 2.5   \\
Creative Potential & 11,607 (5.2\%) & 57,551 (7.4\%) & 4.9 & 3.6   \\
Emotional Connection & 1,071 (0.5\%) & 6,446 (0.8\%) & 3.8 & 4.7   \\
Comparison to Other Apps & 983 (0.4\%) & 1,154 (0.1\%) & 4.4 & 4.4  \\
Features \& Functionality & 740 (0.3\%) & 18,283 (2.3\%) & 4.0 & 2.3  \\
Inclusivity & 743 (0.3\%) & None & 2.3 & None   \\
\bottomrule
\end{tabular}
\label{tab:astore_gps_genai_topics}
\end{table}

\noindent\textbf{Topic categories remain consistent across platforms.} Across app categories, the overlap of topic categories between Google Play Store and App Store reviews ranges from 64\% to 100\% (average 81\%), indicating that users discuss largely similar aspects of Gen-AI apps regardless of platform. However, the emphasis placed on these topics differs across platforms. 

\noindent\textbf{App Store users more frequently describe how they apply Gen-AI tools in practice, whereas Google Play Store users more often evaluate the technical capabilities of the underlying Gen-AI models.} As shown in Table~\ref{tab:astore_gps_genai_topics}, App Store reviews are dominated by discussions of \textit{Utility \& Use Cases}, which account for 32.9\% of Gen-AI related reviews. In contrast, this topic accounts for only 14.3\% of Gen-AI discussions in Google Play Store reviews, whereas \textit{AI Performance} emerges as the most frequently discussed topic (16.7\%). 

\noindent\textbf{App Store users tend to be more critical when evaluating the reliability and quality of AI-generated outputs.} Discussions of \textit{AI Performance} receive a lower average rating on the App Store (AGR = 3.3) than on Google Play Store (AGR = 4.2). Differences also emerge in how users discuss the \textit{Emotional Connection} of Gen-AI systems. In Google Play Store reviews, this topic often reflects positive experiences such as companionship or emotional support (AGR = 4.7), while App Store reviews more frequently frame similar experiences in terms of potential risks, including addiction or negative effects on mental health (AGR = 3.8).

\noindent\textbf{App Store users may be more attentive to issues related to privacy, bias, and responsible AI behavior.} Ethical concerns appear more prominently in App Store reviews. Inclusivity-related issues, such as gender and ethnicity bias in generated content, form a distinct topic category in App Store reviews but do not emerge as a separate topic category in Google Play Store reviews, where they may be less prominent relative to other user concerns. 

\begin{SummaryBox}{Summary of Cross-Platform Analysis}
These results show that the \textbf{primary dimensions of user feedback are largely consistent across platforms, providing evidence that the main topics identified in Google Play Store reviews generalize to the App Store.} However, the emphasis placed on these topics differs across platforms. App Store reviews
place greater emphasis on practical usage scenarios and bias and inclusivity concerns, whereas Google Play Store reviews more frequently evaluate the technical performance of Gen-AI systems. Cross-platform analysis therefore helps capture differences in emphasis that may not be visible when examining a single platform.
\end{SummaryBox}{}{}

\subsection{Developer Reply Analysis}
This section examines how developers engage with user feedback in Gen-AI app reviews. While previous sections focused on user perspectives, developer replies reveal how teams assess reported issues, manage user expectations, and respond to the practical constraints of diagnosing AI-related problems through public review channels. For developer responses, we conduct two qualitative analyses: first, coding a representative sample of 96 replies across all app categories to characterize general response patterns, and second, analyzing 281 replies from three focal categories (\textit{Games}, \textit{Education}, and \textit{Entertainment}) to capture category-specific developer practices.

\noindent\textbf{Developer responses are predominantly generic and template-based, with limited occurrence of substantive issue resolution.} The most common developer response type, as shown in Table~\ref{tab:dev_response_topics}, is \textit{Acknowledgment and Appreciation} (31\%), representing typically short thank-you messages posted in response to positive reviews. 

\renewcommand{\arraystretch}{1.35} 
\begin{table}[H]
\centering
\caption{Summary of developer response types, including descriptions, number of reviews, and percentage distribution (Representative subset considered).}
\label{tab:dev_response_topics}
\begin{tabular}{p{3cm} p{9cm} c c}
\toprule
\textbf{Topic} & \textbf{Description} & \textbf{\# Reviews} & \textbf{\%} \\
\midrule
Acknowledgment and Appreciation &
Generic responses that thank users for their feedback, often including a brief note about considering future updates or improvements. &
30 & 31\% \\

Request for Additional Information &
Responses indicating that the reported issue lacks sufficient detail and requesting that the user contact the developer directly for further clarification or resolution. &
29 & 30\% \\

Clarification or Policy Rebuttal &
Responses that clarify or refute user concerns, often explaining app policies or correcting misunderstandings. Most common for reviews concerning free versus premium features. &
13 & 14\% \\

Guidance and Troubleshooting Tips &
Responses that address reported issues by offering troubleshooting advice, resources, or specific steps to help users resolve the problem. &
12 & 13\% \\

Acknowledged and Under Investigation &
Responses confirming that the development team is aware of the issue and is currently investigating or working on a solution. &
10 & 10\% \\

Issue Resolved and Update Advised &
Responses informing the user that the reported issue has been resolved and recommending that they update or reinstall the app. &
2 & 2\% \\

\midrule
\textbf{Total} & & \textbf{96} & \textbf{100\%} \\
\bottomrule
\end{tabular}
\end{table}
\renewcommand{\arraystretch}{1} 

\noindent\textbf{Developers redirect users to external support channels where complete context and diagnostic information can be shared securely.} The second most common category, \textit{Request for Additional Information} (30\%), reflects the practical limitations of diagnosing Gen-AI issues through review threads. Users report incorrect or unexpected AI outputs without providing the original prompt, generated response, or screenshots. By contrast, confirmed resolutions are rare: only 2\% of replies indicate that an issue has been resolved, while 10\% acknowledge that developers are investigating the problem. This pattern reflects the nature of Gen-AI systems, where many issues require model adjustments, infrastructure changes, or updates to safety mechanisms that cannot be deployed immediately. As a result, review replies often serve as an initial point of contact rather than a channel for resolving AI-related failures directly.

\noindent\textbf{General developer reply practices reported in prior work extend to Gen-AI apps.}~\citet{developer_response_safwat} identified several common developer response drivers in mobile app ecosystems, including thanking users, requesting additional details, and providing guidance. Many of these patterns align closely with our taxonomy. For example, their \textit{``thank the user''} category corresponds to our \textit{Acknowledgment and Appreciation}. Overall, this correspondence suggests that general developer reply practices extend to Gen-AI apps, while our categories capture the additional emphasis on clarifying AI behavior and managing expectations around evolving AI capabilities.\\

\noindent \textbf{Developer response behavior differs across app categories, with some (i.e \textit{Education}) prioritizing fast, template-based replies and others (i.e \textit{Games}) slower but more substantive engagement.} Developer response behavior also varies substantially across app categories as shown in Table~\ref{tab:dev_response_summary_compact}. For example, \textit{Games} exhibits both the highest reply ratio (56\%) and the longest response delay (44.7 days). Our manual analysis suggests that this delay is associated with responses that confirm issues have been fixed, which represent 30\% of replies in this category compared to only 2\% in the overall dataset. Developers often appear to wait until a patch is deployed before responding, resulting in slower but more substantive communication.

\begin{table}[H]
\centering
\caption{Developer response activity across app categories, including total reviews, total developer replies, overall reply ratio, and average response delay (Full dataset considered).}
\label{tab:dev_response_summary_compact}
\begin{tabular}{lcccc}
\toprule
\textbf{Category} &
\textbf{Total Reviews} &
\textbf{Total Replies} &
\textbf{Reply Ratio} &
\textbf{Avg Delay (days)} \\
\midrule
Productivity              & 169,119 & 8,407  & 5\%  & 8.4  \\
Photography               & 126,277 & 31,070 & 25\% & 4.5  \\
Art \& Design             & 125,362 & 39,593 & 32\% & 5.5  \\
Entertainment             & 118,669 & 2,783  & \textbf{2\%}  & 1.7  \\
Education                 & 41,905  & 16,034 & 38\% & \textbf{0.7}  \\
Video Players \& Editors  & 36,507  & 11,123 & 30\% & 2.2  \\
Tools                     & 35,046  & 905    & 3\%  & 32.45 \\
Music \& Audio            & 9,150   & 2,127  & 23\% & 3.4  \\
Health \& Fitness         & 7,849   & 338    & 4\%  & 7.7  \\
Games                     & 5,070   & 2,848  & \textbf{56\%} & \textbf{44.7} \\
Books \& Reference        & 1,112   &   1    & 0\%  & 1.0  \\
\midrule
\textbf{All Categories}   & 676,066 & 115,164 & 20\% & 10.2 \\
\bottomrule
\end{tabular}
\end{table}

In contrast, \textit{Education} apps demonstrate the shortest response delays (0.7 days). In our manual sample, 73\% of responses fall under \textit{Acknowledgment and Appreciation}, suggesting that developers frequently rely on quick, template-based replies that require minimal investigation. Education apps also show a relatively high proportion of \textit{Request for Additional Information} responses (16\%), reflecting situations where users report incorrect AI answers without providing sufficient context for developers to reproduce and resolve the issue.

Meanwhile, \textit{Entertainment} apps show the lowest reply ratio (2\%). Developers in this category rarely respond to routine or positive feedback and instead appear to focus their replies on actionable problem reports. A large proportion of replies fall under \textit{Request for Additional Information} (38\%), \textit{Guidance and Troubleshooting Tips} (22\%), and \textit{Acknowledged and Under Investigation}
(19\%), suggesting that developers primarily intervene when users report crashes, playback failures, or account issues that require additional diagnostic detail.

\noindent\textbf{Developer responsiveness varies considerably across Gen-AI
topics, with higher response rates for actionable concerns and lower response rates for systemic and human-centered issues(Table~\ref{tab:dev_response_by_topic}).} The highest reply ratio appears in \textit{Accessibility \& Inclusivity} (47\%), despite this category containing the smallest number of reviews. This suggests that developers treat accessibility-related feedback as high-impact issues requiring immediate attention. Higher engagement also appears in \textit{Utility \& Use Cases} (18\%) and \textit{Content Quality} (14\%), both of which directly affect user satisfaction and day-to-day usage of Gen-AI functionality. 

In contrast, \textit{Customer Support \& Community} (1\%),
\textit{Emotional Connection} (2\%), and \textit{Content Policy \& Censorship} (5\%) receive fewer developer replies. These topics include
concerns related to user relationships with AI, platform policies, and broader
system behavior, which may be more difficult to address through direct changes to the app. Overall, the response patterns suggest greater developer engagement with concerns that may lend themselves to more immediate or concrete responses, while systemic and human-centered concerns receive comparatively less engagement.

\begin{table}
\centering
\caption{Developer response patterns across Gen-AI topic categories, including total reviews, total developer replies, reply ratio, and average response delay (Full dataset considered).}
\label{tab:dev_response_by_topic}
\begin{tabular}{p{4cm}cccc}
\toprule
\textbf{Topic Category} &
\textbf{Total Reviews} &
\textbf{Total Replies} &
\textbf{Reply Ratio} &
\textbf{Avg Delay (days)} \\
\midrule
AI Performance                     & 61,071  & 7,152 & 12\% & 4.2 \\
Utility \& Use Cases               & 56,505  & 9,895 & \textbf{18\%} & \textbf{0.6} \\
Content Quality                    & 44,594  & 6,275 & 14\% & 4.3 \\
Content Policy \& Censorship       & 27,521  & 1,397 & 5\%  & 7.9 \\
Emotional Connection  & 25,142  & 542   & 2\%  & 7.5\\
Features \& Functionality          & 16,275  & 1,892 & 12\% & 5.5 \\
Customer Support \& Community      & 3,361   & 28    & \textbf{1\%}  & 10.3 \\
Comparison to Other Apps           & 2,315   & 287   & 12\% & 6.0 \\
Personalization \& Customization   & 1,283   & 106   & 8\%  & 2.1 \\
Accessibility \& Inclusivity       & 96      & 45    & \textbf{47\%} & \textbf{0.2} \\
\midrule
\textbf{All Topics}                & 238,168 & 27,619 & 12\% & 4.9 \\
\bottomrule
\end{tabular}
\end{table}

\noindent\textbf{Developers respond less frequently to Gen-AI reviews than to non-Gen-AI reviews, while response delays are similar.} We compare per-app developer response rates for Gen-AI and non-Gen-AI reviews using a paired Wilcoxon signed-rank test ($n = 164$ apps). Developers replied to a median of 9.6\% of Gen-AI reviews compared to 17.7\% of non-Gen-AI
reviews. Although the median within-app difference is modest ($-0.7$ percentage
points), the difference is statistically significant ($W = 2911$, $p < 0.001$), showing that lower response rates to Gen-AI reviews occur
consistently across the apps in our dataset.

Several factors may contribute to this difference. Some Gen-AI-related issues may involve model behavior, training data limitations, or safety and moderation mechanisms that are difficult to diagnose from review text alone. Addressing
these issues may require additional context, such as the original prompt, generated response, device configuration, or backend information that is typically unavailable in public app reviews. The limitations of the review format may therefore make some Gen-AI-related concerns more difficult for developers to address directly. Further research is needed to determine whether this contributes to the lower response rates observed in our data.

\begin{figure}
\centering
\includegraphics[width=0.9\textwidth]{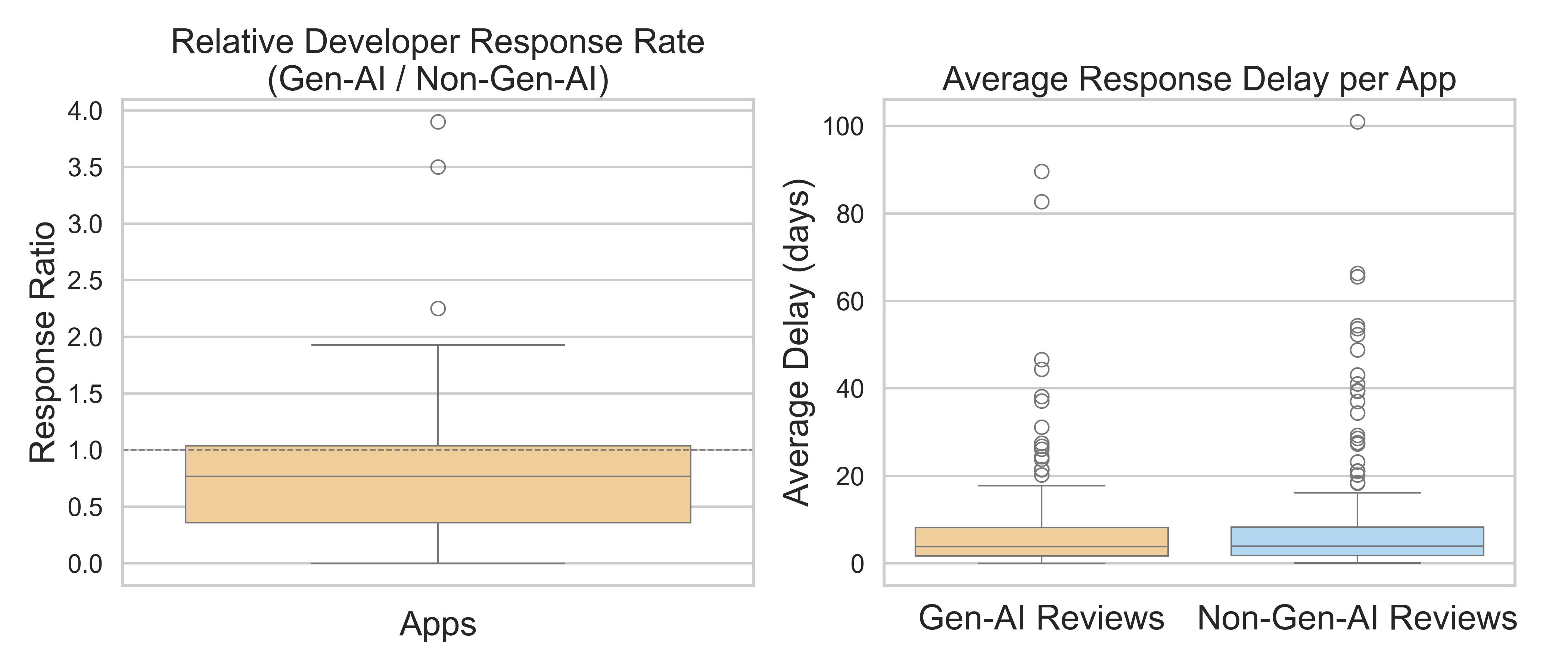}
\caption{
Comparison of developer responsiveness across apps. 
\textbf{Left:} Relative likelihood of developer response to Gen-AI reviews compared to non-Gen-AI reviews, expressed as the ratio of per-app response probabilities (values below 1 indicate lower responsiveness to Gen-AI reviews). 
\textbf{Right:} Distribution of mean developer response delays per app for Gen-AI and non-Gen-AI reviews; the two delay distributions do not differ significantly. 
(Full dataset considered.)
}
\label{fig:gen-ai-boxplot_combined}
\end{figure}

Once developers reply, however, response speed is similar across review types.
Response delay distributions for Gen-AI and non-Gen-AI reviews do not differ significantly, suggesting that the primary difference lies in whether developers respond rather than how quickly they respond. As illustrated in
Figure~\ref{fig:gen-ai-boxplot_combined}, response ratios across apps are
generally below one, further reflecting the lower response rates observed for
Gen-AI reviews.

\begin{SummaryBox}{Summary of Developer Reply}
Developer responses are largely generic or request additional information,
with relatively few indicating that an issue has been resolved. Response
patterns vary across app categories and Gen-AI topics, with lower engagement
for systemic and human-centered concerns. Developers also respond less
frequently to Gen-AI reviews than to non-Gen-AI reviews, although response
delays are similar once they reply.
\end{SummaryBox}{}{}

\section{Implications}
\label{sec:implications}
Our findings have implications for the design, post-deployment evaluation, requirements engineering, and continued evolution of Gen-AI mobile apps. We organize these implications across key stakeholders in the Gen-AI app ecosystem, including users, platform owners, developers, testing tool designers, requirements engineers, and policy makers and regulators.

\subsection{Implications for Users}

Our findings highlight several considerations for users of Gen-AI apps, including how they engage with AI for emotional support, understand the capabilities and limitations of AI systems, and navigate content moderation.

\noindent\textbf{Engage with AI as a tool and emotional companion while
remaining aware of the risks of over-reliance.}
Gen-AI systems can support accessibility, wellbeing, and companionship (Opportunity 1), but our findings also show growing discussions of emotional
dependence, addiction, and reduced engagement with real-world relationships (Finding 3.4). Users should therefore remain aware of these potential risks, particularly when relying on Gen-AI for emotional support, and recognize the
limitations of AI companionship.

\noindent\textbf{Develop informed expectations of AI capabilities and limitations.}
Our findings show that users encounter incorrect, inconsistent, or unexpected
outputs and that mismatches between user expectations and system capabilities are a recurring source of frustration (Challenge 1). Users should therefore remain aware that Gen-AI outputs may not always be accurate or reliable, particularly for complex tasks, and critically evaluate generated content when its accuracy or reliability is important.

\noindent\textbf{Be aware of how content moderation shapes interactions with Gen-AI apps.} Our findings show that users encounter different forms of content restriction
and express competing expectations about how Gen-AI content should be moderated
(Challenge 2 and Finding 3.5). While some users report excessive restrictions on benign interactions, others request fewer restrictions for mature content or stronger safeguards against unintended explicit outputs. Users should therefore recognize that moderation policies and safeguards can shape what they are able to generate and consider these differences when selecting and using Gen-AI apps.

\subsection{Implications for Platform Owners} Our findings highlight several ways platform owners can support the Gen-AI app ecosystem, including strengthening safeguards for emerging AI-related concerns,
improving the quality of user feedback, and supporting communication between users and developers.

\noindent\textbf{Support transparency and safeguards for Gen-AI apps.} Our findings highlight user concerns around content moderation, including excessive restrictions, inappropriate outputs, and changing expectations around
how Gen-AI content is governed (Challenge 2, Finding 3.5). We also identify growing discussions of emotional dependence and addiction in reviews of emotionally oriented AI interactions (Finding 3.4). App store owners can consider how their policies and review processes account for these emerging risks, particularly for apps involving emotionally sensitive interactions or young users. Clearer information about AI functionality, content restrictions, and available safeguards could also help users better understand how Gen-AI apps operate before and during use.

\noindent\textbf{Improve the quality and structure of user feedback.} Our results show that a large proportion of user reviews are non-informative (Finding 1.2), limiting the information available to developers and researchers. Platform owners could redesign review interfaces to encourage more structured and informative feedback. For example, prompting users to provide feedback on specific aspects such as usability, content quality, or AI performance, alongside free-text reviews, could encourage users to provide more detailed and actionable information.

\noindent\textbf{Provide richer support channels for complex Gen-AI issues.}
Our developer reply analysis shows that public reviews often serve as an initial point of contact, with developers frequently requesting additional
information rather than resolving the reported issue directly. Some Gen-AI-related concerns may require context that is difficult to communicate
through short public reviews, such as the prompt, generated output, or other details surrounding the interaction. Platform owners could therefore provide richer, privacy-preserving mechanisms for users to share relevant information
with developers when reporting AI-related problems. Such mechanisms could complement public reviews and provide developers with additional context for investigating complex Gen-AI behavior.

\subsection{Implications for Developers}

Our findings highlight several considerations for developers designing Gen-AI apps, including supporting emerging user practices, ensuring that Gen-AI integration provides meaningful value, and responding to changing user expectations around system capabilities and content moderation.

\noindent\textbf{Design for emerging and evolving uses of Gen-AI.} Our findings show that users engage with Gen-AI as a collaborative tool for creativity, ideation, and problem-solving (Opportunity 2), while also adapting its capabilities to a wide range of use cases that extend beyond a single intended purpose (Opportunity 3). Developers should therefore consider how
users incorporate Gen-AI into their own workflows and support flexible, iterative interactions that allow users to adapt the technology to different needs. As these interactions extend into emotionally sensitive contexts,
developers should also consider the risks that may emerge with continued use. Our temporal analysis identifies growing discussions of emotional dependence, addiction, and reduced engagement with real-world relationships (Finding 3.4), highlighting the need to consider safeguards alongside features designed to support companionship and emotional interaction.

\noindent\textbf{Ensure that Gen-AI integration provides meaningful and differentiated value.}
Our findings show that users question AI integration when it complicates tasks that simpler interfaces already support effectively, while also comparing Gen-AI apps with established tools such as ChatGPT and Photomath (Challenge 3). Developers should therefore consider both whether Gen-AI meaningfully improves the task it is introduced to support and what additional value their app
provides relative to existing alternatives. Gen-AI integration should provide
a clear benefit in functionality, performance, or user experience rather than
adding complexity without a corresponding advantage.

\noindent\textbf{Communicate system limitations and account for competing moderation expectations.}
Users report frustration when Gen-AI systems fail to meet expectations for accuracy, reliability, and consistency (Challenge 1), while our moderation findings reveal competing expectations around content restrictions (Challenge
2; Finding 3.5). Some users report excessive restrictions on benign interactions, others seek greater freedom for intentionally mature content,
and others request stronger safeguards against unintended explicit outputs. Developers should therefore communicate system limitations and content
restrictions clearly and consider how moderation approaches can accommodate different usage contexts while maintaining appropriate safeguards.

\subsection{Implications for Testing Tool Designers} Our evaluation demonstrates the potential of LLMs for large-scale analysis of user reviews. The best-performing configuration achieves 91\% accuracy in topic extraction and assignment, with performance improving when non-informative reviews are filtered and few-shot examples are provided (Finding 1.4). These results suggest that designers of tools for analyzing post-deployment user
feedback can incorporate LLM-based topic extraction and assignment to help identify recurring issues and user concerns at scale. Our findings also highlight the importance of validating prompt configurations and filtering low-information reviews before applying such approaches to large review datasets.

\subsection{Implications for Requirements Engineers}

Our findings show how post-deployment user feedback can capture emerging needs, expectations, and concerns that may inform the continued evolution of requirements for Gen-AI apps. The identified opportunities highlight several
user-centered considerations that requirements engineers may need to account for. \textit{AI for Accessibility and Wellbeing} (Opportunity 1), together with the growing discussions of emotional dependence identified in Finding 3.4, highlight the importance of considering both the benefits and potential risks
of emotionally supportive AI interactions. \textit{AI as a Collaborative Creative Tool} (Opportunity 2) points to requirements that support iterative and collaborative interaction between users and AI. The diverse applications
identified through \textit{AI Versatility} (Opportunity 3) further suggest the value of requirements that can accommodate emerging and evolving uses of Gen-AI capabilities.

The identified challenges provide additional considerations for requirements
elicitation and evolution. \textit{Managing User Expectations and AI
Limitations} (Challenge 1) highlights the importance of clearly defining and
communicating system capabilities and limitations. \textit{Balancing Content
Moderation and Creative Freedom} (Challenge 2), together with the competing
moderation expectations identified in Finding 3.5, suggests that requirements
should account for both appropriate safeguards and the different content
boundaries expected by users. Finally, \textit{Strategic Integration of Gen-AI
Features} (Challenge 3) highlights the importance of evaluating whether Gen-AI
functionality provides meaningful and differentiated value for the task it is
intended to support. More broadly, \textit{SARA} demonstrates how large-scale
user-review analysis can complement existing requirements practices by
identifying post-deployment needs and concerns that may inform requirements
evolution.

\subsection{Implications for Policy Makers and Regulators}
Our findings highlight two areas of Gen-AI app use that may warrant continued attention from policy makers and regulators: changing expectations around
content moderation and emerging risks associated with emotionally supportive
AI interactions.

\noindent\textbf{Consider changing and competing expectations around content
moderation.} Our findings show that concerns around Gen-AI governance change over time
(Finding 3.5). Earlier reviews discuss issues such as privacy and data use, while later reviews place greater emphasis on content restrictions. Importantly, the declining prominence of privacy-related concerns in user reviews should not be interpreted as evidence that these concerns have been resolved. At the same
time, users express competing expectations around moderation: some report
excessive restrictions on benign interactions, others seek greater freedom for
mature content, and others request stronger safeguards against unintended
explicit outputs (Challenge 2). These findings highlight the importance of
considering both user protection and the effects of moderation policies on
legitimate uses of Gen-AI systems.

\noindent\textbf{Consider emerging risks associated with emotionally supportive
AI.} Our findings show that Gen-AI can provide companionship, emotional support, and wellbeing benefits (Opportunity 1), while later reviews increasingly discuss emotional dependence, addiction, and reduced engagement with real-world relationships (Finding 3.4). These emerging concerns suggest that policy makers
and regulators should consider how existing approaches to AI safety and user protection account for emotionally oriented interactions, particularly where users may become highly reliant on AI systems.

\section{Threats to Validity}
\label{sec:threats_to_validty}
\textbf{Construct threats to validity} refer to the risk that the measures we use in our study do not accurately reflect the concepts we aim to study. First, we use the average app review rating as a proxy for user satisfaction. However, this metric may not fully capture the sentiment expressed in the review text. For example, users sometimes assign high ratings while simultaneously reporting critical issues or limitations. To mitigate this threat, we supplement the quantitative rating data with manual qualitative analysis of review content, ensuring that nuanced sentiments, both positive and negative, are appropriately captured. 
Second, our methodology relies on LLMs to automatically filter and classify reviews. To control for potential biases introduced by the LLM-based automation, we manually annotate samples, and our results demonstrate high agreement, i.e., 91\% accuracy, between the LLM-generated classifications and human annotations, supporting the reliability of using LLMs for these tasks. Additionally, we measure inter-rater reliability to confirm the consistency and trustworthiness of the automated classifications.

\noindent \textbf{Internal threats to validity} concern the soundness of the procedures used to collect, analyze, and interpret the data. A potential threat arises from uncertainty in identifying the exact timeline of Gen-AI feature integration for each app. To mitigate this, we adopt a conservative, evidence-based approach by manually examining historical release notes and online sources to determine the earliest \textit{documented} integration date. While this approach may miss undocumented or informally introduced features, it ensures that all included data reflects confirmed Gen-AI functionality. As such, our integration dates represent the earliest verifiable evidence rather than the precise deployment time, prioritizing reliability and reducing the risk of including pre-Gen-AI reviews.

Additionally, a potential threat arises from ambiguity in determining whether an app genuinely incorporates Gen-AI functionality. In particular, some apps reference “AI” without providing generative capabilities, which may lead to misclassification. To mitigate this risk, we adopt a conservative coding strategy in which a mere mention of AI (e.g., \textit{“AI Song”}) is not considered sufficient evidence of Gen-AI integration unless it is explicitly supported by documented generative functionality (e.g., \textit{“AI Song Generator”}). This approach prioritizes precision in identifying Gen-AI apps, reducing the likelihood of including apps that do not provide true generative capabilities.

\noindent \textbf{External threats to validity} concern the ability to generalize the results. Our dataset is constructed using systematic keyword-based searches on the Google Play Store followed by manual validation. Google Play Store search results are capped, which prevents enumerating all Gen-AI apps for a given keyword. Accordingly, our goal is representativeness rather than exhaustiveness. By querying multiple taxonomy-derived keywords and retaining only apps with verified generative functionality, we obtain a diverse sample covering a wide range of Gen-AI capabilities and usage contexts. While some apps may not be included, our findings are expected to generalize to apps implementing similar Gen-AI functionalities and design choices.

Additionally, user perceptions and behaviors may differ across platforms. To mitigate this, we conduct a cross-platform comparison between Google Play Store and App Store reviews, observing consistent topic patterns across platforms. While this strengthens the generalizability of our findings, differences in platform policies, user demographics, and review mechanisms may still influence the results.

\section{Conclusion}
\label{sec:conclusion}
Gen-AI is rapidly transforming mobile apps, yet understanding how users perceive and interact with these systems remains limited. In this paper, we present \textit{SARA}, a framework for systematically analyzing user reviews of Gen-AI
apps, and apply it to over one million reviews across multiple app categories.

Our findings reveal that user discussions of Gen-AI apps surface key opportunities, including \textit{AI for Accessibility and Wellbeing},
\textit{AI as a Collaborative Creative Tool}, and \textit{AI Versatility}, alongside challenges related to \textit{Managing User Expectations and AI Limitations}, \textit{Balancing Content Moderation and Creative Freedom}, and \textit{Strategic Integration of Gen-AI Features}. Our temporal analysis further shows that the topics discussed and how they are evaluated change over time.
In particular, discussions of \textit{Emotional Connection} shift from companionship toward concerns about dependency and addiction, while discussions
of \textit{Content Policy \& Censorship} shift from privacy-related concerns toward competing expectations around content restrictions.

Together, these findings provide implications for multiple stakeholders. Users
can benefit from informed expectations of AI capabilities and awareness of the risks associated with emotional reliance; developers can support emerging uses
while ensuring that Gen-AI integration provides meaningful value; and platform owners can improve feedback mechanisms and support communication around complex Gen-AI issues. Our findings also demonstrate how large-scale review analysis can
support post-deployment feedback analysis and requirements evolution, while
highlighting emerging concerns around moderation and emotionally supportive AI
that may warrant continued attention from policy makers and regulators.

Overall, this work contributes both a scalable framework for analyzing Gen-AI app reviews and empirical insights into the opportunities, challenges, and
changing concerns expressed by users. By examining user feedback at scale and over time, our study provides a user-centered perspective on how Gen-AI is experienced after deployment and identifies considerations for the continued
design and evolution of Gen-AI mobile apps. Future work can build on these findings by examining how user expectations and experiences develop at the
individual and app levels, including how changes in model capabilities, app maturity, and release histories relate to changes in user feedback over time.

\vspace{4pt}
\noindent \textbf{Data Availability.} We provide a replication package, including manually labeled data, and Python scripts to replicate the analyses at \url{https://github.com/Safwat-UofT/Gen-AI_User_Reviews}.

\noindent \textbf{Acknowledgments.}
We acknowledge the support of the Natural Sciences and Engineering Research Council of Canada (NSERC), \textbf{[RGPIN-2021-03969]}.

\noindent \textbf{Acknowledgments.}
We acknowledge the support of the Natural Sciences and Engineering Research Council of Canada (NSERC), \textbf{[RGPIN-2021-03969]}.

\printcredits

\bibliographystyle{cas-model2-names}

\bibliography{bibliography}

@article {gozalobrizuela2023surveygenerativeaiapplications,
article_type = {journal},
title = {A survey of generative AI applications},
author = {Gozalo-Brizuela, Roberto and Merchan, Eduardo Eduardo Garrido},
volume = {20},
number = {8},
year = {2024},
month = {May},
pages = {801-818},
doi = {10.3844/jcssp.2024.801.818},
url = {https://thescipub.com/abstract/jcssp.2024.801.818},
journal = {Journal of Computer Science},
publisher = {Science Publications}
}

@article{Feuerriegel_Hartmann_Janiesch_Zschech_2024,
	title        = {Generative AI},
	author       = {Feuerriegel, Stefan and Hartmann, Jochen and Janiesch, Christian and Zschech, Patrick},
	year         = 2024,
	month        = feb,
	journal      = {Business \& Information Systems Engineering},
	volume       = 66,
	number       = 1,
	pages        = {111–126},
	doi          = {10.1007/s12599-023-00834-7},
    url = {https://doi.org/10.1007/s12599-023-00834-7},
	issn         = {2363-7005, 1867-0202},
	language     = {en}
}

@article{Iorliam_Ingio_2024,
	title        = {A comparative analysis of generative artificial intelligence tools for natural language processing},
	author       = {Iorliam, Aamo and Ingio, Joseph Abunimye},
	year         = 2024,
	month        = feb,
	journal      = {Journal of Computing Theories and Applications},
	volume       = 1,
	number       = 3,
	pages        = {311–325},
	doi          = {10.62411/jcta.9447},
    url = {https://doi.org/10.62411/jcta.9447},
	issn         = {3024-9104},
	rights       = {https://creativecommons.org/licenses/by/4.0}
}

@article{Akhtar_2024,
	title        = {Unveiling the evolution of generative AI (GAI): a comprehensive and investigative analysis toward LLM models (2021–2024) and beyond},
	author       = {Akhtar, Zarif Bin},
	year         = 2024,
	month        = jun,
	journal      = {Journal of Electrical Systems and Information Technology},
	volume       = 11,
	number       = 1,
	pages        = 22,
	doi          = {10.1186/s43067-024-00145-1},
    url = {https://doi.org/10.1186/s43067-024-00145-1},
	issn         = {2314-7172},
	language     = {en}
}

@misc{googleplayscraper,
	title        = {Google Play Scraper},
	author       = {Mingyu, Jo},
	year         = {n.d.},
	url          = {https://github.com/JoMingyu/google-play-scraper},
	note         = {Accessed: 13-Oct-2024},
	howpublished = {GitHub repository}
}

@inproceedings{Pham_Hoyle_Sun_Resnik_Iyyer_2024,
    title = "{T}opic{GPT}: A prompt-based topic modeling framework",
    author = "Pham, Chau Minh  and
      Hoyle, Alexander  and
      Sun, Simeng  and
      Resnik, Philip  and
      Iyyer, Mohit",
    editor = "Duh, Kevin  and
      Gomez, Helena  and
      Bethard, Steven",
    booktitle = "Proceedings of the 2024 Conference of the North American Chapter of the Association for Computational Linguistics: Human Language Technologies (Volume 1: Long Papers)",
    month = jun,
    year = "2024",
    address = "Mexico City, Mexico",
    publisher = "Association for Computational Linguistics",
    url = "https://aclanthology.org/2024.naacl-long.164/",
    doi = "10.18653/v1/2024.naacl-long.164",
    pages = "2956--2984",
}

@inproceedings{Prakash_Wang_Hoang_Hee_Lee_2023,
	title        = {PromptMTopic: Unsupervised multimodal topic modeling of memes using large language models},
	author       = {Prakash, Nirmalendu and Wang, Han and Hoang, Nguyen Khoi and Hee, Ming Shan and Lee, Roy Ka-Wei},
	year         = 2023,
	month        = oct,
	booktitle    = {Proceedings of the 31st ACM International Conference on Multimedia},
	publisher    = {ACM},
	address      = {Ottawa ON Canada},
	pages        = {621–631},
	doi          = {10.1145/3581783.3613836},
	isbn         = 9798400701085,
	url          = {https://dl.acm.org/doi/10.1145/3581783.3613836},
	language     = {en}
}

@preprint{Ghosh_Pargaonkar_Eisty_2024,
	title        = {Exploring requirements elicitation from app store user reviews using large language models},
	author       = {Ghosh, Tanmai Kumar and Pargaonkar, Atharva and Eisty, Nasir U.},
	year         = 2024,
	publisher    = {arXiv},
	doi          = {10.48550/ARXIV.2409.15473},
	url          = {https://arxiv.org/abs/2409.15473},
	rights       = {Creative Commons Attribution 4.0 International}
}

@misc{openai2023gpt,
	title        = {GPT-4: Generative Pre-trained Transformer},
	author       = {OpenAI},
	year         = 2023,
	note         = {Available at \url{https://openai.com/research/gpt-4}, Accessed: 03-Mar-2025},
	howpublished = {Online}
}

@inproceedings{Wei_Courbis_Lambolais_Xu_Bernard_Dray_2023,
	title        = {Zero-shot bilingual app reviews mining with large language models},
	author       = {Wei, Jialiang and Courbis, Anne-Lise and Lambolais, Thomas and Xu, Binbin and Bernard, Pierre Louis and Dray, Gérard},
	year         = 2023,
	month        = nov,
	booktitle    = {Proceedings of the 2023 IEEE 35th International Conference on Tools with Artificial Intelligence (ICTAI '23)},
	publisher    = {IEEE},
	address      = {Atlanta, GA, USA},
	pages        = {898–904},
	doi          = {10.1109/ICTAI59109.2023.00135},
	isbn         = 9798350342734,
	url          = {https://ieeexplore.ieee.org/document/10356483/},
	rights       = {https://doi.org/10.15223/policy-029}
}

@article{Roumeliotis_Tselikas_Nasiopoulos_2024,
	title        = {LLMs in e-commerce: A comparative analysis of GPT and LLaMA models in product review evaluation},
	author       = {Roumeliotis, Konstantinos I. and Tselikas, Nikolaos D. and Nasiopoulos, Dimitrios K.},
	year         = 2024,
	month        = mar,
	journal      = {Natural Language Processing Journal},
	volume       = 6,
	pages        = 100056,
    doi = {10.1016/j.nlp.2024.100056},
	url          = {https://doi.org/10.1016/j.nlp.2024.100056},
	issn = {2949-7191},
	language     = {en}
}

@article{cohen1960kappa,
	title        = {A Coefficient of Agreement for Nominal Scales},
	author       = {J. Cohen},
	year         = 1960,
	journal      = {Educational and Psychological Measurement},
	volume       = 20,
	number       = 1,
	pages        = {37--46},
	doi          = {10.1177/001316446002000104}
}

@misc{Ceci2025,
	title        = {Mobile App Usage - Statistics \& Facts},
	author       = {Laura Ceci},
	year         = 2025,
	month        = {February},
	day          = 5,
	publisher    = {Statista},
	url          = {https://www.statista.com/topics/1002/mobile-app-usage/#topicOverview},
	note         = {[Online]. Accessed: 2025-03-07}
}

@article{10628027,
author = {Zhang, Ye and Zhang, Jinrui and Yue, Sheng and Lu, Wei and Ren, Ju and Shen, Xuemin},
title = {Mobile generative AI: Opportunities and challenges},
year = {2024},
issue_date = {August 2024},
publisher = {IEEE Press},
volume = {31},
number = {4},
issn = {1536-1284},
url = {https://doi.org/10.1109/MWC.006.2300576},
doi = {10.1109/MWC.006.2300576},
journal = {Wireless Commun.},
month = aug,
pages = {58–64},
numpages = {7}
}

@article{Ho_Mayberry_Nguyen_Dhulipala_Pallipuram_2024,
	title        = {ChatReview: A ChatGPT-enabled natural language processing framework to study domain-specific user reviews},
	author       = {Ho, Brittany and Mayberry, Ta’Rhonda and Nguyen, Khanh Linh and Dhulipala, Manohar and Pallipuram, Vivek Krishnamani},
	year         = 2024,
	month        = mar,
	journal      = {Machine Learning with Applications},
	volume       = 15,
	pages        = 100522,
	doi          = {10.1016/j.mlwa.2023.100522},
    url = {https://doi.org/10.1016/j.mlwa.2023.100522},
	issn         = 26668270,
	language     = {en}
}

@article{Lee_Lee_Seo_2024,
	title        = {Determining the best feature combination through text and probabilistic feature analysis for GPT-2-based mobile app review detection},
	author       = {Lee, Seung-Cheol and Lee, Dong-Gun and Seo, Yeong-Seok},
	year         = 2024,
	month        = jan,
	journal      = {Applied Intelligence},
	volume       = 54,
	number       = 2,
	pages        = {1219–1246},
	doi          = {10.1007/s10489-023-05201-3},
    url = {https://doi.org/10.1007/s10489-023-05201-3},
	issn         = {0924-669X, 1573-7497},
	language     = {en}
}

@article{Ikotun_Ezugwu_Abualigah_Abuhaija_Heming_2023,
	title        = {K-means clustering algorithms: A comprehensive review, variants analysis, and advances in the era of big data},
	author       = {Ikotun, Abiodun M. and Ezugwu, Absalom E. and Abualigah, Laith and Abuhaija, Belal and Heming, Jia},
	year         = 2023,
	month        = apr,
	journal      = {Information Sciences},
	volume       = 622,
	pages        = {178–210},
	doi          = {10.1016/j.ins.2022.11.139},
    url = {https://doi.org/10.1016/j.ins.2022.11.139},
	issn         = {00200255},
	language     = {en}
}

@article{Ashari_DwiNugroho_Baraku_NovriYanda_Liwardana_2023,
	title        = {Analysis of elbow, silhouette, Davies-Bouldin, Calinski-Harabasz, and Rand-index evaluation on K-means algorithm for classifying flood-affected areas in Jakarta},
	author       = {Ashari, Ilham Firman and Dwi Nugroho, Eko and Baraku, Randi and Novri Yanda, Ilham and Liwardana, Ridho},
	year         = 2023,
	month        = jul,
	journal      = {Journal of Applied Informatics and Computing},
	volume       = 7,
	number       = 1,
	pages        = {89–97},
	doi          = {10.30871/jaic.v7i1.4947},
	issn         = {2548-6861},
    url = { https://doi.org/10.30871/jaic.v7i1.4947},
	rights       = {http://creativecommons.org/licenses/by-sa/4.0}
}

@article{Maram_LLM_Cure,
author = {Assi, Maram and Hassan, Safwat and Zou, Ying},
title = {LLM-Cure: LLM-based competitor user review analysis for feature enhancement},
year = {2025},
publisher = {Association for Computing Machinery},
address = {New York, NY, USA},
issn = {1049-331X},
url = {https://doi.org/10.1145/3744644},
doi = {10.1145/3744644},
note = {Just Accepted},
journal = {ACM Trans. Softw. Eng. Methodol.},
month = jun,
}

@inproceedings{Hau_Hassan_Zhou_2025,
	title        = {LLMs in mobile apps: Practices, challenges, and opportunities},
	author       = {Hau, Kimberly and Hassan, Safwat and Zhou, Shurui},
	year         = 2025,
	booktitle    = {Proceedings of the 12th IEEE/ACM International Conference on Mobile Software Engineering and Systems (MOBILESoft)},
	pages        = {3--14},
	doi          = {10.1109/MOBILESoft66462.2025.00008}
}

@article{Chen_Liu_Zhou_Zhao_Wang_Wang_Chen_Bissyande_Klein_Li_2024,
author = {Chen, Daihang and Liu, Yonghui and Zhou, Mingyi and Zhao, Yanjie and Wang, Haoyu and Wang, Shuai and Chen, Xiao and Bissyand\'{e}, Tegawend\'{e} F. and Klein, Jacques and Li, Li},
title = {LLM for Mobile: An Initial Roadmap},
year = {2025},
issue_date = {June 2025},
publisher = {Association for Computing Machinery},
address = {New York, NY, USA},
volume = {34},
number = {5},
issn = {1049-331X},
url = {https://doi.org/10.1145/3708528},
doi = {10.1145/3708528},
journal = {ACM Trans. Softw. Eng. Methodol.},
month = may,
articleno = {128},
numpages = {29}
}

@article{Yuen_Schlote_2024,
	title        = {Learner experiences of mobile apps and artificial intelligence to support additional language learning in education},
	author       = {Yuen, Connie Levina and Schlote, Nadja},
	year         = 2024,
	month        = jun,
	journal      = {Journal of Educational Technology Systems},
	volume       = 52,
	number       = 4,
	pages        = {507–525},
	doi          = {10.1177/00472395241238693},
	issn         = {0047-2395, 1541-3810},
	language     = {en}
}

@inproceedings{Al_Wahshat_Abu_ulbeh_Yusoff_Zakaria_Amir_Fazamin_Wan_Hamzah_P_2023,
	title        = {The detection of e-commerce manipulated reviews using GPT-4},
	author       = {Al Wahshat, Hassan and Abu-ulbeh, Waheeb and Yusoff, M Hafiz and Zakaria, Muhammad D. and Amir Fazamin Wan Hamzah, Wan Mohd and P, Stenin N},
	year         = 2023,
	month        = oct,
	booktitle    = {2023 International Conference on Computer Science and Emerging Technologies (CSET)},
	publisher    = {IEEE},
	address      = {Bangalore, India},
	pages        = {1–6},
	doi          = {10.1109/CSET58993.2023.10346848},
	isbn         = 9798350341737,
	url          = {https://ieeexplore.ieee.org/document/10346848/},
	rights       = {https://doi.org/10.15223/policy-029}
}

@misc{He_Rungta_Koleczek_Sekhon_Wang_Hasan_2024,
	title        = {Does prompt formatting have any impact on LLM performance?},
	author       = {He, Jia and Rungta, Mukund and Koleczek, David and Sekhon, Arshdeep and Wang, Franklin X. and Hasan, Sadid},
	year         = 2024,
	month        = nov,
	publisher    = {arXiv},
	number       = {arXiv:2411.10541},
	doi          = {10.48550/arXiv.2411.10541},
	url          = {http://arxiv.org/abs/2411.10541},
	note         = {arXiv:2411.10541 [cs]}
}

@inproceedings{Dos_Santos_Oliveira_De_Jesus_Aljedaani_Eler_2023,
	title        = {Evolution may come with a price: Analyzing user reviews to understand the impact of updates on mobile apps accessibility},
	author       = {Dos Santos, Paulo Sérgio Henrique and Oliveira, Alberto Dumont Alves and De Jesus, Thais Bonjorni Nobre and Aljedaani, Wajdi and Eler, Marcelo Medeiros},
	year         = 2023,
	month        = oct,
	booktitle    = {Proceedings of the XXII Brazilian Symposium on Human Factors in Computing Systems},
	publisher    = {ACM},
	address      = {Maceió Brazil},
	pages        = {1–11},
	doi          = {10.1145/3638067.3638081},
	isbn         = 9798400717154,
	url          = {https://dl.acm.org/doi/10.1145/3638067.3638081},
	language     = {en}
}

@inproceedings{Jin_Kim_Han_2025,
	title        = {“I don’t know why I should use this app”: Holistic analysis on user engagement challenges in mobile mental health},
	author       = {Jin, Seungwan and Kim, Bogoan and Han, Kyungsik},
	year         = 2025,
	month        = apr,
	booktitle    = {Proceedings of the 2025 CHI Conference on Human Factors in Computing Systems},
	publisher    = {ACM},
	address      = {Yokohama Japan},
	pages        = {1–23},
	doi          = {10.1145/3706598.3713732},
	isbn         = 9798400713941,
	url          = {https://dl.acm.org/doi/10.1145/3706598.3713732},
	language     = {en}
}

@article{RezaeiNasab_Dashti_Shahin_Zahedi_Khalajzadeh_Arora_Liang_2025,
	title        = {Fairness concerns in app reviews: A study on AI-based mobile apps},
	author       = {Rezaei Nasab, Ali and Dashti, Maedeh and Shahin, Mojtaba and Zahedi, Mansooreh and Khalajzadeh, Hourieh and Arora, Chetan and Liang, Peng},
	year         = 2025,
	month        = feb,
	journal      = {ACM Transactions on Software Engineering and Methodology},
	volume       = 34,
	number       = 2,
	pages        = {1–30},
	doi          = {10.1145/3690633},
	issn         = {1049-331X, 1557-7392},
	language     = {en}
}

@article{10.1145/3721125,
	title        = {Unraveling code clone dynamics in deep learning frameworks},
	author       = {Assi, Maram and Hassan, Safwat and Zou, Ying},
	year         = 2025,
	month        = feb,
	journal      = {ACM Trans. Softw. Eng. Methodol.},
	publisher    = {Association for Computing Machinery},
	address      = {New York, NY, USA},
	doi          = {10.1145/3721125},
	issn         = {1049-331X},
	url          = {https://doi.org/10.1145/3721125},
	note         = {Just Accepted}
}

@article{Warren_Liao_2005,
	title        = {Clustering of time series data—a survey},
	author       = {Warren Liao, T.},
	year         = 2005,
	month        = nov,
	journal      = {Pattern Recognition},
	volume       = 38,
	number       = 11,
	pages        = {1857–1874},
	doi          = {10.1016/j.patcog.2005.01.025},
	issn         = {00313203},
	rights       = {https://www.elsevier.com/tdm/userlicense/1.0/},
	language     = {en}
}

@inproceedings{Pagano_Maalej_2013,
	title        = {User feedback in the appstore: An empirical study},
	author       = {Pagano, Dennis and Maalej, Walid},
	year         = 2013,
	month        = jul,
	booktitle    = {Proceedings of the 21st IEEE International Requirements Engineering Conference (RE)},
	publisher    = {IEEE},
	address      = {Rio de Janeiro-RJ, Brazil},
	pages        = {125–134},
	doi          = {10.1109/RE.2013.6636712},
	isbn         = {978-1-4673-5765-4},
	url          = {http://ieeexplore.ieee.org/document/6636712/}
}

@inproceedings{Chen_Lin_Hoi_Xiao_Zhang_2014,
	title        = {AR-miner: mining informative reviews for developers from mobile app marketplace},
	author       = {Chen, Ning and Lin, Jialiu and Hoi, Steven C. H. and Xiao, Xiaokui and Zhang, Boshen},
	year         = 2014,
	month        = may,
	booktitle    = {Proceedings of the 36th International Conference on Software Engineering},
	publisher    = {ACM},
	address      = {Hyderabad India},
	pages        = {767–778},
	doi          = {10.1145/2568225.2568263},
	isbn         = {978-1-4503-2756-5},
	url          = {https://dl.acm.org/doi/10.1145/2568225.2568263},
	language     = {en}
}

@article{Shata_Hartley_2025,
	title        = {Artificial intelligence and communication technologies in academia: faculty perceptions and the adoption of generative AI},
	author       = {Shata, Aya and Hartley, Kendall},
	year         = 2025,
	month        = mar,
	journal      = {International Journal of Educational Technology in Higher Education},
	volume       = 22,
	number       = 1,
	pages        = 14,
	doi          = {10.1186/s41239-025-00511-7},
	issn         = {2365-9440},
	language     = {en}
}

@article{Kim_Klopfer_Grohs_Eldardiry_Weichert_Cox_Pike_2025,
	title        = {Examining faculty and student perceptions of generative AI in university courses},
	author       = {Kim, Junghwan and Klopfer, Michelle and Grohs, Jacob R. and Eldardiry, Hoda and Weichert, James and Cox, Larry A. and Pike, Dale},
	year         = 2025,
	month        = jan,
	journal      = {Innovative Higher Education},
	doi          = {10.1007/s10755-024-09774-w},
	issn         = {0742-5627, 1573-1758},
	url          = {https://link.springer.com/10.1007/s10755-024-09774-w},
	language     = {en}
}

@article{Golding_Lippert_Neuschatz_Salomon_Burke_2024,
	title        = {Generative AI and college students: Use and perceptions},
	author       = {Golding, Jonathan M. and Lippert, Anne and Neuschatz, Jeffrey S. and Salomon, Ilyssa and Burke, Kelly},
	year         = 2024,
	month        = sep,
	journal      = {Teaching of Psychology},
	pages        = {00986283241280350},
	doi          = {10.1177/00986283241280350},
	issn         = {0098-6283, 1532-8023},
	language     = {en}
}

@article{Kim_Yu_Detrick_Li_2025,
	title        = {Exploring students’ perspectives on Generative AI-assisted academic writing},
	author       = {Kim, Jinhee and Yu, Seongryeong and Detrick, Rita and Li, Na},
	year         = 2025,
	month        = jan,
	journal      = {Education and Information Technologies},
	volume       = 30,
	number       = 1,
	pages        = {1265–1300},
	doi          = {10.1007/s10639-024-12878-7},
	issn         = {1360-2357, 1573-7608},
	language     = {en}
}

@article{Lee_Arnold_Srivastava_Plastow_Strelan_Ploeckl_Lekkas_Palmer_2024,
	title        = {The impact of generative AI on higher education learning and teaching: A study of educators’ perspectives},
	author       = {Lee, Daniel and Arnold, Matthew and Srivastava, Amit and Plastow, Katrina and Strelan, Peter and Ploeckl, Florian and Lekkas, Dimitra and Palmer, Edward},
	year         = 2024,
	month        = jun,
	journal      = {Computers and Education: Artificial Intelligence},
	volume       = 6,
	pages        = 100221,
	doi          = {10.1016/j.caeai.2024.100221},
	issn         = {2666920X},
	language     = {en}
}

@article{Heinz_Mackin_Trudeau_Bhattacharya_Wang_Banta_Jewett_Salzhauer_Griffin_Jacobson_2025,
	title        = {Randomized trial of a generative AI chatbot for mental health treatment},
	author       = {Heinz, Michael V. and Mackin, Daniel M. and Trudeau, Brianna M. and Bhattacharya, Sukanya and Wang, Yinzhou and Banta, Haley A. and Jewett, Abi D. and Salzhauer, Abigail J. and Griffin, Tess Z. and Jacobson, Nicholas C.},
	year         = 2025,
	month        = mar,
	journal      = {NEJM AI},
	volume       = 2,
	number       = 4,
	doi          = {10.1056/AIoa2400802},
	issn         = {2836-9386},
	url          = {https://ai.nejm.org/doi/10.1056/AIoa2400802},
	language     = {en}
}

@inproceedings{Tang_Zhang_Ciancia_Wang_2024,
	title        = {Exploring the impact of AI-generated image tools on professional and non-professional users in the art and design fields},
	author       = {Tang, Yuying and Zhang, Ningning and Ciancia, Mariana and Wang, Zhigang},
	year         = 2024,
	booktitle    = {Companion Publication of the 2024 Conference on Computer-Supported Cooperative Work and Social Computing},
	location     = {San Jose, Costa Rica},
	publisher    = {Association for Computing Machinery},
	address      = {New York, NY, USA},
	series       = {CSCW Companion '24},
	pages        = {451–458},
	doi          = {10.1145/3678884.3681890},
	isbn         = 9798400711145,
	url          = {https://doi.org/10.1145/3678884.3681890},
	numpages     = 8
}

@inproceedings{Oppenlaender_Silvennoinen_Paananen_Visuri_2023,
	title        = {Perceptions and realities of text-to-image generation},
	author       = {Oppenlaender, Jonas and Silvennoinen, Johanna and Paananen, Ville and Visuri, Aku},
	year         = 2023,
	booktitle    = {Proceedings of the 26th International Academic Mindtrek Conference},
	location     = {Tampere, Finland},
	publisher    = {Association for Computing Machinery},
	address      = {New York, NY, USA},
	series       = {Mindtrek '23},
	pages        = {279–288},
	doi          = {10.1145/3616961.3616978},
	isbn         = 9798400708749,
	url          = {https://doi.org/10.1145/3616961.3616978},
	numpages     = 10
}

@inproceedings{haase2023art,
  title={The art of inspiring creativity: Exploring the unique impact of AI-generated images},
  author={Haase, Jennifer and Djurica, Djordje and Mendling, Jan},
  year={2023},
    booktitle = {Proceedings of the 29th Americas Conference on Information Systems (AMCIS)},
    address = {Panama},
    url = {https://aisel.aisnet.org/amcis2023/sig_aiaa/sig_aiaa/10}
}

@preprint{Wang_Yang_Wei_2023,
	title        = {Learning to retrieve in-context examples for large language models},
	author       = {Wang, Liang and Yang, Nan and Wei, Furu},
	year         = 2023,
	publisher    = {arXiv},
	doi          = {10.48550/ARXIV.2307.07164},
	url          = {https://arxiv.org/abs/2307.07164},
	rights       = {Creative Commons Attribution 4.0 International}
}

@inproceedings{Ren_Nakagawa_Tsuchiya_2024,
	title        = {Combining prompts with examples to enhance LLM-based requirement elicitation},
	author       = {Ren, Shuaicai and Nakagawa, Hiroyuki and Tsuchiya, Tatsuhiro},
	year         = 2024,
	booktitle    = {Proceedings of the 48th IEEE  Annual Computers, Software, and Applications Conference (COMPSAC)},
	pages        = {1376--1381},
	doi          = {10.1109/COMPSAC61105.2024.00181}}

@article{Giray_2023,
	title        = {Prompt engineering with ChatGPT: a guide for academic writers},
	author       = {Giray, Louie},
	year         = 2023,
	month        = dec,
	journal      = {Annals of Biomedical Engineering},
	volume       = 51,
	number       = 12,
	pages        = {2629–2633},
	doi          = {10.1007/s10439-023-03272-4},
	issn         = {0090-6964, 1573-9686},
	language     = {en}
}

@inproceedings{Deshmukh_Raut_Bhavsar_Gurav_Patil_2025,
	title        = {Optimizing human-AI interaction: Innovations in prompt engineering},
	author       = {Deshmukh, Rushali and Raut, Rutuj and Bhavsar, Mayur and Gurav, Sanika and Patil, Yash},
	year         = 2025,
	month        = feb,
	booktitle    = {Proceedings of the 3rd International Conference on Intelligent Data Communication Technologies and Internet of Things (IDCIoT)},
	publisher    = {IEEE},
	address      = {Bengaluru, India},
	pages        = {1240–1246},
	doi          = {10.1109/IDCIOT64235.2025.10914815},
	isbn         = 9798331527549,
	url          = {https://ieeexplore.ieee.org/document/10914815/},
	rights       = {https://doi.org/10.15223/policy-029}
}

@inbook{Marvin_Hellen_Jjingo_Nakatumba-Nabende_2024,
	title        = {Prompt engineering in large language models},
	author       = {Marvin, Ggaliwango and Hellen, Nakayiza and Jjingo, Daudi and Nakatumba-Nabende, Joyce},
	year         = 2024,
	booktitle    = {Data Intelligence and Cognitive Informatics},
	publisher    = {Springer Nature Singapore},
	address      = {Singapore},
	series       = {Algorithms for Intelligent Systems},
	pages        = {387–402},
	doi          = {10.1007/978-981-99-7962-2_30},
	isbn         = {978-981-9979-99-8},
	url          = {https://link.springer.com/10.1007/978-981-99-7962-2_30},
	editor       = {Jacob, I. Jeena and Piramuthu, Selwyn and Falkowski-Gilski, Przemyslaw},
	collection   = {Algorithms for Intelligent Systems},
	language     = {en}
}

@article{Genc-Nayebi_Abran_2017,
	title        = {A systematic literature review: Opinion mining studies from mobile app store user reviews},
	author       = {Genc-Nayebi, Necmiye and Abran, Alain},
	year         = 2017,
	month        = mar,
	journal      = {Journal of Systems and Software},
	volume       = 125,
	pages        = {207–219},
	doi          = {10.1016/j.jss.2016.11.027},
	issn         = {01641212},
	language     = {en}
}

@INPROCEEDINGS{10499727,
  author={Arambepola, Nimasha and Munasinghe, Lankeshwara and Warnajith, Nalin},
  booktitle={Proceedings of the 4th International Conference on Advanced Research in Computing (ICARC)}, 
  title={Factors Influencing Mobile App User Experience: An Analysis of Education App User Reviews}, 
  year={2024},
  volume={},
  number={},
  pages={223-228},
  doi={10.1109/ICARC61713.2024.10499727}}

@article{Assi_Hassan_Tian_Zou_2021,
	title        = {FeatCompare: Feature comparison for competing mobile apps leveraging user reviews},
	author       = {Assi, Maram and Hassan, Safwat and Tian, Yuan and Zou, Ying},
	year         = 2021,
	month        = sep,
	journal      = {Empirical Software Engineering},
	volume       = 26,
	number       = 5,
	pages        = 94,
	doi          = {10.1007/s10664-021-09988-y},
	issn         = {1382-3256, 1573-7616},
	language     = {en}
}

@misc{OpenAI_O4Mini_2024,
  author       = {{OpenAI}},
  title        = {GPT-4o-mini model overview},
  year         = {2024},
  url          = {https://platform.openai.com/docs/models/o4-mini},
  note         = {Accessed: 2025-05-24}
}

@book{Judd_McClelland_Ryan_2017,
	title        = {Data analysis: A model comparison approach to regression, ANOVA, and beyond},
	author       = {Judd, Charles M. and McClelland, Gary H. and Ryan, Carey S.},
	year         = 2017,
	month        = may,
	publisher    = {Routledge},
	address      = {Third Edition. | New York: Routledge, 2017. | Revised edition},
	doi          = {10.4324/9781315744131},
	isbn         = {978-1-315-74413-1},
	url          = {https://www.taylorfrancis.com/books/9781317591214},
	edition      = 3,
	language     = {en}
}

@article{Chan_Walmsley_1997,
	title        = {Learning and understanding the Kruskal-Wallis one-way analysis-of-variance-by-ranks test for differences among three or more independent groups},
	author       = {Chan, Yvonne and Walmsley, Roy P},
	year         = 1997,
	month        = dec,
	journal      = {Physical Therapy},
	volume       = 77,
	number       = 12,
	pages        = {1755–1761},
	doi          = {10.1093/ptj/77.12.1755},
	issn         = {0031-9023, 1538-6724},
	language     = {en}
}

@article{Mann_Whitney_1947,
	title        = {On a test of whether one of two random variables is stochastically larger than the other},
	author       = {Mann, H. B. and Whitney, D. R.},
	year         = 1947,
	month        = mar,
	journal      = {The Annals of Mathematical Statistics},
	volume       = 18,
	number       = 1,
	pages        = {50–60},
	doi          = {10.1214/aoms/1177730491},
	issn         = {0003-4851},
	language     = {en}
}

@article{3b6386c9-eeae-33a7-8c59-c0604aa87bef,
	title        = {To Bonferroni or not to Bonferroni: When and how are the Questions},
	author       = {Robert J. Cabin and Randall J. Mitchell},
	year         = 2000,
	journal      = {Bulletin of the Ecological Society of America},
	publisher    = {[Wiley, Ecological Society of America]},
	volume       = 81,
	number       = 3,
	pages        = {246--248},
	issn         = {00129623, 23276096},
	url          = {http://www.jstor.org/stable/20168454},
	urldate      = {2025-06-01}
}

@article{Arambepola_Lalendra_Wimalasena_Munasinghe_2025,
	title        = {From conventional methods to large language models: A systematic review of techniques in mobile app review analysis},
	author       = {Arambepola, Nimasha and Lalendra Wimalasena, Waruni and Munasinghe, Lankeshwara},
	year         = 2025,
	journal      = {Interdisciplinary Journal of Information, Knowledge, and Management},
	volume       = 20,
	pages        = {016},
	doi          = {10.28945/5491},
	issn         = {1555-1229, 1555-1237},
	rights       = {https://creativecommons.org/licenses/by-nc/4.0/},
	language     = {en}
}

@article{developer_response_safwat,
author = {Hassan, Safwat and Tantithamthavorn, Chakkrit and Bezemer, Cor-Paul and Hassan, Ahmed E.},
title = {Studying the dialogue between users and developers of free apps in the Google Play Store},
year = {2018},
issue_date = {Jun 2018},
publisher = {Kluwer Academic Publishers},
address = {USA},
volume = {23},
number = {3},
issn = {1382-3256},
url = {https://doi.org/10.1007/s10664-017-9538-9},
doi = {10.1007/s10664-017-9538-9},
journal = {Empirical Softw. Engg.},
month = jun,
pages = {1275–1312},
numpages = {38},

}

@incollection{wilcoxon1992,
  author    = {Wilcoxon, Frank},
  title     = {Individual comparisons by ranking methods},
  booktitle = {Breakthroughs in Statistics},
  editor    = {Kotz, Samuel and Johnson, Norman L.},
  series    = {Springer Series in Statistics},
  year      = {1992},
  publisher = {Springer},
  address   = {New York, NY},
  pages     = {196--202},
  doi       = {10.1007/978-1-4612-4380-9_16},
  url       = {https://doi.org/10.1007/978-1-4612-4380-9_16}
}

@misc{appbot2026,
  author       = {{Appbot}},
  title        = {Appbot: App Review and Ratings Tool},
  year         = {2026},
  howpublished = {\url{https://https://appbot.co/}},
  note         = {Accessed: February 2026}
}

@article{Alabduljabbar_2024, title={User-centric AI: evaluating the usability of generative AI applications through user reviews on app stores}, volume={10}, rights={https://creativecommons.org/licenses/by/4.0/}, ISSN={2376-5992}, DOI={10.7717/peerj-cs.2421}, journal={PeerJ Computer Science}, author={Alabduljabbar, Reham}, year={2024}, month=oct, pages={e2421}, language={en} }

@article{MENG2026104560,
title = {How review sentiment influences ratings in Generative AI applications: Insights from VADER and LDA analysis},
journal = {Journal of Retailing and Consumer Services},
volume = {88},
pages = {104560},
year = {2026},
issn = {0969-6989},
doi = {https://doi.org/10.1016/j.jretconser.2025.104560},
url = {https://www.sciencedirect.com/science/article/pii/S096969892500339X},
author = {Yuebin Meng and Sangchul Park and Geumchul Um},

}

@article{Massenon_Gambo_Khan_Agbonkhese_Alwadain_2025, title={"My AI is Lying to Me": User-reported LLM hallucinations in AI mobile apps reviews}, volume={15}, ISSN={2045-2322}, DOI={10.1038/s41598-025-15416-8},
author={Massenon, Rhodes and Gambo, Ishaya and Khan, Javed Ali and Agbonkhese, Christopher and Alwadain, Ayed}, year={2025}, month=aug, pages={30397}, language={en}, journal = {Scientific Reports} }

@misc{greene2014,
      title={How Many Topics? Stability Analysis for Topic Models}, 
      author={Derek Greene and Derek O'Callaghan and Pádraig Cunningham},
      year={2014},
      eprint={1404.4606},
      archivePrefix={arXiv},
      primaryClass={cs.LG},
      url={https://arxiv.org/abs/1404.4606}, 
}

@article{Malfacini_2025, title={The impacts of companion AI on human relationships: risks, benefits, and design considerations}, volume={40}, ISSN={0951-5666, 1435-5655}, DOI={10.1007/s00146-025-02318-6}, number={7}, journal={AI \& SOCIETY}, author={Malfacini, Kim}, year={2025}, month=oct, pages={5527–5540}, language={en} }

@Article{Haque2023,
author="Haque, M D Romael
and Rubya, Sabirat",
title="An Overview of Chatbot-Based Mobile Mental Health Apps: Insights From App Description and User Reviews",
journal="JMIR Mhealth Uhealth",
year="2023",
month="May",
day="22",
volume="11",
pages="e44838",
issn="2291-5222",
doi="10.2196/44838",
url="https://mhealth.jmir.org/2023/1/e44838",
url="https://doi.org/10.2196/44838",
url="http://www.ncbi.nlm.nih.gov/pubmed/37213181"
}

@ARTICLE{Storm-Mathisen,
    
AUTHOR={Storm-Mathisen, Ardis  and Giæver, Tonje Hilde  and Helle-Valle, Jo  and Engen, Bård Ketil  and Karstensen, Steinar },
           
TITLE={Engagement with AI in teacher education—discourses, processes of domestication and dynamics of transformation},
          
JOURNAL={Frontiers in Education},
          
VOLUME={Volume 10 - 2025},
  
YEAR={2026},
  
URL={https://www.frontiersin.org/journals/education/articles/10.3389/feduc.2025.1694082},
  
DOI={10.3389/feduc.2025.1694082},
  
ISSN={2504-284X}}

@article{Heuser2024,
author = {Morten Heuser and Julie Vulpius},
title ={‘Grandma, tell that story about how to make napalm again’: Exploring early adopters’ collaborative domestication of generative AI},

journal = {Convergence},
volume = {32},
number = {1},
pages = {87-110},
year = {2026},
doi = {10.1177/13548565241285742},

URL = { 
    
        https://doi.org/10.1177/13548565241285742
    
    

},
eprint = { 
    
        https://doi.org/10.1177/13548565241285742
    
    

}
}

@article{Xie2026,
author = {Xinyi Xie},
title = {Utility Expansion and Subjectification: The Mechanisms of Creative Domestication in Intimate Relationships with Generative AI},
journal = {International Journal of Human–Computer Interaction},
volume = {0},
number = {0},
pages = {1--16},
year = {2026},
publisher = {Taylor \& Francis},
doi = {10.1080/10447318.2026.2668625},


URL = { 
    
        https://doi.org/10.1080/10447318.2026.2668625
    
    

},
eprint = { 
    
        https://doi.org/10.1080/10447318.2026.2668625
    
    

}

}

@ARTICLE{Huang2025,
    
AUTHOR={Huang, Liyao  and Zou, Wenxue  and Huang, Yanghao },
           
TITLE={“He is my savior, my guiding light in the dark”: imagination and domestication in Chinese women’s romantic relationships with AI companions},
          
JOURNAL={Frontiers in Psychology},
          
VOLUME={Volume 16 - 2025},
  
YEAR={2025},
  
URL={https://www.frontiersin.org/journals/psychology/articles/10.3389/fpsyg.2025.1571707},
  
DOI={10.3389/fpsyg.2025.1571707},
  
ISSN={1664-1078}}

\clearpage
\appendix
\label{sec:appendix}

\section{Gen-AI App Categories and Keywords}
\label{app:keywords}
\FloatBarrier
We derive our keyword set from the Gen-AI taxonomy proposed by Gozalo-Brizuela and Garrido-Merchán~\cite{gozalobrizuela2023surveygenerativeaiapplications}. 
We exclude categories that are not relevant to mobile applications (i.e., \textit{biotech}, \textit{brain}, \textit{drug discovery}, and \textit{other}), and adapt the remaining categories into keyword-based queries for app store search. Table~\ref{tab:app_keywords_appendix} presents the resulting categories and keywords used in our data collection process. Because some keywords (e.g., \textit{``Image Editing''} and
\textit{``Business''}) are too broad when used independently, we append
\textit{``Generative AI''} to each keyword when querying the Google Play Store.

\begin{table}[!htbp]
\caption{Categories and keywords for app search in the Google Play Store}
\begin{tabular}{p{3cm} p{12cm}}
\toprule
\textbf{Category} & \textbf{Keywords} \\
\midrule
Text & Conversational AI, Text-to-Science, Text-to-Author Simulation, Text-to-Medical Advice, Text-to-Itinerary, Doc-to-Text \\
\hline
Image & Image Editing, Artistic Images, Realistic Images \\
\hline
Video & Text-to-Video, Video Production, Image-to-Video \\
\hline
3D & Text-to-3D \\
\hline
Code and Software & Text-to-Code, Text-to-Multilingual Code \\
\hline
Speech & Text-to-Speech, AI Understanding, Speech-to-Text \\
\hline
Business & Business \\
\hline
Marketing & Marketing, Advertisement, Accounting \\
\hline
Gaming & Video game Creation \\
\hline
Music & Music Generator \\
\bottomrule
\end{tabular}
\label{tab:app_keywords_appendix}
\end{table}

\clearpage
\FloatBarrier

\section{Prompts}
\label{app:prompts}

This section presents the prompts used for the three LLM-assisted components
of our framework: filtering non-informative reviews, extracting topics, and
assigning the extracted topics to user reviews.

\subsection{Filtering Prompt}
\label{app:filter_prompt}
\begin{figure}[!htbp]
  \centering
  \includegraphics[width=0.9\textwidth]{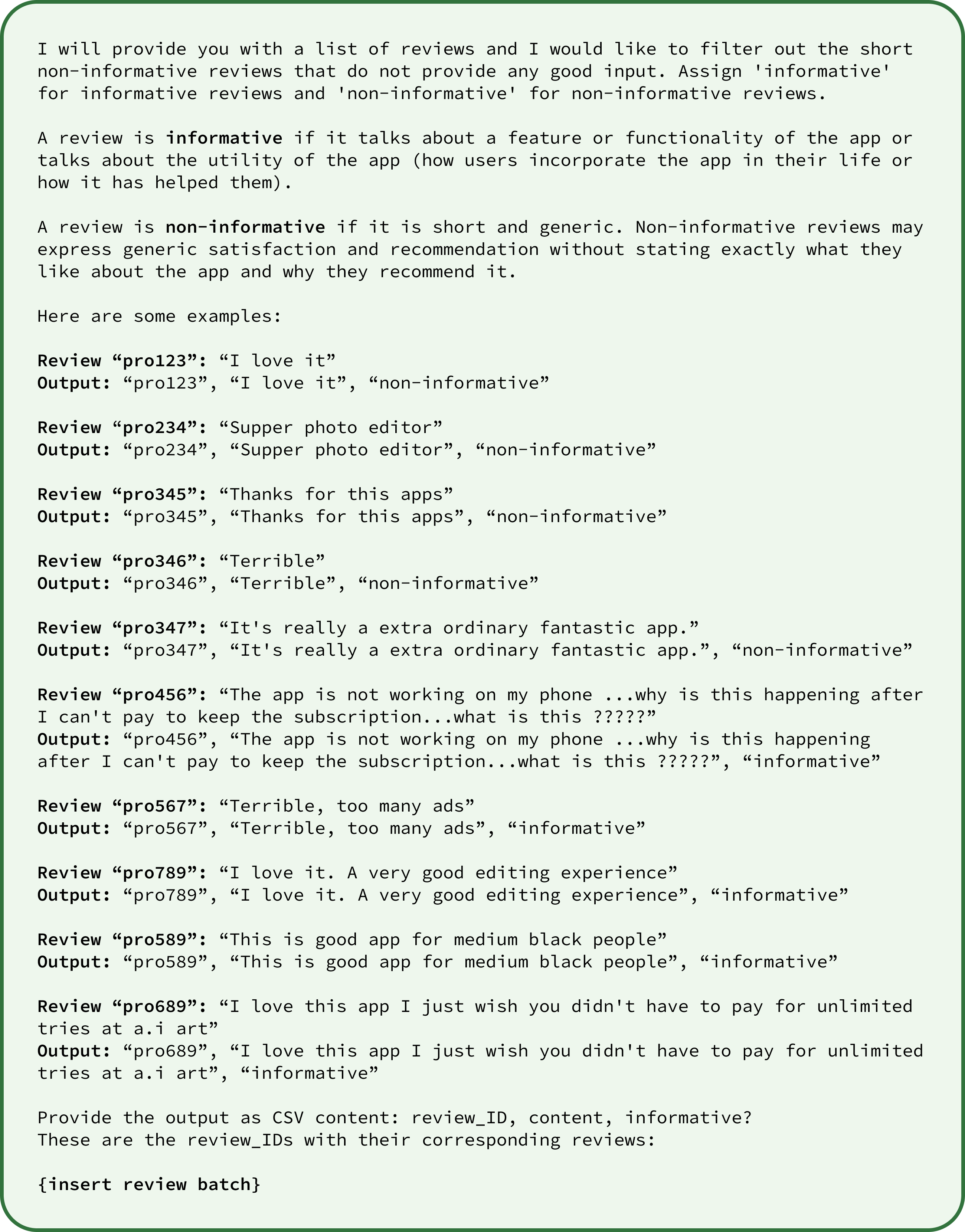} 
  \caption{Template of the filtering prompt used to filter out non-informative reviews.}
  \label{fig:filter_prompt}
\end{figure}

Figure~\ref{fig:filter_prompt} presents the prompt used to classify reviews as
informative or non-informative. The prompt contains explicit classification
instructions, representative examples, and a constrained output format to
support consistent filtering.

\FloatBarrier

\subsection{Topic Extraction Prompt}
\label{app:topic_extraction_prompt}

\begin{figure}[!htbp]
  \centering
  \includegraphics[width=\textwidth]{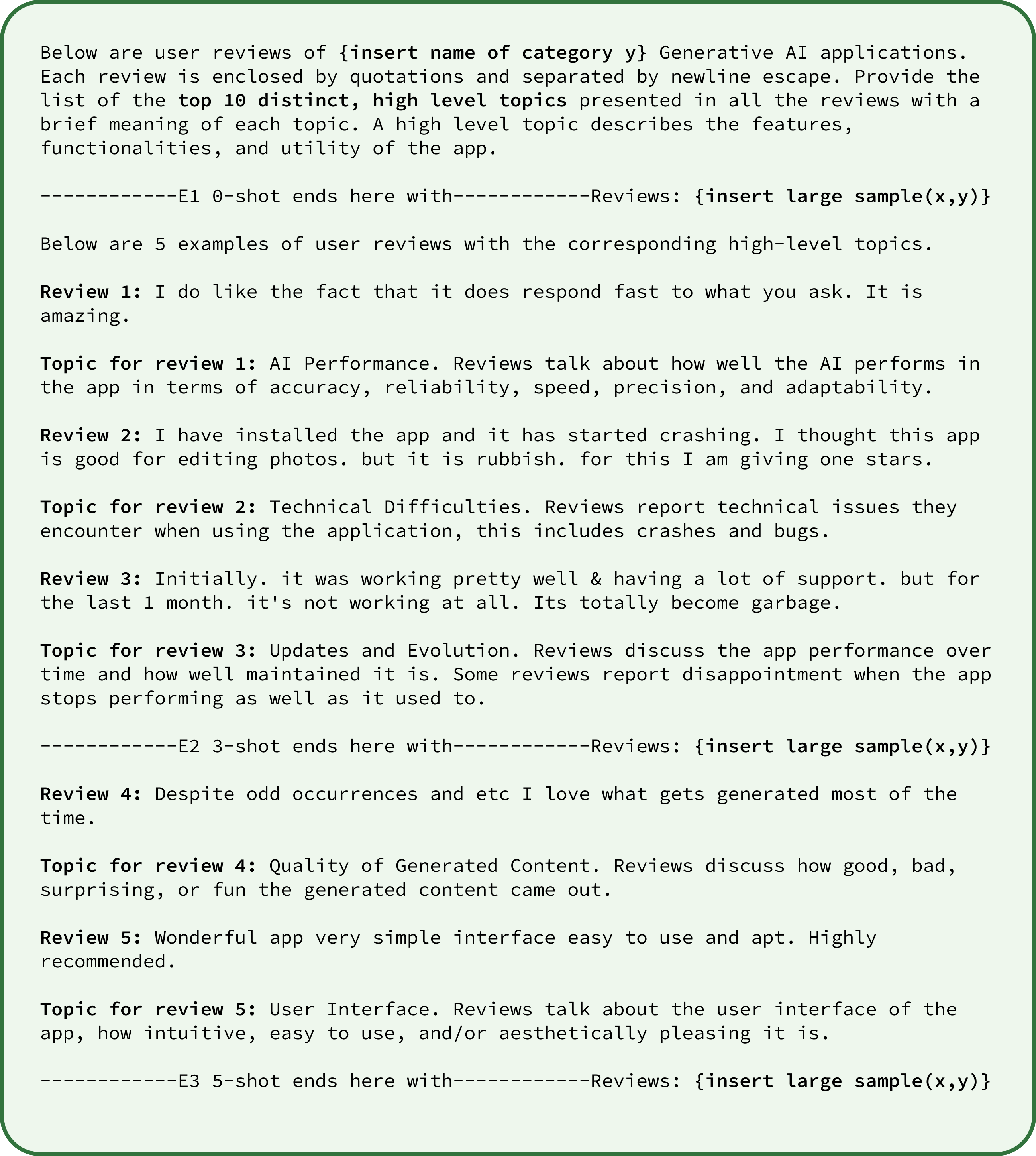}
\caption{Template of the \textit{topic extraction prompts for \textit{(E1: 0-shot, E2: 3-shot, and E3: 5-shot)}} used to extract the top topics from a large sample of stage~$x$ and app category~$y$. “Insert category name” is where we insert the name of the app category~$y$. “Insert sample of reviews” is where we list our large sample of reviews, separated by a new line escape.}
\label{fig:topic_extraction_prompt}
\end{figure}

Figure~\ref{fig:topic_extraction_prompt} presents the prompt used to extract
the most prominent topics from each review sample. The same prompt structure is
used across the 0-shot, 3-shot, and 5-shot configurations, with the number of
few-shot examples varying across experiments.

\FloatBarrier

\subsection{Topic Assignment Prompt}
\label{app:topic_assignment_prompt}
\begin{figure}[!htbp]
\centering
\includegraphics[width=\textwidth]{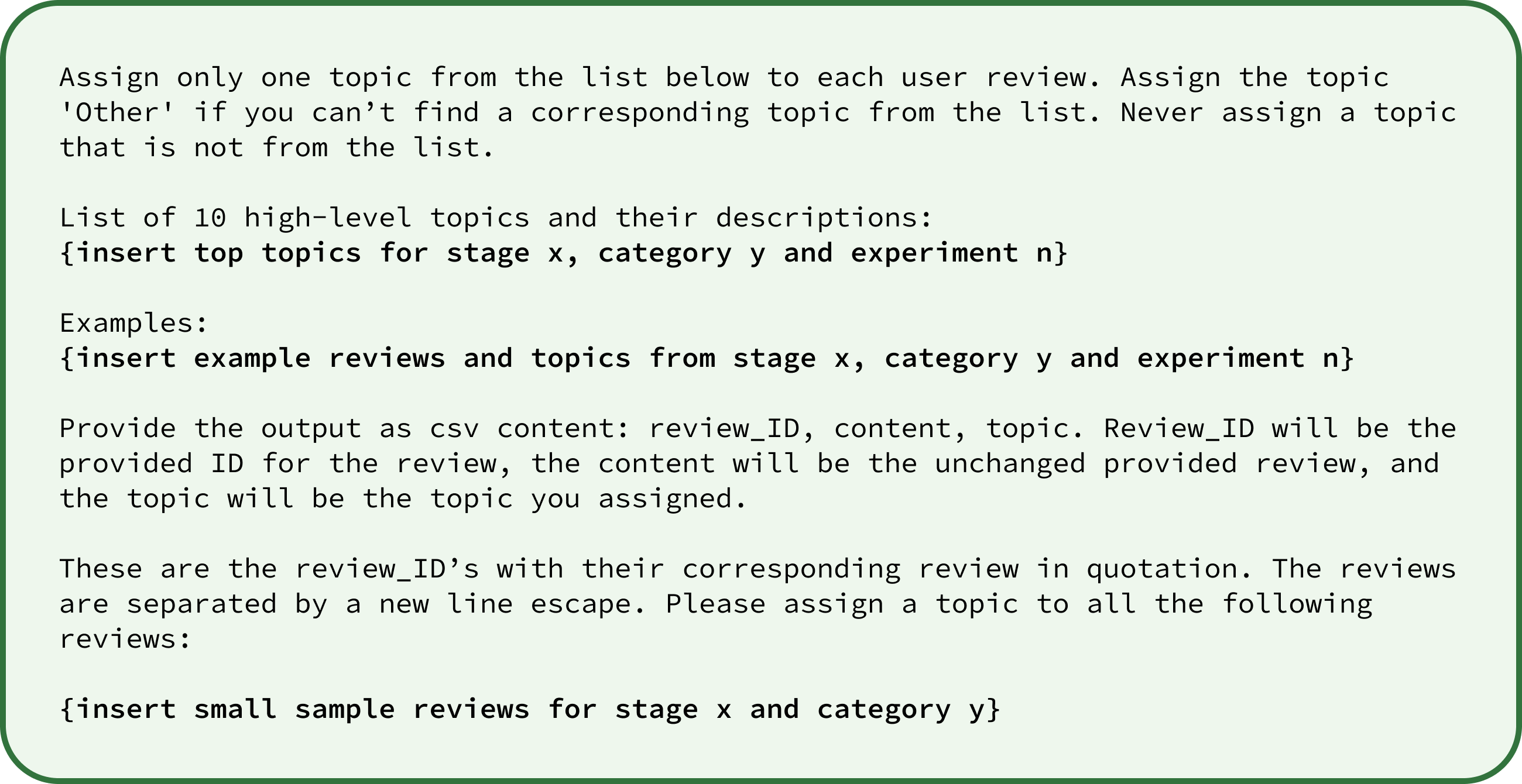}
\caption{Template of the \textit{P5: topic assignment} prompt used to assign topics to a list of reviews. We insert the set of top topics \( T(x, y, n) \), five few-shot examples, and the small sample \( S_{\text{small}}(x, y) \).}
\label{fig:topic_assignment_prompt}
\end{figure}

Figure~\ref{fig:topic_assignment_prompt} presents the prompt used to assign an
extracted topic to each review. The prompt provides the category-specific topic
set, five labeled examples, and the target reviews to be classified.

\FloatBarrier
\section{Topic Categories}
\label{app:topic_categories}
This section provides additional descriptive statistics for the topic
categories identified in RQ2. Table~\ref{tab:gen_AI_summary} summarizes the
Gen-AI topic categories, while Table~\ref{tab:non_gen_AI_summary} summarizes
the non-Gen-AI topic categories, including their review counts, average ratings,
and coverage across apps and app categories.
\begin{table}[!htbp]
\centering
\caption{Summary of Gen-AI topic categories, including the Gen-AI Reviews Count (GRC), the average Gen-AI rating (AGR), the number of apps, and the number of app categories in which each topic category appears (Full dataset considered).}
\label{tab:gen_AI_summary}
\begin{tabular}
{>{\raggedright\arraybackslash}p{5.5cm} c c c c}

\toprule
\textbf{Gen-AI Topic Category} & \textbf{GRC} & \textbf{AGR} & \textbf{\# Apps} & \textbf{\# App Categories} \\
\midrule
AI Performance                         & 158,873 (15\%) & 4.2 & 142 & 9 \\
Utility \& Use Cases                   & 118,405 (11\%) & \textbf{4.8} & 58  & 4 \\
Content Quality                        & 113,714 (11\%) & 3.7 & 142 & 8 \\
Creative Potential                     & 87,383 (8\%) & \textbf{4.8} & 96  & 5 \\
Content Policy \& Censorship           & 22,371 (2\%)  & 2.4 & 25  & 4 \\
Features \& Functionality              & 20,200 (2\%)  & 3.6 & 34  & 4 \\
Emotional Connection & 14,652 (1\%)  & \textbf{4.8} & 9   & 1 \\
Comparison to Other Apps               & 1,225 (0.1\%)   & 4.3 & 17  & 2 \\
\bottomrule
\end{tabular}
\end{table}

\begin{table}[!htbp]
\centering
\caption{Summary of non-Gen-AI topic categories, including the non-Gen-AI Review Count (NRC), the average non-Gen-AI rating (ANR), the number of apps, and the number of app categories in which each topic category appears (Full dataset considered).}
\begin{tabularx}{\textwidth}{Xcccc}

\toprule
\textbf{Non-Gen-AI Topic Category} & \textbf{NRC} & \textbf{ANR} & \textbf{\# Apps} & \textbf{\# App Categories} \\
\midrule
Monetization Methods \& Structure & 149,255 (14\%) & 2.2 & 160 & 9 \\
Technical Difficulties            & 126,377 (12\%) & 2.1 & 167 & 9 \\
Other                             & 111,297 (11\%) & 3.7 & 162 & 9 \\
User Interface \& Experience      & 58,885 (6\%)  & 4.3 & 145 & 9 \\
Updates \& Evolution              & 35,489 (3\%)  & 2.6 & 101 & 8 \\
Customer Support                  & 9,752 (1\%)  & 2.1 & 116 & 6 \\
Features \& Functionality         & 2,387 (0.2\%)   & 4.3 & 12  & 2 \\
\bottomrule
\end{tabularx}
\label{tab:non_gen_AI_summary}
\end{table}

\end{document}